\documentclass[12pt]{article}

\usepackage{jheppub, bm, wrapfig,float,array}
\usepackage[utf8]{inputenc}
\numberwithin{equation}{section}
\renewcommand\afterTocRuleSpace{\bigskip}

\usepackage{amsfonts}
\usepackage{amsmath}
\usepackage{color}
\usepackage{graphicx}
\usepackage{float}
\usepackage[T1]{fontenc}
\usepackage[utf8]{inputenc}
\usepackage{alphabeta}
\usepackage{fancyvrb}
\usepackage[dvipsnames]{xcolor}
\usepackage{bold-extra}
\usepackage{lmodern}

\usepackage{tikz}
\usetikzlibrary{arrows}
\usetikzlibrary{shapes.geometric,calc,arrows, positioning,shapes.misc,decorations.markings}
\tikzset{
  big arrow/.style={
    decoration={markings,mark=at position 1 with {\arrow[scale=2,#1]{>}}},
    postaction={decorate},
    shorten >=0.4pt},
  big arrow/.default=black}

\newcommand{\bea}{\begin{eqnarray}}
\newcommand{\eea}{\end{eqnarray}}
\newcommand{\be}{\begin{equation}}
\newcommand{\ee}{\end{equation}}
\newcommand{\bit}{\begin{itemize}}
\newcommand{\eit}{\end{itemize}}
\newcommand{\ben}{\begin{enumerate}}
\newcommand{\een}{\end{enumerate}}

\renewcommand{\ni}{\noindent}

\newcommand{\half}{\frac{1}{2}}

\newcommand{\Z}{{\mathbb Z}}

\renewcommand{\P}{{\mathbb P}}

\newcommand{\cF}{\mathcal{F}}

\newcommand{\cI}{\mathcal{I}}

\newcommand{\cN}{\mathcal{N}}
\newcommand{\cO}{\mathcal{O}}

\newcommand{\fe}{\mathfrak{e}}
\newcommand{\ff}{\mathfrak{f}}
\newcommand{\fg}{\mathfrak{g}}

\newcommand{\su}{\mathfrak{su}}
\renewcommand{\sp}{\mathfrak{sp}}
\newcommand{\so}{\mathfrak{so}}
\renewcommand{\u}{\mathfrak{u}}

\newcommand{\bF}{{\mathbb F}}
\newcommand{\dP}{\mathbf{dP}}

\title{On the classification of $5d$ SCFTs}

\author{Lakshya Bhardwaj}

\affiliation{Department of Physics, Harvard University\\
17 Oxford St, Cambridge, MA 02138, USA}

\abstract{We determine all $5d$ SCFTs upto rank three by studying RG flows of $5d$ KK theories. Our analysis reveals the existence of new rank one and rank two $5d$ SCFTs not captured by previous classifications. In addition to that, we provide for the first time a systematic and conjecturally complete classification of rank three $5d$ SCFTs. Our methods are based on a recently studied geometric description of $5d$ KK theories, thus demonstrating the utility of these geometric descriptions. It is straightforward, though computationally intensive, to extend this work and systematically classify $5d$ SCFTs of higher ranks (greater than or equal to four) by using the geometric description of $5d$ KK theories.
}

\begin{document}

\maketitle

\section{Introduction and Conclusions} \label{intro}
Since the successful classification of $6d$ SCFTs \cite{Heckman:2013pva,Heckman:2015bfa,Bhardwaj:2015xxa,Bhardwaj:2019hhd}, there has been considerable interest in classifying $5d$ SCFTs \cite{DelZotto:2017pti,Xie:2017pfl,Jefferson:2017ahm,Jefferson:2018irk,Bhardwaj:2018yhy,Bhardwaj:2018vuu,Apruzzi:2018nre,Apruzzi:2019vpe,Apruzzi:2019opn,Apruzzi:2019enx,Bhardwaj:2019fzv} (see also \cite{Closset:2018bjz}). In this regard, an interesting conjecture was made \cite{Jefferson:2018irk} by observing that there seems to be an upper bound on the number of matter hypermultiplets \cite{Jefferson:2017ahm} that can be carried by a supersymmetric $5d$ gauge theory for it to have a UV completion. At the tip of the bound, the UV completion is a $6d$ SCFT, and below the bound, the UV completion is a $5d$ SCFT. Based on this observation, it was conjectured that it should be possible to obtain all $5d$ SCFTs by systematically integrating out BPS particles from $6d$ SCFTs compactified on a circle\footnote{Notice that the way this conjecture has been phrased, it not only applies to $5d$ SCFTs having an effective gauge theory description, but also to $5d$ SCFTs not having such a gauge theory description. An example of such a $5d$ SCFT is the theory ``$\su(2)$ with minus one number of fundamental hypers'' which can be obtained by compactifying M-theory on a local $\P^2$.}.

This conjecture was tested successfully in a geometric context in \cite{Jefferson:2018irk}. There a classification of shrinkable smooth local Calabi-Yau threefolds was performed such that compactifying M-theory on such a threefold would give rise to a $5d$ SCFT of \emph{rank less than or equal to two}. The RG flows associated with integrating out BPS particles translate to geometric operations involving flops and blowdowns on this threefold. It was then shown that all such Calabi-Yau threefolds can be obtained from a handful of ``parent'' Calabi-Yau threefolds via flops and blowdowns. A parent threefold is not shrinkable and, in fact, compactifying M-theory on it produces a $5d$ KK theory, which is another name for a $6d$ SCFT compactified on a circle possibly with a twist by a discrete global symmetry around the circle.

Motivated by the successful test of this conjecture, a number of recent works \cite{Bhardwaj:2018yhy,Bhardwaj:2018vuu,Bhardwaj:2019fzv} (see also \cite{Apruzzi:2018nre,Apruzzi:2019vpe,Apruzzi:2019opn,Apruzzi:2019enx}) undertook the task of determining the Calabi-Yau threefold\footnote{For generic KK theories, it is a smooth threefold. For some exceptional KK theories, the Calabi-Yau threefold may not be smooth and/or the compactification of M-theory on the threefold might not be completely geometric. See \cite{Bhardwaj:2019fzv} for more details. In this paper, we will use the word ``geometry'' without distinguishing whether the Calabi-Yau threefold is smooth or singular, and whether the compactification requires extra non-geometric ingredients or not. In the cases where extra non-geometric ingredients are involved, the effect of such ingredients can be captured in the difference between the set of generators of the Mori cone (i.e. the set of holomorphic curves) of the threefold and the set of fundamental BPS particles in the resulting $5d$ theory.} associated to each $5d$ KK theory. According to the conjecture, these Calabi-Yau threefolds act as parent threefolds for the ``descendant'' Calabi-Yau threefolds associated to $5d$ SCFTs, where the descendant threefolds can be determined from the parent threefolds by performing sequences of flops and blowdowns on the parent threefolds. The main goal of this paper is to explicitly carry out such a procedure to determine the Calabi-Yau threefolds associated to all $5d$ SCFTs of \emph{rank less than or equal to three}, thus extending the results of \cite{Jefferson:2018irk}.

This work provides for the first time a systematic and conjecturally complete classification of $5d$ SCFTs of rank three. The contents of this paper can also be viewed as an illustration of the general procedure by which one can extract the identities of all $5d$ SCFTs starting from the results of \cite{Bhardwaj:2018yhy,Bhardwaj:2018vuu,Bhardwaj:2019fzv}. In principle, there is no problem in extending the methods used in this work to obtain the classification of $5d$ SCFTs for any arbitrary rank. However, this task becomes increasingly complex in a computational sense as the rank is increased.

Our approach can be termed as ``top-down'', distinguishing it from the ``bottom-up'' approach of \cite{Jefferson:2018irk} where $5d$ SCFTs were determined by building shrinkable Calabi-Yau threefolds in a bottom-up fashion. Our top-down approach instead starts from the Calabi-Yau threefolds associated to $5d$ KK theories, upon which flops and blowdowns are performed to reach Calabi-Yau threefolds associated to $5d$ SCFTs. Using this top-down approach, we revisit the classification of rank one and rank two $5d$ SCFTs which was already undertaken in \cite{Jefferson:2018irk} using the bottom-up approach. The top-down approach uncovers the existence of a few new rank one and rank two $5d$ SCFTs not accounted in \cite{Jefferson:2018irk}. These $5d$ SCFTs are (\ref{r1new}), (\ref{su2TL}), (\ref{su2TL2}), (\ref{r2new}) and (\ref{su1su1T3L}).

\section{Rank one}\label{R1}
Notice that integrating out matter hypermultiplets from a $5d$ gauge theory does not change the rank of the theory. This generalizes to the fact that integrating out BPS particles from a $5d$ theory does not change its rank. Thus the KK theories relevant to the classification of rank one $5d$ SCFTs themselves have rank one. So the starting point to the classification of rank one $5d$ SCFTs is the classification of rank one $5d$ KK theories.

The rank of a $5d$ KK theory can be determined from the tensor branch description of the associated $6d$ SCFT. Recall that a $6d$ SCFT is described on its tensor branch by a $6d$ gauge theory interacting with a collection of tensor multiplets. If the $6d$ SCFT is compactified on a circle without any twist, then both the $6d$ vector multiplets and $6d$ tensor multiplets descend to $5d$ vector multiplets and hence the rank $r$ of the resulting $5d$ KK theory can be written as
\be
r=t+g
\ee
where $t$ is the number of tensor multiplets arising on the tensor branch and $g$ is the rank of $6d$ gauge algebra arising on the tensor branch. 

The possible twists of $6d$ SCFTs were studied in \cite{Bhardwaj:2019fzv}. The description of the most general twist can be found in Section 2 of \cite{Bhardwaj:2019fzv}. The most general twist can be described as a permutation $S$ of tensor multiplets combined with an outer automorphism $\cO$ of the $6d$ gauge algebra such that this combination $S\cO$ is a discrete symmetry of the corresponding $6d$ SCFT. When the $6d$ SCFT is compactified on a circle with the twist $S\cO$, then the different $5d$ vector multiplets arising from the reduction of $6d$ tensor multiplets are identified with each other according to the action of $S$, and the $5d$ vector multiplets arising from the reduction of $6d$ vector multiplets are identified according to the action of $\cO$. Thus, the rank $r$ of the resulting $5d$ KK theory can be written as
\be\label{rank}
r=T+G
\ee
where $T$ is the number of orbits of the permutation $S$ and $G$ is the rank of algebra left invariant by $\cO$.

Since every $6d$ SCFT has at least one tensor multiplet, a rank one $5d$ KK theory can only arise from $6d$ SCFTs which do not carry any $6d$ gauge algebra on their tensor branch. All the rank one $5d$ KK theories can be determined to be
\be\label{sp0}
\begin{tikzpicture}
\node at (-0.5,0.4) {1};
\node at (-0.45,0.9) {$\sp(0)^{(1)}$};
\end{tikzpicture}
\ee
\be\label{su1}
\begin{tikzpicture}
\node at (-0.5,0.4) {2};
\node at (-0.45,0.9) {$\su(1)^{(1)}$};
\end{tikzpicture}
\ee
\be\label{su1T}
\begin{tikzpicture}
\node (v1) at (-0.5,0.4) {2};
\node at (-0.45,0.9) {$\su(1)^{(1)}$};
\draw (v1) .. controls (-1.5,-0.5) and (0.5,-0.5) .. (v1);
\end{tikzpicture}
\ee
where we have used the graphical notation for $5d$ KK theories developed in Section 2 of \cite{Bhardwaj:2019fzv}:
\bit
\item (\ref{sp0}) denotes the KK theory arising from the untwisted compactification of the $6d$ SCFT arising from an empty $-1$ curve\footnote{In a field theoretic language, we say that there is a fundamental BPS string in the theory whose self Dirac pairing is $+1$. The string arises from a D3 brane wrapping the $-1$ curve, and the self-intersection of the curve is identified with negative of the self Dirac pairing of the string.} in F-theory, commonly known as the E-string theory. The label $1$ denotes that $-1$ curve is used and the label $\sp(0)$ denotes that the gauge algebra living on the $-1$ curve is trivial. The superscript $(1)$ in the label $\sp(0)^{(1)}$ denotes that there is no gauge algebra outer automorphism involved, which is obvious in this case since a trivial gauge algebra cannot have any outer automorphisms. 
\item Similarly, (\ref{su1}) denotes the KK theory arising from the untwisted compactification of the $6d$ SCFT arising from an empty $-2$ curve in F-theory, commonly known as the $A_1$ $(2,0)$ theory. Here $\su(1)$ denotes that the gauge algebra is trivial and the superscript again denotes the non-existence of an outer automorphism. 
\item (\ref{su1T}) denotes the KK theory arising from a twisted compactification of the $6d$ SCFT
\be\label{A2}
\begin{tikzpicture}
\node (v1) at (-0.5,0.45) {2};
\node at (-0.45,0.9) {$\su(1)$};
\begin{scope}[shift={(1.5,0)}]
\node (v2) at (-0.5,0.45) {2};
\node at (-0.45,0.9) {$\su(1)$};
\end{scope}
\draw  (v1) edge (v2);
\end{tikzpicture}
\ee
arising from two $-2$ curves intersecting each other at one point\footnote{The fact that the two curves intersect at one point translates to the fact that the two fundamental BPS strings arising from these two curves have a mutual Dirac pairing of $-1$.}, such that both of the curves carry empty gauge algebra. This $6d$ SCFT is commonly known as the $A_2$ $(2,0)$ theory. The discrete symmetry associated to the twist permutes the two tensor multiplets arising from the two $-2$ curves. The graph (\ref{su1T}) is simply a folding of the graph (\ref{A2}) induced by the exchange of the two nodes in (\ref{A2}). The loop in (\ref{su1T}) is the image of the edge in (\ref{A2}) after the folding. The superscript on $\su(1)$ in (\ref{su1T}) again denotes that there is no outer automorphism.
\eit

It is convenient to organize the KK theories by their number of mass parameters. Each time a BPS particle is integrated out, the number of mass parameters decreases by one. So, in a sense, the most number of RG flows are produced by KK theories with the most number of mass parameters. Quite often, the RG flows from theories containing less number of mass parameters lead to the same $5d$ SCFTs obtained via RG flows from KK theories with more number of mass parameters. The number of mass parameters $M$ for a KK theory can be written as
\be\label{M}
M=F+1
\ee
where $F$ is the rank of the subalgebra of the flavor symmetry algebra of the associated $6d$ SCFT left invariant by the twist. The extra mass parameter not included in $F$ is given by the inverse of the radius of compactification $R$.

It is well-known that the E-string theory has an $\fe_8$ flavor symmetry, so the number of mass parameters for the KK theory (\ref{sp0}) is
\be
M=9
\ee
The flavor symmetry for $A_1$ $(2,0)$ theory is $\su(2)$ giving
\be
M=2
\ee
for the KK theory (\ref{su1}). Finally, the flavor symmetry for $A_2$ $(2,0)$ theory is $\su(2)$. This symmetry is carried by a non-compact curve intersecting one of the two $-2$ curves. Thus it is not possible to preserve the full $\su(2)$ after the twist, but a $\u(1)$ subalgebra can be preserved. The F-theory configuration involves two non-compact curves each carrying an $I_1$ singularity and each intersecting a different $-2$ curve. This configuration is invariant under the discrete symmetry interchanging the two $-2$ curves. Thus, we have
\be
M=2
\ee
for the KK theory (\ref{su1T}). 

\ni Now we move onto a study of RG flows of these three KK theories.

\subsection{$M=9$}\label{R1M9}
The Calabi-Yau threefold associated to (\ref{sp0}) is \cite{Jefferson:2018irk,Bhardwaj:2018yhy,Bhardwaj:2018vuu,Bhardwaj:2019fzv} a local neighborhood of the surface
\be\label{sp0KK}
\begin{tikzpicture} [scale=1.9]
\node (v2) at (-2.2,-0.5) {$\bF^{8}_{1}$};
\end{tikzpicture}
\ee
which denotes an eight-point blowup of the Hirzebruch surface $\bF_1$. See Appendix A of \cite{Bhardwaj:2019fzv} for a quick review on Hirzebruch and del Pezzo surfaces. 

The RG flows of (\ref{sp0}) are captured by the blowdowns of $-1$ curves in the surface $\bF_1^8$ above. The key point is that a $-1$ curve in a geometry can always be written as a blowup on a surface in some isomorphism frame of the geometry. For example, except the blowups $x_i$ ($i=1,\cdots,8$), some of the other $-1$ curves in (\ref{sp0KK}) are $f-x_i$ and $e$. To see that a curve $f-x_i$ (for a fixed $i$) can be written as a blowup, one can perform the isomorphism $\bF_1^8$ to $\bF_0^8$ given by
\begin{align}
e&\to e-x_i \label{F1F0s}\\
f-x_i&\to x_i\\
x_i&\to f-x_i\\
x_j&\to x_j~~~~~~~~\text{for $j\neq i$} \label{F1F0e}
\end{align}
This isomorphism leads to an equivalent geometry
\be\label{sp0KK2}
\begin{tikzpicture} [scale=1.9]
\node (v2) at (-2.2,-0.5) {$\bF^{8}_{0}$};
\end{tikzpicture}
\ee
describing the KK theory (\ref{sp0}). Now if one performs an isomorphism $\bF_0^8\to \bF_1^8$ given by
\begin{align}
e-x_j&\to e \label{F0F1s}\\
f-x_j&\to x_j\\
x_j&\to f-x_j\\
x_k&\to x_k~~~~~~~~\text{for $k\neq j$}\label{F0F1e}
\end{align}
with $j\neq i$, then the geometry is again described by (\ref{sp0KK}). The combination of these two isomorphisms is an automorphism of $\bF_1^8$ that maps a curve $f-x_i$ to a blowup. Similarly, the curve $e$ in $\bF_1^8$ can be written as a blowup by using the isomorphism\footnote{In this paper we denote the del Pezzo surface obtained by blowing $n$ points on $\P^2$ as $\mathbf{dP}^n$ rather than $\mathbf{dP}_n$. The surface $\P^2$ is displayed as $\mathbf{dP}$ without any superscript.} $\bF_1^8\to \dP^9$ given by
\begin{align}
e&\to x_9\label{F1dPs}\\
f&\to l-x_9\\
x_i&\to x_i\label{F1dPe}
\end{align}
leading to an equivalent geometric description
\be\label{sp0KK3}
\begin{tikzpicture} [scale=1.9]
\node (v2) at (-2.2,-0.5) {$\dP^{9}$};
\end{tikzpicture}
\ee
of the KK theory (\ref{sp0}). Combining it with another isomorphism $\dP^9\to \bF_1^8$ given by
\begin{align}
l-x_1&\to f\label{dPF1s}\\
x_1&\to e\\
x_{i+1}&\to x_i\label{dPF1e}
\end{align}
for $1\le i\le 8$ leads to an automorphism on $\bF_1^8$ which converts the curve $e$ to a blowup. One can check that all the other $-1$ curves in $\bF_1^8$ can be converted to a blowup of $\bF_1^8$ by an automorphism generated by combining the isomorphisms discussed above along with the automorphism $\bF_0^8\to\bF_0^8$
\begin{align}
e&\to f\label{F0F0s}\\
f&\to e\\
x_{i}&\to x_i\label{F0F0e}
\end{align}
that simply interchanges $e$ and $f$.

Thus we see that, at the first step, all the RG flows are equivalent to a single RG flow corresponding to the removal of a blowup from (\ref{sp0KK}), leading to a $5d$ SCFT described by a local neighborhood of the surface
\be
\begin{tikzpicture} [scale=1.9]
\node (v2) at (-2.2,-0.5) {$\bF^{7}_{1}$};
\end{tikzpicture}
\ee
In a similar fashion, at the next few steps, the only RG flows are the ones corresponding to removal of more blowups. Thus, at the next few steps, we obtain $5d$ SCFTs described by geometries
\be
\begin{tikzpicture} [scale=1.9]
\node (v2) at (-2.2,-0.5) {$\bF^{8-m}_{1}$};
\end{tikzpicture}
\ee
where $1\le m\le 7$. 

The geometry at $m=7$ has two inequivalent blowdowns. One of them corresponds to blowing down the only blowup $x$ on $\bF_1^1$ and the other one corresponds to blowing down the curve $f-x$. Notice that, unlike the cases with more blowups, there is no automorphism of $\bF_1^1$ which transforms $f-x$ to $x$. Such an automorphism requires the existence of at least one more blowup. Blowing down $x$ leads to a $5d$ SCFT described by the geometry
\be\label{F1}
\begin{tikzpicture} [scale=1.9]
\node (v2) at (-2.2,-0.5) {$\bF_{1}$};
\end{tikzpicture}
\ee
To blow down $f-x$, we can first write it as the blowup in $\bF_0^1$ by using an isomorphism discussed above, and then remove this blowup leading to the geometry
\be\label{F0}
\begin{tikzpicture} [scale=1.9]
\node (v2) at (-2.2,-0.5) {$\bF_{0}$};
\end{tikzpicture}
\ee
which describes a $5d$ SCFT distinct from the $5d$ SCFT described by (\ref{F1}).

Now, the geometry (\ref{F0}) has no $-1$ curves, but the geometry (\ref{F1}) has a $-1$ curve which is the $e$ curve. To blow this down, we can write the $e$ curve as the blowup in $\dP^1$ and then remove this blowup to obtain the geometry described by the surface
\be\label{dP}
\begin{tikzpicture} [scale=1.9]
\node (v2) at (-2.2,-0.5) {$\dP$};
\end{tikzpicture}
\ee
where $\dP$ is our notation for the surface $\P^2$.

Thus $5d$ SCFTs descending from the $5d$ KK theory (\ref{sp0}) are described by the geometries
\be\label{sp0m}
\begin{tikzpicture} [scale=1.9]
\node (v2) at (-2.2,-0.5) {$\mathbf{\bF^{8-m}_{1}}$};
\draw  (-2.6,-0.2) rectangle (-1.8,-0.8);
\end{tikzpicture}
\ee
with $1\le m\le8$,
\be\label{sp08'}
\begin{tikzpicture} [scale=1.9]
\node (v2) at (-2.2,-0.5) {$\mathbf{\bF_{0}}$};
\draw  (-2.5,-0.3) rectangle (-1.9,-0.7);
\end{tikzpicture}
\ee
and
\be\label{sp09}
\begin{tikzpicture} [scale=1.9]
\node (v2) at (-2.2,-0.5) {$\mathbf{\dP}$};
\draw  (-2.5,-0.3) rectangle (-1.9,-0.7);
\end{tikzpicture}
\ee
We put a box around a geometry when it describes a $5d$ SCFT not equivalent to any of the $5d$ SCFTs discussed earlier in the paper. We hope that this will lead to an easy identification of all the inequivalent $5d$ SCFTs, since each different box describes a different $5d$ SCFT. The number of mass parameters for the family of $5d$ SCFTs (\ref{sp0m}) is
\be
M=9-m
\ee
while the number of mass parameters for the $5d$ SCFT described by (\ref{sp08'}) is
\be
M=1
\ee
and for the one described by (\ref{sp09}) is
\be
M=0
\ee

The prepotential\footnote{In this paper, we will ignore the contributions to the prepotential involving mass parameters.} $\cF$ for a rank one $5d$ theory is a single term of the form $x\phi^3$ where $\phi$ is the Coulomb branch parameter. $x$ can be computed \cite{Jefferson:2018irk,Bhardwaj:2018yhy,Bhardwaj:2018vuu,Bhardwaj:2019fzv} by counting the number of blowups on the compact surface in the geometric description of the $5d$ theory. If the surface is a Hirzebruch surface $\bF_n^b$ then $6\cF=(8-b)\phi^3$, and if it is a del Pezzo surface $\dP^b$ then $6\cF=(9-b)\phi^3$. Thus the prepotential for the $5d$ SCFT described by (\ref{sp0m}) is
\be
6\cF=m\phi^3
\ee
The prepotential for the $5d$ SCFT described by (\ref{sp08'}) is
\be
6\cF=8\phi^3
\ee
and the prepotential for the $5d$ SCFT described by (\ref{sp09}) is
\be
6\cF=9\phi^3
\ee

\subsection{$M=2$}\label{R1M2}
The KK theory (\ref{su1}) is described by the geometry \cite{Bhardwaj:2018yhy,Bhardwaj:2018vuu,Bhardwaj:2019fzv}
\be\label{su16d}
\begin{tikzpicture} [scale=1.9]
\node (v2) at (-2.2,-0.5) {$\mathbf{\bF^{1+1}_{0}}$};
\draw (v2) .. controls (-1.4,0) and (-1.4,-1) .. (v2);
\node at (-1.9,-0.2) {\scriptsize{$e$-$x$}};
\node at (-1.9,-0.8) {\scriptsize{$e$-$y$}};
\end{tikzpicture}
\ee
where the loop denotes a self-gluing of $\bF_0^2$. The labels at the ends of the loop denote that the curve $e-x$ is glued to the curve $e-y$ where $x$ and $y$ are the two blowups. Using the automorphism (\ref{F0F0s}--\ref{F0F0e}) that interchanges $e$ and $f$ in $\bF_0$, we can write (\ref{su16d}) as
\be\label{su1KK2}
\begin{tikzpicture} [scale=1.9]
\node (v2) at (-2.2,-0.5) {$\mathbf{\bF^{1+1}_{0}}$};
\draw (v2) .. controls (-1.4,0) and (-1.4,-1) .. (v2);
\node at (-1.9,-0.2) {\scriptsize{$f$-$x$}};
\node at (-1.9,-0.8) {\scriptsize{$f$-$y$}};
\end{tikzpicture}
\ee
Now using the isomorphism (\ref{F0F1s}--\ref{F0F1e}) with blowup $x$, we can write the above geometry as
\be
\begin{tikzpicture} [scale=1.9]
\node (v2) at (-2.2,-0.5) {$\mathbf{\bF^{1+1}_{1}}$};
\draw (v2) .. controls (-1.4,0) and (-1.4,-1) .. (v2);
\node at (-1.9,-0.2) {\scriptsize{$x$}};
\node at (-1.9,-0.8) {\scriptsize{$f$-$y$}};
\end{tikzpicture}
\ee
Finally using the isomorphism (\ref{F1F0s}--\ref{F1F0e}) with the blowup $y$, we can write the geometry associated to the KK theory (\ref{su1}) in the isomorphism frame
\be\label{su1KK}
\begin{tikzpicture} [scale=1.9]
\node (v2) at (-2.2,-0.5) {$\mathbf{\bF^{1+1}_{0}}$};
\draw (v2) .. controls (-1.4,0) and (-1.4,-1) .. (v2);
\node at (-1.9,-0.2) {\scriptsize{$x$}};
\node at (-1.9,-0.8) {\scriptsize{$y$}};
\end{tikzpicture}
\ee
It can be shown that all of the $-1$ curves in the above geometry are equivalent either to the blowup $x$ or to the curve $f-x$ by using automorphisms composed out of the isomorphisms (\ref{F1F0s}--\ref{F1F0e}), (\ref{F0F1s}--\ref{F0F1e}), (\ref{F1dPs}--\ref{F1dPe}), (\ref{dPF1s}--\ref{dPF1e}) and (\ref{F0F0s}--\ref{F0F0e}).

Let us first consider the blowdown of $f-x$, which can be written as the blowup $x$ in the isomorphism frame given by (\ref{su1KK2}). Since $f-x$ and $f-y$ are glued to each other in (\ref{su1KK2}), their volumes must be same, implying that the volumes of $x$ and $y$ must be same. So, blowing down $x$ in (\ref{su1KK2}) blows down $y$ along with it. But $x$ and $y$ intersect the gluing curves $f-x$ and $f-y$, and hence their blowdown continues into a flop transition creating two new blowups $x'$ and $y'$. Since $x$ intersects the gluing curve $f-x$ at one point, the flop of $x$ transforms the gluing curves $f-x$ to $(f-x)+x=f$ and the gluing curve $f-y$ is transformed to $f-y-x'$. Thus at this intermediate step, $f-x'-y$ is glued to $f$. Now, the flop of $y$ subsequently transforms the gluing curve $f-y-x'$ to $f-x'$ and the gluing curve $f$ to $f-y'$. The geometry after the flop is
\be
\begin{tikzpicture} [scale=1.9]
\node (v2) at (-2.2,-0.5) {$\mathbf{\bF^{1+1}_{0}}$};
\draw (v2) .. controls (-1.4,0) and (-1.4,-1) .. (v2);
\node at (-1.8,-0.1) {\scriptsize{$f$-$x'$}};
\node at (-1.8,-0.9) {\scriptsize{$f$-$y'$}};
\end{tikzpicture}
\ee
which is identical to (\ref{su1KK2}).

Now, we turn our attention to the blowdown of $x$ in (\ref{su1KK}). Since $x$ is glued to $y$, both $x$ and $y$ are blown down together. At the end of this blowdown, the self-gluing is removed since the gluing curves participating in the self-gluing have been blown down. The resulting geometry is
\be
\begin{tikzpicture} [scale=1.9]
\node (v2) at (-2.2,-0.5) {$\bF_{0}$};
\end{tikzpicture}
\ee
and describes a $5d$ SCFT already discovered while studying RG flows of the KK theory (\ref{sp0}) in the last subsection. See the discussion around (\ref{F0}). There are no more remaining $-1$ curves, and hence no new $5d$ SCFTs are found by studying the RG flows of (\ref{su1}).

There is another rank one KK theory with $M=2$, which is (\ref{su1T}). This KK theory is described by the geometry \cite{Bhardwaj:2019fzv}
\be\label{su1TKK}
\begin{tikzpicture} [scale=1.9]
\node (v2) at (-2.2,-0.5) {$\mathbf{\bF^{1+1}_{1}}$};
\draw (v2) .. controls (-1.4,0) and (-1.4,-1) .. (v2);
\node at (-1.9,-0.2) {\scriptsize{$x$}};
\node at (-1.9,-0.8) {\scriptsize{$y$}};
\end{tikzpicture}
\ee
The different $-1$ curves in this geometry are all equivalent to $x$, $f-x$ and $e$. The blowdown of $f-x$ is a flop transition giving back the same geometry as above. The blowdown of $x$ removes the self-gluing and produces the geometry (\ref{F1}) already discovered.

To blow down the $e$ curve we use the isomorphism (\ref{F1dPs}--\ref{F1dPe}) to write (\ref{su1TKK}) as
\be
\begin{tikzpicture} [scale=1.9]
\node (v2) at (-2.2,-0.5) {$\mathbf{\dP^{1+1+1}}$};
\draw (v2) .. controls (-1.4,0) and (-1.4,-1) .. (v2);
\node at (-1.9,-0.2) {\scriptsize{$x$}};
\node at (-1.9,-0.8) {\scriptsize{$y$}};
\end{tikzpicture}
\ee
such that one of the blowups does not participate in the self-gluing. The blowdown of $e$ curve in (\ref{su1TKK}) corresponds to blowdown of this blowup in the above geometry. Carrying out the blowdown produces the geometry
\be\label{r1new}
\begin{tikzpicture} [scale=1.9]
\node (v2) at (-2.2,-0.5) {$\mathbf{\dP^{1+1}}$};
\draw (v2) .. controls (-1.4,0) and (-1.4,-1) .. (v2);
\node at (-1.9,-0.2) {\scriptsize{$x$}};
\node at (-1.9,-0.8) {\scriptsize{$y$}};
\draw  (-2.7,0) rectangle (-1.3,-1);
\end{tikzpicture}
\ee
which is a new rank one $5d$ SCFT not discussed in the literature before\footnote{It is also possible to see the existence of this $5d$ SCFT by using a brane construction for the KK theory (\ref{su1T}). We thank Hee-Cheol Kim for a private discussion on this point.}. The only possible flow now removes the self-gluing from (\ref{r1new}) giving rise to the $5d$ SCFT described by (\ref{dP}) discovered earlier. For a Hirzebruch surface $\bF_n^{b+2s}$ with $2s$ blowups out of $b+2s$ blowups participating in $s$ number of self-gluings, the prepotential is $6\cF=(8-8s-b)\phi^3$. For a del Pezzo surface $\dP^{b+2s}$ with $2s$ blowups out of $b+2s$ blowups participating in $s$ number of self-gluings, the prepotential is $6\cF=(9-8s-b)\phi^3$.

Let us now discuss some aspects of the new $5d$ SCFT described by the geometry (\ref{r1new}). This SCFT has number of mass parameters
\be
M=1
\ee
and prepotential
\be
6\cF=\phi^3
\ee
It can be informally thought of as an ``$\su(2)$ gauge theory with minus one hypers in fundamental and one hyper in adjoint'' since (\ref{su1TKK}) describes a $5d$ $\su(2)$ gauge theory with one hyper in adjoint, from which we have removed a $-1$ curve which \emph{formally} corresponds to the removal of a fundamental. From this point of view, the existence of this $5d$ SCFT makes sense: 

\ni It is known that a $5d$ $\cN=1$ $\su(2)$ pure gauge theory with $\theta=\pi$ is a $5d$ SCFT described by geometry (\ref{F1}). There is an RG flow from this $5d$ SCFT to another $5d$ SCFT described by geometry (\ref{dP}). The RG flow corresponds formally to the removal of a fundamental and hence the resulting $5d$ SCFT described by (\ref{dP}) is often called ``$\su(2)$ gauge theory with minus one hypers in fundamental''. Now, it is known that $5d$ $\cN=1$ $\su(2)$ gauge theory with an adjoint hyper and $\theta=\pi$ is the $5d$ KK theory (\ref{su1T}). Thus, it must be possible to formally remove a fundamental from this theory and flow to a $5d$ SCFT. This is precisely the $5d$ SCFT described by (\ref{r1new}).

A subtle aspect of the geometry (\ref{r1new}) is that it is non-shrinkable in the sense defined in \cite{Jefferson:2018irk}. The curve $l-x-y$ is a generator of the Mori cone and has negative volume on the Coulomb branch. This is problematic since in a traditional M-theory compactification, M2 brane wrapping a generator of the Mori cone gives rise to a fundamental\footnote{We define a fundamental BPS particle to be a BPS particle that cannot arise as a bound state of other BPS particles.} BPS particle. The fact that $l-x-y$ has negative volume would imply that the corresponding BPS particle has negative mass on the Coulomb branch of the $5d$ theory.

However, as discussed in detail in \cite{Bhardwaj:2019fzv}, not every generator of the Mori cone of (\ref{su1TKK}) leads to a fundamental BPS particle. This is probably due to the effect some non-geometric ingredient in the M-theory compactification. This situation is analogous to the situation in the six-dimensional compactifications in the frozen phase of F-theory \cite{Bhardwaj:2018jgp} where, unlike the traditional non-frozen six-dimensional compactifications of F-theory, not every generator of the Mori cone of the base of the threefold used to compactify F-theory leads to a fundamental BPS string (via the wrapping of a D3 brane on it).

The Mori cone of (\ref{su1TKK}) is generated by\footnote{Only the homology class of the curves in the full threefold is recorded in the definition of Mori cone. Thus, $f-x$ and $f-y$ are equal in the Mori cone since $x$ is identified with $y$ due to self-gluing of the surface.} $e$, $h-x-y$, $f-x=f-y$ and $x=y$. However, the set of fundamental BPS particles is proposed to be identified instead with $e$, $2h-x-2y$, $f-x=f-y$, and $x=y$. From this, we can deduce the set of fundamental BPS particles for (\ref{r1new}) should be identified with $2l-x-2y$, $l-x=l-y$, and $x=y$. It can be easily checked that all of these curves have non-negative volume on the Coulomb branch (without turning on mass parameters), and hence the Coulomb branch physics is consistent, thus resolving the puzzle raised above.

Although we have discussed how the dictionary between geometry and physics should be modified for this slightly non-traditional M-theory compactification, we have not explored the precise physical mechanism behind this effect. It is plausible that it is related to frozen singularities and/or some discrete fluxes in M-theory \cite{Witten:1997bs,deBoer:2001wca,Tachikawa:2015wka}. We leave a more detailed study to future work.

\section{Rank two}\label{R2}
The relationship (\ref{rank}) implies that for $r=2$ we have the following possibilities:
\bit
\item $T=1,G=1$
\item $T=2,G=0$
\eit
In the class $T=G=1$, we have the following $5d$ KK theories:
\bit
\item \be\label{sp1}
\begin{tikzpicture}
\node at (-0.5,0.4) {1};
\node at (-0.45,0.9) {$\sp(1)^{(1)}$};
\end{tikzpicture}
\ee
which describes the untwisted compactification of the $6d$ SCFT carrying $\sp(1)$ gauge algebra on a $-1$ curve. The $6d$ theory carries $10$ hypers in fundamental of $\sp(1)$. Hence there is a rank ten flavor symmetry implying that
\be
M=11
\ee
is the number of mass parameters carried by the KK theory (\ref{sp1}). See the discusion around (\ref{M}).
\item \be\label{su2}
\begin{tikzpicture}
\node at (-0.5,0.4) {2};
\node at (-0.45,0.9) {$\su(2)^{(1)}$};
\end{tikzpicture}
\ee
which describes the untwisted compactification of the $6d$ SCFT carrying $\su(2)$ gauge algebra on a $-2$ curve. Even though the theory carries four hypers of $\su(2)$, it is known that the rank of flavor symmetry algebra\footnote{We suspect that this is because a $\u(1)$ subalgebra of the naive $\so(8)$ flavor symmetry algebra is anomalous with the anomaly proportional to the $\Z_2$ valued $6d$ theta angle for $\su(2)$. This dovetails nicely with the fact that even when all four fundamentals of $\su(2)$ are gauged by an $\su(4)$, the theta angle of $\su(2)$ is still physically irrelevant. An anomaly of the above form will explain the absence of theta angle. We thank Gabi Zafrir for a useful discussion on this point. It would be interesting to verify whether this suspicion is correct.} is three \cite{Heckman:2015bfa,Ohmori:2015pia}, thus
\be
M=4
\ee
for the KK theory (\ref{su2}).
\item \be\label{su3T}
\begin{tikzpicture}
\node at (-0.5,0.4) {$k$};
\node at (-0.45,0.9) {$\su(3)^{(2)}$};
\end{tikzpicture}
\ee
for $1\le k \le 3$, which describes a twisted compactification of the $6d$ SCFT carrying $\su(3)$ on $-k$ curve. The kind of twist is depicted by the superscript $(2)$ in the label $\su(3)^{(2)}$ which denotes that an outer automorphism of order 2 is acting on the $\su(3)$ gauge algebra as one goes around the circle. The invariant subalgebra of $\su(3)$ under the outer automorphism is $\sp(1)$ which implies that indeed $G=1$ for (\ref{su3T}). The $6d$ SCFT has $18-6k$ hypers in fundamental of $\su(3)$, which are exchanged with each other in pairs \cite{Bhardwaj:2019fzv} under the outer automorphism. This means that after the reduction we obtain $9-3k$ hypers in fundamental of $\sp(1)$ and hence
\be
M=10-3k
\ee
for (\ref{su3T}).
\item \be\label{su2T}
\begin{tikzpicture}
\node (v1) at (-0.5,0.4) {2};
\node at (-0.45,0.9) {$\su(2)^{(1)}$};
\draw (v1) .. controls (-1.5,-0.5) and (0.5,-0.5) .. (v1);
\end{tikzpicture}
\ee
which denotes the KK theory obtained by compactifying the $6d$ SCFT
\be
\begin{tikzpicture}
\node (v1) at (-0.5,0.45) {2};
\node at (-0.45,0.9) {$\su(2)$};
\begin{scope}[shift={(1.5,0)}]
\node (v2) at (-0.5,0.45) {2};
\node at (-0.45,0.9) {$\su(2)$};
\end{scope}
\draw  (v1) edge (v2);
\end{tikzpicture}
\ee
with an exchange of the two $-2$ curves (along with the $\su(2)$ gauge algebras living over them) as one goes around the circle. The matter spectrum of the $6d$ SCFT is a hyper in bifundamental plus two extra hypers in fundamental carried by each $\su(2)$. The bifundamental gives rise to a $\u(1)$ flavor symmetry and the extra fundamentals give rise to a $\su(2)\oplus\su(2)$ flavor symmetry. The discrete symmetry exchanging the two $\su(2)$ gauge algebras exchanges the two $\su(2)$ flavor symmetry algebra, while preserving the $\u(1)$ flavor symmetry. Thus, the KK theory (\ref{su2T}) has
\be
M=3
\ee
\eit
In the class $T=2,G=0$, we have the following KK theories:
\bit
\item \be\label{sp0su1}
\begin{tikzpicture}
\node (v1) at (-0.5,0.4) {1};
\node at (-0.45,0.9) {$\sp(0)^{(1)}$};
\begin{scope}[shift={(2,0)}]
\node (v2) at (-0.5,0.4) {2};
\node at (-0.45,0.9) {$\su(1)^{(1)}$};
\end{scope}
\draw  (v1) edge (v2);
\end{tikzpicture}
\ee
which denotes an untwisted compactification of the $6d$ SCFT arising from an empty $-1$ curve intersecting an empty $-2$ curve at a single point. This $6d$ SCFT is commonly known as the rank two E-string theory. It is known to have an $\fe_8\oplus\su(2)$ flavor symmetry. This can be understood from the facts observed in the last section that the theory arising from an empty $-1$ curve has $\fe_8$ symmetry and the theory arising from an empty $-2$ curve has $\su(2)$ flavor symmetry. Thus,
\be
M=10
\ee
for the KK theory (\ref{sp0su1}).
\item \be\label{su1su1}
\begin{tikzpicture}
\node (v1) at (-0.5,0.4) {2};
\node at (-0.45,0.9) {$\su(1)^{(1)}$};
\begin{scope}[shift={(2,0)}]
\node (v2) at (-0.5,0.4) {2};
\node at (-0.45,0.9) {$\su(1)^{(1)}$};
\end{scope}
\draw  (v1) edge (v2);
\end{tikzpicture}
\ee
which is the untwisted compactification of the $A_2$ $(2,0)$ theory (\ref{A2}). It is known to have an $\su(2)$ flavor symmetry and thus
\be
M=2
\ee
for (\ref{su1su1}).
\item \be\label{su1su1L}
\begin{tikzpicture}
\node (v1) at (-0.5,0.4) {$2$};
\node at (-0.45,0.9) {$\su(1)^{(1)}$};
\begin{scope}[shift={(2,0)}]
\node (v2) at (-0.5,0.4) {$2$};
\node at (-0.45,0.9) {$\su(1)^{(1)}$};
\end{scope}
\draw  (v1) -- (v2);
\draw (v1) .. controls (-1.5,-0.5) and (0.5,-0.5) .. (v1);
\end{tikzpicture}
\ee
which comes from the twisted compactification of the $A_4$ $(2,0)$ theory
\be
\begin{tikzpicture}
\node (v1) at (-0.5,0.45) {2};
\node at (-0.45,0.9) {$\su(1)$};
\begin{scope}[shift={(1.5,0)}]
\node (v2) at (-0.5,0.45) {2};
\node at (-0.45,0.9) {$\su(1)$};
\end{scope}
\begin{scope}[shift={(3,0)}]
\node (v3) at (-0.5,0.45) {2};
\node at (-0.45,0.9) {$\su(1)$};
\end{scope}
\begin{scope}[shift={(4.5,0)}]
\node (v4) at (-0.5,0.45) {2};
\node at (-0.45,0.9) {$\su(1)$};
\end{scope}
\draw  (v1) edge (v2);
\draw  (v3) edge (v2);
\draw  (v3) edge (v4);
\end{tikzpicture}
\ee
by the twist exchanging the two $-2$ curves at the two ends with each other while simultaneously exchanging the two $-2$ curves in the middle:
\be
\begin{tikzpicture}
\node (v1) at (-0.5,0.45) {2};
\node at (-0.45,0.9) {$\su(1)$};
\begin{scope}[shift={(1.5,0)}]
\node (v2) at (-0.5,0.45) {2};
\node at (-0.45,0.9) {$\su(1)$};
\end{scope}
\begin{scope}[shift={(-1.5,0)}]
\node (v0) at (-0.5,0.45) {2};
\node at (-0.45,0.9) {$\su(1)$};
\end{scope}
\begin{scope}[shift={(3,0)}]
\node (v3) at (-0.5,0.45) {2};
\node at (-0.45,0.9) {$\su(1)$};
\end{scope}
\draw  (v1) -- (v2);
\draw  (v0) edge (v1);
\draw  (v2) edge (v3);
\draw[<->,red,thick] (v0) .. controls (-1.5,-1.5) and (2,-1.5) .. (v3);
\draw[<->,red,thick] (v1) .. controls (-0.5,-0.5) and (1,-0.5) .. (v2);
\end{tikzpicture}
\ee
As for the KK theory (\ref{su1T}), a $\u(1)$ subalgebra of the $\su(2)$ flavor symmetry algebra of the $A_4$ $(2,0)$ theory can be preserved under this twist leading to
\be
M=2
\ee
\item \be\label{su1su1T2}
\begin{tikzpicture}
\node (v1) at (-0.5,0.4) {2};
\node at (-0.45,0.9) {$\su(1)^{(1)}$};
\begin{scope}[shift={(2,0)}]
\node (v2) at (-0.5,0.4) {2};
\node at (-0.45,0.9) {$\su(1)^{(1)}$};
\end{scope}
\node (v3) at (0.5,0.4) {\tiny{$2$}};
\draw  (v1) edge (v3);
\draw  [->](v3) -- (v2);
\end{tikzpicture}
\ee
This theory arises from the following twist of $A_3$ $(2,0)$ theory
\be\label{A3}
\begin{tikzpicture}
\node (v1) at (-0.5,0.45) {2};
\node at (-0.45,0.9) {$\su(1)$};
\begin{scope}[shift={(1.5,0)}]
\node (v2) at (-0.5,0.45) {2};
\node at (-0.45,0.9) {$\su(1)$};
\end{scope}
\begin{scope}[shift={(-1.5,0)}]
\node (v0) at (-0.5,0.45) {2};
\node at (-0.45,0.9) {$\su(1)$};
\end{scope}
\draw  (v1) -- (v2);
\draw  (v0) edge (v1);
\draw[<->,red,thick] (v0) .. controls (-1.5,-1) and (0.5,-1) .. (v2);
\end{tikzpicture}
\ee
and has
\be
M=2
\ee
Observe that the graph (\ref{su1su1T2}) is a folding of the graph associated to $A_3$ $(2,0)$ theory by the action (\ref{A3}). The tiny label $2$ in the middle of the directed edge in the graph (\ref{su1su1T2}) denotes that there are two directed edges.
\item Finally, we have
\be\label{su1su1T3}
\begin{tikzpicture}
\node (v1) at (-0.5,0.4) {2};
\node at (-0.45,0.9) {$\su(1)^{(1)}$};
\begin{scope}[shift={(2,0)}]
\node (v2) at (-0.5,0.4) {2};
\node at (-0.45,0.9) {$\su(1)^{(1)}$};
\end{scope}
\node (v3) at (0.5,0.4) {\tiny{$3$}};
\draw  (v1) edge (v3);
\draw  [->](v3) -- (v2);
\end{tikzpicture}
\ee
which arises by twisting the $D_4$ $(2,0)$ theory by the following action
\be\label{D4}
\begin{tikzpicture}
\node (v1) at (-0.5,0.45) {2};
\node (v4) at (-0.45,0.9) {$\su(1)$};
\begin{scope}[shift={(1.5,0)}]
\node (v2) at (-0.5,0.45) {2};
\node (v5) at (-0.45,0.9) {$\su(1)$};
\end{scope}
\begin{scope}[shift={(-1.5,0)}]
\node (v0) at (-0.5,0.45) {2};
\node (v6) at (-0.45,0.9) {$\su(1)$};
\end{scope}
\begin{scope}[shift={(0.05,1.7)}]
\node (v3) at (-0.5,0.45) {2};
\node at (-0.45,0.9) {$\su(1)$};
\end{scope}
\draw  (v1) -- (v2);
\draw  (v0) edge (v1);
\draw[->,red,thick] (v0) .. controls (-1.5,-1) and (0.5,-1) .. (v2);
\draw[->,red,thick] (v5) .. controls (1,1.7) and (0.4,2.1) .. (v3);
\draw[<-,red,thick] (v6) .. controls (-1.9,1.7) and (-1.4,2.1) .. (v3);
\draw  (v3) edge (v4);
\end{tikzpicture}
\ee
It can be shown in the F-theory setup that a $\u(1)$ flavor algebra can be preserved under this twist, and hence this KK theory has
\be
M=2
\ee
Again, observe that the graph (\ref{su1su1T3}) is a folding of the graph associated to $D_4$ $(2,0)$ theory by the action (\ref{D4}).
\eit
Now we move onto a study of RG flows of these KK theories.

\subsection{$M=11$}
According to \cite{Bhardwaj:2019fzv}, the KK theory (\ref{sp1}) has geometry given by
\be\label{sp16d}
\begin{tikzpicture} [scale=1.9]
\node (v1) at (-4.25,-0.5) {$\bF^{10}_1$};
\node (v2) at (-2.2,-0.5) {$\bF_0$};
\draw  (v1) edge (v2);
\node at (-3.7,-0.4) {\scriptsize{$2h$-$\sum x_i$}};
\node at (-2.6,-0.4) {\scriptsize{$2e$+$f$}};
\end{tikzpicture}
\ee
which denotes two surfaces $\bF^{10}_1$ and $\bF_0$ intersecting with each other. The intersection is described as a gluing of the two surfaces. A single edge between the two surfaces denotes that there is a single gluing. The labels at the end of the edge denote that the curve $2e+f$ in $\bF_0$ is glued to the curve $2h-\sum x_i$ (where the sum is over all the ten blowups) in $\bF_1^{10}$.

Using the isomorphism (\ref{F0F0s}--\ref{F0F0e}) on the right surface, we can rewrite the geometry (\ref{sp16d}) as
\be
\begin{tikzpicture} [scale=1.9]
\node (v1) at (-4.25,-0.5) {$\bF^{10}_1$};
\node (v2) at (-2.2,-0.5) {$\bF_0$};
\draw  (v1) edge (v2);
\node at (-3.7,-0.4) {\scriptsize{$2h$-$\sum x_i$}};
\node at (-2.6,-0.4) {\scriptsize{$e$+$2f$}};
\end{tikzpicture}
\ee
Notice that the gluing curve inside the right surface has changed appropriately. Now performing the isomorphism (\ref{F1F0s}--\ref{F1F0e}) on the left surface, we can rewrite the above geometry as
\be\label{g0}
\begin{tikzpicture} [scale=1.9]
\node (v1) at (-4.25,-0.5) {$\bF^{10}_1$};
\node (v2) at (-2.2,-0.5) {$\bF_0$};
\draw  (v1) edge (v2);
\node at (-3.6,-0.4) {\scriptsize{$2e$+$f$-$\sum x_i$}};
\node at (-2.6,-0.4) {\scriptsize{$e$+$2f$}};
\end{tikzpicture}
\ee
Interchanging $e$ and $f$ on the left surface and performing the isomorphism (\ref{F0F1s}--\ref{F0F1e}) on the left surface we obtain
\be
\begin{tikzpicture} [scale=1.9]
\node (v1) at (-4.25,-0.5) {$\bF^{9+1}_1$};
\node (v2) at (-2.2,-0.5) {$\bF_0$};
\draw  (v1) edge (v2);
\node at (-3.6,-0.4) {\scriptsize{$h$+$f$-$\sum x_i$}};
\node at (-2.6,-0.4) {\scriptsize{$e$+$2f$}};
\end{tikzpicture}
\ee
where we have divided the ten blowups on the left surface into a set of nine blowups denoted by $x_i$ and one blowup denoted by $y$. Performing (\ref{F1F0s}--\ref{F1F0e}) on the left surface using the blowup $x_9$, we obtain
\be
\begin{tikzpicture} [scale=1.9]
\node (v1) at (-4.25,-0.5) {$\bF^{8+2}_0$};
\node (v2) at (-2.2,-0.5) {$\bF_0$};
\draw  (v1) edge (v2);
\node at (-3.6,-0.4) {\scriptsize{$e$+$f$-$\sum x_i$}};
\node at (-2.6,-0.4) {\scriptsize{$e$+$2f$}};
\end{tikzpicture}
\ee
where the ten blowups on the left surface have been divided into two sets of eight and two blowups respectively, with the blowups in the first set denoted by $x_i$ and the blowups in second set denoted by $y_i$. Now applying (\ref{F0F1s}--\ref{F0F1e}), (\ref{F1F0s}--\ref{F1F0e}) and (\ref{F0F1s}--\ref{F0F1e}) in that sequence on the left surface we reach the geometry
\be\label{sp16d2}
\begin{tikzpicture} [scale=1.9]
\node (v1) at (-4.25,-0.5) {$\bF^{5+5}_1$};
\node (v2) at (-2.2,-0.5) {$\bF_0$};
\draw  (v1) edge (v2);
\node at (-3.7,-0.4) {\scriptsize{$e$-$\sum x_i$}};
\node at (-2.6,-0.4) {\scriptsize{$e$+$2f$}};
\end{tikzpicture}
\ee

Now we will apply isomorphisms which generalize the isomorphism (\ref{F0F1s}--\ref{F0F1e}). These isomorphisms take $\bF_n^1\to\bF_{n+1}^1$ and are given by\footnote{Notice that the blowup used in this isomorphism is a non-generic blowup for $n\ge1$, since the isomorphism involves the existence of the curve $e-x$ whose self-intersection is less than $-1$.}
\begin{align}
e-x&\to e\\
f-x&\to x\\
x&\to f-x\\
\end{align}
From now on, we will denote this isomorphism sending $\bF_n^1$ to $\bF_{n+1}^1$ by $\cI_n$. The isomorphism (\ref{F1F0s}--\ref{F1F0e}) can be noticed to equal $\cI_{0}^{-1}$.

Applying $\cI_1$ to the first surface of (\ref{sp16d2}) using one of the blowups appearing in the gluing curve $e-\sum x_i$, we obtain
\be
\begin{tikzpicture} [scale=1.9]
\node (v1) at (-4.25,-0.5) {$\bF^{4+6}_2$};
\node (v2) at (-2.2,-0.5) {$\bF_0$};
\draw  (v1) edge (v2);
\node at (-3.7,-0.4) {\scriptsize{$e$-$\sum x_i$}};
\node at (-2.6,-0.4) {\scriptsize{$e$+$2f$}};
\end{tikzpicture}
\ee
Let us successively applying $\cI_2,\cI_3,\cI_4,\cI_5$ in that order to the left surface of the above geometry every time choosing a blowup in the gluing curve living in the left surface. This leads us to our desired frame to represent the starting geometry (\ref{sp16d})
\be\label{sp1KK}
\begin{tikzpicture} [scale=1.9]
\node (v1) at (-4.25,-0.5) {$\bF^{10}_6$};
\node (v2) at (-2.2,-0.5) {$\bF_0$};
\draw  (v1) edge (v2);
\node at (-3.9,-0.4) {\scriptsize{$e$}};
\node at (-2.6,-0.4) {\scriptsize{$e$+$2f$}};
\end{tikzpicture}
\ee
All of the $-1$ curves in the above geometry are equivalent either to a blowup $x_i$ or to a curve of the form $f-x_i$ in the left surface. Blowing down some of the blowups, we obtain a series of $5d$ SCFTs described by
\be\label{sp1Lm}
\begin{tikzpicture} [scale=1.9]
\node (v1) at (-4.25,-0.5) {$\bF^{10-m}_6$};
\node (v2) at (-2.2,-0.5) {$\bF_0$};
\draw  (v1) edge (v2);
\node at (-3.8,-0.4) {\scriptsize{$e$}};
\node at (-2.6,-0.4) {\scriptsize{$e$+$2f$}};
\node at (-0.8,-0.5) {$1\le m\le10$};
\draw  (-4.7,-0.2) rectangle (-0.1,-1.6);
\node at (-2.4,-0.9) {$M=11-m$};
\node at (-2.4,-1.3) {$6\cF=(m-2)\phi_L^3+8\phi_R^3-18\phi_L\phi_R^2+12\phi_R\phi_L^2$};
\end{tikzpicture}
\ee
where from now on we will display the number of mass parameters $M$ and the prepotential $6\cF$ (in the phase described by the displayed geometry) of the $5d$ SCFT inside the box as well. Let us describe how the prepotential is computed. For a rank two theory, we can call the Coulomb branch parameter associated to the left surface as $\phi_L$ and the Coulomb branch parameter associated to the right surface as $\phi_R$. Then there are four terms appearing in the prepotential, namely $\phi_L^3$, $\phi_R^3$, $\phi_L^2\phi_R$ and $\phi_R^2\phi_L$. As discussed in Sections \ref{R1M9} and \ref{R1M2}, the coefficients of $\phi_L^3$ and $\phi_R^3$ in $6\cF$ are computed by counting blowups and self-gluings on the left and right surfaces respectively. The coefficient of $\phi_L\phi_R^2$ in $6\cF$ is computed by $3C_{L;R}^2$ where $C_{L;R}$ is the curve living in the left surface gluing it to the  right surface. In the geometry (\ref{sp1Lm}) $C_{L;R}$ is the $e$ curve in the left surface $\bF_6^{10-m}$. Similarly, the coefficient of $\phi_R\phi_L^2$ in $6\cF$ is computed by $3C_{R;L}^2$ where $C_{R;L}$ is the curve living in the right surface gluing it to the left surface. In the geometry (\ref{sp1Lm}) $C_{R;L}$ is the curve $e+2f$ in the right surface $\bF_0$.

The blow down of a curve of the form $f-x_i$ living in the left surface of (\ref{sp1KK}) generates a flop transition converting (\ref{sp1KK}) to
\be
\begin{tikzpicture} [scale=1.9]
\node (v1) at (-4.25,-0.5) {$\bF^{9}_5$};
\node (v2) at (-2.2,-0.5) {$\bF^1_0$};
\draw  (v1) edge (v2);
\node at (-3.9,-0.4) {\scriptsize{$e$}};
\node at (-2.7,-0.4) {\scriptsize{$e$+$2f$-$x$}};
\end{tikzpicture}
\ee
which can be rewritten by performing $\cI_1$ on the right surface as
\be\label{sp1KK2}
\begin{tikzpicture} [scale=1.9]
\node (v1) at (-4.25,-0.5) {$\bF^{9}_5$};
\node (v2) at (-2.2,-0.5) {$\bF^1_1$};
\draw  (v1) edge (v2);
\node at (-3.9,-0.4) {\scriptsize{$e$}};
\node at (-2.6,-0.4) {\scriptsize{$h$+$f$}};
\end{tikzpicture}
\ee
In going from (\ref{sp1KK}) to (\ref{sp1KK2}), we have effectively ``moved a blowup from the left surface to the right surface''. 

The $-1$ curves in (\ref{sp1KK2}) are all equivalent to either $x_i, f-x_i$ in the left surface or $x,f-x,e$ in the right surface. Blowing down $f-x$ in the right surface implements an inverse flop transition to the one we just discussed taking us back to (\ref{sp1KK}). Blowing down $m$ number of blowups in the left surface simply provides a different flop frame for the geometry (\ref{sp1Lm}), since after blowing down the $m$ blowups we could do the flop transition generated by $f-x$ in the second surface converting the geometry to (\ref{sp1Lm}). Blowing down $x$ in the second surface produces the geometry
\be
\begin{tikzpicture} [scale=1.9]
\node (v1) at (-4.25,-0.5) {$\bF^{9}_5$};
\node (v2) at (-2.2,-0.5) {$\bF_1$};
\draw  (v1) edge (v2);
\node at (-3.9,-0.4) {\scriptsize{$e$}};
\node at (-2.6,-0.4) {\scriptsize{$h$+$f$}};
\end{tikzpicture}
\ee
which is flop equivalent to the $m=1$ geometry in (\ref{sp1Lm}) as can be seen by moving all of the blowups from the left surface to the right surface and then exchanging the two surfaces. So the above geometry does not give rise to a new $5d$ SCFT not discovered earlier in this paper. 

However, blowing down $x$ in the right surface of (\ref{sp1KK2}) and then blowing down $m$ number of blowups in the left surface of (\ref{sp1KK2}), we discover a series of $5d$ SCFTs described by the geometries
\be\label{sp1LmR}
\begin{tikzpicture} [scale=1.9]
\node (v1) at (-4.25,-0.5) {$\bF^{9-m}_5$};
\node (v2) at (-2.2,-0.5) {$\bF_1$};
\draw  (v1) edge (v2);
\node at (-3.8,-0.4) {\scriptsize{$e$}};
\node at (-2.6,-0.4) {\scriptsize{$h$+$f$}};
\node at (-0.8,-0.5) {$1\le m\le9$};
\draw  (-4.7,-0.2) rectangle (-0.2,-1.6);
\node at (-2.4,-0.9) {$M=10-m$};
\node at (-2.4,-1.3) {$6\cF=(m-1)\phi_L^3+8\phi_R^3-15\phi_L\phi_R^2+9\phi_R\phi_L^2$};
\end{tikzpicture}
\ee
We can say that the SCFTs (\ref{sp1Lm}) are generated by integrating out $m$ blowups from the left and the SCFTs (\ref{sp1LmR}) are generated by integrating out one blowup from the right and $m$ blowups from the left. Similarly, now we can integrate out two blowups from the right and $m$ blowups from the left to obtain another series of $5d$ SCFTs described by
\be\label{sp1LmR2}
\begin{tikzpicture} [scale=1.9]
\node (v1) at (-4.25,-0.5) {$\bF^{8-m}_4$};
\node (v2) at (-2.2,-0.5) {$\bF_0$};
\draw  (v1) edge (v2);
\node at (-3.8,-0.4) {\scriptsize{$e$}};
\node at (-2.6,-0.4) {\scriptsize{$e$+$f$}};
\node at (-0.8,-0.5) {$2\le m\le8$};
\draw  (-4.7,-0.2) rectangle (-0.2,-1.6);
\node at (-2.3,-0.9) {$M=9-m$};
\node at (-2.3,-1.3) {$6\cF=m\phi_L^3+8\phi_R^3-12\phi_L\phi_R^2+6\phi_R\phi_L^2$};
\end{tikzpicture}
\ee
The lower bound on $m$ in (\ref{sp1LmR2}) is placed so that we do not overcount the same $5d$ SCFT. One can check that the geometry
\be
\begin{tikzpicture} [scale=1.9]
\node (v1) at (-4.25,-0.5) {$\bF^{7}_4$};
\node (v2) at (-2.2,-0.5) {$\bF_0$};
\draw  (v1) edge (v2);
\node at (-3.9,-0.4) {\scriptsize{$e$}};
\node at (-2.6,-0.4) {\scriptsize{$e$+$f$}};
\end{tikzpicture}
\ee
obtained by substituting $m=1$ in (\ref{sp1LmR2}) is flop equivalent to the geometry obtained by substituting $m=2$ in (\ref{sp1LmR}). The $5d$ SCFTs corresponding to (\ref{sp1LmR2}) have

Continuing in a similar fashion, we can integrate out $p$ blowups from the right before integrating out $m$ blowups from the left to obtain a series of $5d$ SCFTs given by geometries
\be\label{sp1LmRp}
\begin{tikzpicture} [scale=1.9]
\node (v1) at (-4.25,-0.5) {$\bF^{10-m-p}_{6-p}$};
\node (v2) at (-2.2,-0.5) {$\bF_{p-4}$};
\draw  (v1) edge (v2);
\node at (-3.7,-0.4) {\scriptsize{$e$}};
\node at (-2.6,-0.4) {\scriptsize{$e$}};
\node at (-0.8,-0.3) {$p\le m\le 10-p$};
\draw  (-5.3,0) rectangle (0.5,-1.7);
\node at (-0.8,-0.7) {$3\le p\le 5$};
\node at (-2.4,-1) {$M=11-m-p$};
\node at (-2.4,-1.4) {$6\cF=(m+p-2)\phi_L^3+8\phi_R^3+3(p-6)\phi_L\phi_R^2+3(4-p)\phi_R\phi_L^2$};
\end{tikzpicture}
\ee
The bounds on $m$ and $p$ in (\ref{sp1LmRp}) have been placed in such a fashion that we do not over-count flop equivalent geometries. For $p=3$, the right hand surface is $\bF_{-1}$ which is by definition isomorphic to the surface $\bF_1$ with the isomorphism $\bF_{-1}\to\bF_1$ given by
\begin{align}
e&\to h\\
h&\to e\\
f&\to f\\
\end{align}
For example, the gluing curve in the right surface for $p=3$ is the $e$ curve of $\bF_{-1}$ which is by definition the $h$ curve of $\bF_1$ and has self-intersection $+1$. We will continue to use the surface $\bF_{-1}$ throughout this paper.

Notice that we still haven't discussed the blowdown of the $e$ curve in (\ref{sp1KK2}). Its blow down leads to a flop transition resulting in the geometry
\be

\ee
which can also be written as
\be
%
\ee
Furthermore, removing $m$ blowups from the left surface, we obtain the geometry
\be
%
\ee
As long as $m\le9$, the above geometry is flop equivalent to (\ref{sp1Lm}). To see this, notice that, for $m\le9$, there is a blowup remaining on the left surface which can be moved onto the right surface to yield the geometry
\be\label{sp1Lm2}
%
\ee
which after application of $\cI_{0}^{-1}$ (on the right surface by using the blowup $x$) becomes
\be
%
\ee
Blowing down $x$ and interchanging $e$ and $f$ on the right surface leads us to the geometry (\ref{sp1Lm}). However, when $m=10$, then (\ref{sp1Lm2}) is not flop equivalent to (\ref{sp1Lm}), thus giving rise to a different $5d$ SCFT not accounted above
\be\label{sp1L9L'}
%
\ee
The $e$ curve in the right surface of (\ref{sp1L9L'}) can be blowndown to give rise to another $5d$ SCFT given by
\be\label{sp1L9L'R}
%
\ee
The above theory (\ref{sp1L9L'R}) can also be obtained by integrating out (towards left) the $e$ curve in second surface of (\ref{sp1LmR}) for $m=9$. For lower values of $m$, such an operation does not give rise to any new $5d$ SCFTs not discussed in the paper before.

Similarly, there is a blowdown of $e$ curve in the geometries having $p=3$ in (\ref{sp1LmRp}) that has not been addressed so far. These geometries are
\be
%
\ee
Blowing down the $e$ curve in the second surface leads to new $5d$ SCFTs described by geometries
\be\label{sp1LmR3Rt}
%
\ee
It can be checked that it is impossible to integrate out any blowups from the right in (\ref{sp1LmR3Rt}). For $m=3$, the above geometry (\ref{sp1LmR3Rt}) is
\be\label{sp1L3R3Rt}
%
\ee
which has a $-1$ curve not equivalent to either the blowups $x_i$ or to $f-x_i$. This curve is $h-\sum x_i$. To consider the blowdown of this curve let us apply $\cI_2^{-1},\cI_1^{-1},\cI_0^{-1},\cI_{0}$ in that order to the left surface every time choosing a blowup not appearing in the gluing curve. This rewrites the above geometry as
\be\label{sp1L3R3Rt2}
%
\ee
The curve $h-\sum x_i$ in the left surface of (\ref{sp1L3R3Rt}) becomes the $e$ curve in the left surface of (\ref{sp1L3R3Rt2}). Blowing it down, we obtain a $5d$ SCFT described by
\be\label{sp1L3LtR3Rt}
%
\ee
There are no further flows from this geometry. 

\subsection{$M=10$}
The geometry associated to the KK theory (\ref{sp0su1}) is \cite{Bhardwaj:2018vuu,Bhardwaj:2019fzv}
\be\label{sp0su16d}
%
\ee
where the two blowups on the right surface have been divided into two sets containing one blowup each. The blowup in the first set is called $x$ and the blowup in the set is called $y$. This is a notation we will follow throughout this paper. Whenever the blowups are divided into sets, we will denote the blowups in the first set by $x_i$, in the second set by $y_i$, in the third set by $z_i$, in the fourth set by $w_i$ etc.

Exchanging $e$ and $f$ on the second surface and using a blowup on the first surface to convert $\dP^9$ into $\bF_1^8$, we obtain
\be
%
\ee
Applying $\cI_0$ on the right surface using the blowup $x$, and the applying $\cI_0^{-1}$ on the right surface using the blowup $y$, we can rewrite the above geometry as
\be\label{g1}
%
\ee
Now flopping $x\sim y$ living in the second surface, we obtain
\be\label{g2}
%
\ee
One way to understand the appearance of $-2y$ in the gluing curve on the left side in the above geometry is as follows: The curve $e+f-x-y$ has genus zero without the presence of self-gluing, but genus one when $x\sim y$. Correspondingly, the curve $2h+f-\sum x_i$ also has genus one. When we remove the genus of the gluing curve on right side decreases by one. Correspondingly, the genus of gluing curve must decrease by one. This can be achieved by transforming the left gluing curve to $2h+f-\sum x_i-2y$.

Now implementing $\cI_0^{-1}$ on the left surface using the blowup $y$, we get
\be
%
\ee
Now, just as we passed from (\ref{g0}) to (\ref{sp1KK}), we can pass from the above geometry to the geometry
\be
%
\ee
Moving all the blowups to the right surface and exchanging the two surfaces we obtain
\be\label{sp0su1KK}
%
\ee
which is the our final desired frame to represent the geometry (\ref{sp0su16d}).

Integrating out a blowup towards right from (\ref{sp0su1KK}) produces the $5d$ SCFT
\be
%
\ee
which is already accounted in (\ref{sp1Lm}). Similar statements will hold true for all geometries that follow in this subsection: Every time a $-1$ curve is integrated out from the right surface, the resulting $5d$ SCFT has already been accounted. Thus, we only have to integrate out $-1$ curves from the left surface.

Removing blowups from the left surface in (\ref{sp0su1KK}), we obtain the following $5d$ SCFTs
\be\label{sp0su1Lm}
%
\ee
Consider the geometry at $m=7$
\be
%
\ee
Moving one blowup to the right and exchanging $e$ with $f$ in the resulting right surface, we can rewrite the above as
\be
%
\ee
Now, as already discussed in the last subsection we can move the blowup on the left surface to the right surface and then back to the left surface such that the geometry is transformed to
\be
%
\ee
Removing the blowup on the left surface we obtain a $5d$ SCFT described by
\be
%
\ee

At level $m=1$ of (\ref{sp0su1Lm}) the geometry is
\be
%
\ee
which has a $-1$ curve $h-\sum x_i$ which can be converted to $e$ curve of left surface by writing the above geometry in the following isomorphism frame
\be
%
\ee
Blowing down the $e$ curve we obtain a $5d$ SCFT described by geometry
\be
%
\ee
Integrating out blowups from the right leads to theories (\ref{sp1LmR3Rt}).

\subsection{$M=7$}
The geometry associated to the KK theory (\ref{su3T}) for $k=1$ is \cite{Bhardwaj:2019fzv}
\be\label{su3T16d}
%
\ee
where the label $6$ in the middle of the loop around the first surface denotes that there are six self-gluings given by $x_i\sim y_i$. Flopping all of these and exchanging $e$ with $f$ in the second surface gives
\be
%
\ee
Applying $\cI_0$ on the second surface using $x_6$ gives
\be
%
\ee
Applying $\cI_0^{-1}$ on the second surface using $x_5$ gives
\be
%
\ee
Applying $\cI_0$ on the second surface using $x_4$ gives
\be
%
\ee
Applying $\cI_0^{-1}$ on the second surface using $x_3$ and exchanging $e$ with $f$ gives
\be
%
\ee
Applying $\cI_0$ on the second surface using $x_2$ gives
\be
%
\ee
Applying $\cI_0^{-1}$ on the second surface using $x$ and exchanging $e$ with $f$ gives
\be
%
\ee
Now we send all the blowups onto the left surface to obtain the desired frame to represent (\ref{su3T16d})
\be
%
\ee
As in last subsection, integrating out any $-1$ curve to the right does not produce any new $5d$ SCFTs not accounted already in the paper. Thus we can focus on integrating out $-1$ curves from the left only.

Removing $m$ blowups gives a series of $5d$ SCFTs described by
\be\label{su3TLm}
%
\ee
When $m=3$, we can send two of the blowups onto the right surface and then exchange $e$ with $f$ on the right surface to obtain
\be
%
\ee
Now, as explained in last subsection, a way to integrate out the blowup remaining on the left surface produces a $5d$ SCFT
\be
%
\ee

\subsection{$M=4$}
The KK theory (\ref{su3T}) with $k=2$ has geometry given by \cite{Bhardwaj:2019fzv}
\be
%
\ee
Performing operations similar to the ones described in last subsection, we can rewrite the above geometry as
\be
%
\ee
from which it is easy to see that there are no new $5d$ SCFTs produced by this KK theory.

The KK theory (\ref{su2}) has geometry given by \cite{Bhardwaj:2018yhy,Bhardwaj:2018vuu,Bhardwaj:2019fzv}
\be\label{su2KK}
%
\ee
where the subscript $2$ in the middle of the edge denotes that there are two gluing curves between the two surfaces. The labels placed at the ends of edge identify these two gluing curves in the order in which they appear. Thus, $e$ in left surface is glued to $e$ in right surface, and $e-\sum x_i$ in left surface is glued to $h$ in right surface.

One can observe that it is not possible to integrate out the blowups from (\ref{su2KK}) since blowing them down always leads to a flop transition. If one tries to remove them from a gluing curve by using the isomorphisms $\cI_n$, they simply appear in the other gluing curve. For example, if we perform $\cI_0$ on the left surface using $x_4$, we obtain the description
\be
\begin{tikzpicture} [scale=1.9]
\node (v1) at (-4.4,-0.5) {$\bF_1^{3+1}$};
\node (v2) at (-1.9,-0.5) {$\bF_2$};
\node at (-3.7,-0.4) {\scriptsize{$h$-$y,e$-$\sum x_i$}};
\node at (-2.3,-0.4) {\scriptsize{$e,h$}};
\node (v3) at (-3,-0.5) {\scriptsize{2}};
\draw  (v1) edge (v3);
\draw  (v3) edge (v2);
\end{tikzpicture}
\ee
As we can see, the blowup $x_4$ became the blowup $y$ and entered into the description of the other gluing curve.

Thus, the only way to obtain an RG flow is to blow down the $e$ curve of an $\bF_1$ in some frame. But this will always appear as one of the gluing curves. For example, first send one of the blowups from the left surface in (\ref{su2KK}) to the right surface
\be
\begin{tikzpicture} [scale=1.9]
\node (v1) at (-4.25,-0.5) {$\bF_0^3$};
\node (v2) at (-1.9,-0.5) {$\bF^1_2$};
\node at (-3.7,-0.4) {\scriptsize{$e,e$-$\sum x_i$}};
\node at (-2.3,-0.4) {\scriptsize{$e,h$-$x$}};
\node (v3) at (-3,-0.5) {\scriptsize{2}};
\draw  (v1) edge (v3);
\draw  (v3) edge (v2);
\end{tikzpicture}
\ee
which can be written, by using $\cI_1^{-1}$ with the blowup $x$ on the right surface, as
\be
\begin{tikzpicture} [scale=1.9]
\node (v1) at (-4.25,-0.5) {$\bF_0^3$};
\node (v2) at (-1.9,-0.5) {$\bF^1_1$};
\node at (-3.7,-0.4) {\scriptsize{$e,e$-$\sum x_i$}};
\node at (-2.3,-0.4) {\scriptsize{$e$-$x,h$}};
\node (v3) at (-3,-0.5) {\scriptsize{2}};
\draw  (v1) edge (v3);
\draw  (v3) edge (v2);
\end{tikzpicture}
\ee
Now, blowing down $x$ sends the blowup back to the left surface but along the other gluing curve, yielding
\be
\begin{tikzpicture} [scale=1.9]
\node (v1) at (-4.25,-0.5) {$\bF_1^4$};
\node (v2) at (-1.9,-0.5) {$\bF_1$};
\node at (-3.7,-0.4) {\scriptsize{$e,h$-$\sum x_i$}};
\node at (-2.3,-0.4) {\scriptsize{$e,h$}};
\node (v3) at (-3,-0.5) {\scriptsize{2}};
\draw  (v1) edge (v3);
\draw  (v3) edge (v2);
\end{tikzpicture}
\ee
We can see that the $-1$ curve $e$ has appeared as a gluing curve. Blowing down the $e$ curve removes one of the gluing curves and produces the $5d$ SCFT
\be
\begin{tikzpicture} [scale=1.9]
\node (v1) at (-4.25,-0.5) {$\dP^4$};
\node (v2) at (-2.2,-0.5) {$\dP$};
\draw  (v1) edge (v2);
\node at (-3.7,-0.4) {\scriptsize{$l$-$\sum x_i$}};
\node at (-2.6,-0.4) {\scriptsize{$l$}};
\end{tikzpicture}
\ee
which has already been found in (\ref{sp1L3LtR3Rt}). Note that we had not considered the possibility of blowing down a gluing curve between two distinct surfaces so far because till now we only encountered examples involving a single gluing curve between two distinct surfaces. Blowing down this single gluing curve would decouple the two surfaces and lead to a direct sum of two rank one SCFTs rather than a rank two SCFT.

\subsection{$M=3$}
The geometry for the KK theory (\ref{su2T}) is \cite{Bhardwaj:2019fzv}
\be
\begin{tikzpicture} [scale=1.9]
\node (v1) at (-5.2,-0.5) {$\bF_1^2$};
\node (v2) at (-2.2,-0.5) {$\bF_0^{1+1}$};
\node at (-4.6,-0.4) {\scriptsize{$h,h$-$\sum x_i$}};
\node at (-3,-0.4) {\scriptsize{$e$+$f$-$x$-$2y,e$-$x$}};
\node (v3) at (-3.8,-0.5) {\scriptsize{2}};
\draw  (v1) edge (v3);
\draw  (v3) edge (v2);
\draw (v2) .. controls (-1.4,0) and (-1.4,-1) .. (v2);
\node at (-1.8,-0.2) {\scriptsize{$x$}};
\node at (-1.8,-0.8) {\scriptsize{$y$}};
\end{tikzpicture}
\ee
which can be rewritten by performing some isomorphisms as
\be\label{su2T6d2}
\begin{tikzpicture} [scale=1.9]
\node (v1) at (-5.2,-0.5) {$\bF_1^{1+1}$};
\node (v2) at (-2.2,-0.5) {$\bF_0^{1+1}$};
\node at (-4.6,-0.4) {\scriptsize{$h,f$-$x$}};
\node at (-3,-0.4) {\scriptsize{$e$+$f$-$x$-$2y,f$-$x$}};
\node (v3) at (-3.8,-0.5) {\scriptsize{2}};
\draw  (v1) edge (v3);
\draw  (v3) edge (v2);
\draw (v2) .. controls (-1.4,0) and (-1.4,-1) .. (v2);
\node at (-1.8,-0.2) {\scriptsize{$x$}};
\node at (-1.8,-0.8) {\scriptsize{$y$}};
\end{tikzpicture}
\ee
Sending the blowup $y$ from the left surface to the right surface, and exchanging the positions of the two surfaces we can write the geometry for the KK theory as
\be\label{su2TKK}
\begin{tikzpicture} [scale=1.9]
\node (v1) at (5.1,-0.5) {$\bF_0^{1}$};
\node (v2) at (2.2,-0.5) {$\bF_1^{1+1+1}$};
\node at (4.6,-0.4) {\scriptsize{$e$+$f,f$-$x$}};
\node at (3,-0.4) {\scriptsize{$h$-$x$-$2y,f$-$x$}};
\node (v3) at (3.8,-0.5) {\scriptsize{2}};
\draw  (v1) edge (v3);
\draw  (v3) edge (v2);
\draw (v2) .. controls (1.4,0) and (1.4,-1) .. (v2);
\node at (1.8,-0.2) {\scriptsize{$x$}};
\node at (1.8,-0.8) {\scriptsize{$y$}};
\end{tikzpicture}
\ee
Integrating out the blowup $z$ on the left surface gives rise to the $5d$ SCFT
\be\label{su2TL}
\begin{tikzpicture} [scale=1.9]
\node (v1) at (5.1,-0.5) {$\bF_0^{1}$};
\node (v2) at (2.2,-0.5) {$\bF_1^{1+1}$};
\node at (4.6,-0.4) {\scriptsize{$e$+$f,f$-$x$}};
\node at (2.9,-0.4) {\scriptsize{$h$-$x$-$2y,f$-$x$}};
\node (v3) at (3.8,-0.5) {\scriptsize{2}};
\draw  (v1) edge (v3);
\draw  (v3) edge (v2);
\draw (v2) .. controls (1.4,0) and (1.4,-1) .. (v2);
\node at (1.8,-0.2) {\scriptsize{$x$}};
\node at (1.8,-0.8) {\scriptsize{$y$}};
\draw  (1.3,0) rectangle (5.4,-1.8);
\node at (3.5,-1.1) {$M=2$};
\node at (3.5,-1.5) {$6\cF=7\phi_R^3-15\phi_L\phi_R^2+9\phi_R\phi_L^2$};
\end{tikzpicture}
\ee
This SCFT is desribed by $5d$ gauge theory with gauge algebra $\su(3)$ at CS level $\half$ and a hyper transforming in two-index symmetric representation of $\su(3)$. The existence of this SCFT has been known in the literature, but this is the first time a geometry has been written down describing this $5d$ SCFT. We propose that the fundamental BPS particles coming from the left surface are associated to curves $h-x-2y$, $e$, $f-x=f-y$ and $x=y$ in the left surface rather than the set of generators of Mori cone of the left surface.

When there are multiple gluing curves between two surfaces then we take the sum of all the gluing curves to compute the prepotential. For example, the term $\phi_L\phi_R^2$ in $6\cF$ associated to (\ref{su2TL}) is computed by $3C_{L;R}^2$ where $C_{L;R}=(h-x-2y)+(f-x)=h+f-2x-2y$.

We can also integrate out the curve $e-x$ living in the right surface of (\ref{su2TL}) towards the left, which gives rise to a new $5d$ SCFT whose existence has not been predicted in the literature. This SCFT is given by the geometry
\be\label{su2TL2}
\begin{tikzpicture} [scale=1.9]
\node (v1) at (5.1,-0.5) {$\bF_1$};
\node (v2) at (2.2,-0.5) {$\bF_2^{1+1}$};
\node at (4.6,-0.4) {\scriptsize{$h$+$f,e$}};
\node at (3,-0.4) {\scriptsize{$e$+$f$-$x$-$2y,f$-$x$}};
\node (v3) at (3.9,-0.5) {\scriptsize{2}};
\draw  (v1) edge (v3);
\draw  (v3) edge (v2);
\draw (v2) .. controls (1.4,0) and (1.4,-1) .. (v2);
\node at (1.8,-0.2) {\scriptsize{$x$}};
\node at (1.8,-0.8) {\scriptsize{$y$}};
\draw  (1.3,0) rectangle (5.4,-1.8);
\node at (3.5,-1.1) {$M=1$};
\node at (3.5,-1.5) {$6\cF=8\phi_R^3-18\phi_L\phi_R^2+12\phi_R\phi_L^2$};
\end{tikzpicture}
\ee
The fundamental BPS particles arising from the left surface are again different from the corresponding Mori cone generators. We propose that the fundamental BPS particles coming from the left surface can be identified with the curves $e+f-x-2y$, $f-x=f-y$ and $x=y$ living in the left surface.

It is also possible to remove the self-gluing. Flopping $x\sim y$ simply sends the self-gluing to the other surface, and hence not useful for this purpose. For example, flopping $x\sim y$ in (\ref{su2T6d2}) gives rise to 
\be
\begin{tikzpicture} [scale=1.9]
\node (v1) at (5.1,-0.5) {$\bF_0$};
\node (v2) at (2,-0.5) {$\bF_1^{1+1+1+1}$};
\node at (4.6,-0.4) {\scriptsize{$e$+$f,f$}};
\node at (3,-0.4) {\scriptsize{$h$-$x$-$2y,f$-$x$-$z$}};
\node (v3) at (3.8,-0.5) {\scriptsize{2}};
\draw  (v1) edge (v3);
\draw  (v3) edge (v2);
\draw (v2) .. controls (1.2,0) and (1.2,-1) .. (v2);
\node at (1.6,-0.2) {\scriptsize{$x$}};
\node at (1.6,-0.8) {\scriptsize{$y$}};
\end{tikzpicture}
\ee
However, flopping $f-x$ and $f-y$ (they flop together since they have same volume) yields a new phase from which it is possible to remove the self-gluing. Let us perform this flop on the geometry (\ref{su2TKK}). Notice that it also flops the curve $f-x$ in the right surface of (\ref{su2TKK}). The geometry after the flop is
\be\label{wf}
\begin{tikzpicture} [scale=1.9]
\node (v1) at (4.5,-0.5) {$\bF_1$};
\node (v2) at (2.2,-0.5) {$\bF_5^{1+1+1}$};
\node at (4.1,-0.4) {\scriptsize{$h$+$f$}};
\node at (2.7,-0.4) {\scriptsize{$e$}};
\draw (v2) .. controls (1.4,0) and (1.4,-1) .. (v2);
\node at (1.8,-0.2) {\scriptsize{$x$}};
\node at (1.8,-0.8) {\scriptsize{$y$}};
\draw  (v2) edge (v1);
\end{tikzpicture}
\ee
We refer the reader to Appendix B of \cite{Bhardwaj:2019fzv} for more details on this flop procedure. From the above geometry we can simply remove the self-gluing by blowing down $x\sim y$ living in the first surface and obtain the geometry with $m=8$ in the series of geometries (\ref{sp1LmR}). Integrating out other $-1$ curves from (\ref{wf}) simply give rise to geometries flop equivalent to those discussed above in this subsection.

\subsection{$M=2$}
The geometry for the KK theory (\ref{su1su1}) is given by \cite{Bhardwaj:2018vuu,Bhardwaj:2019fzv}
\be
\begin{tikzpicture} [scale=1.9]
\node (v1) at (-4.5,-0.5) {$\bF_0^{1+1}$};
\node (v2) at (-2.1,-0.5) {$\bF_0^{1+1}$};
\node at (-4,-0.4) {\scriptsize{$f$-$x$,$x$}};
\node at (-2.6,-0.4) {\scriptsize{$f$-$x$,$x$}};
\draw (v1) .. controls (-5.2,-0.1) and (-5.2,-0.9) .. (v1);
\draw (v2) .. controls (-1.4,-0.1) and (-1.4,-0.9) .. (v2);
\node at (-4.9,-0.2) {\scriptsize{$e$-$x$}};
\node at (-4.9,-0.8) {\scriptsize{$e$-$y$}};
\node at (-1.7,-0.2) {\scriptsize{$e$-$x$}};
\node at (-1.7,-0.8) {\scriptsize{$e$-$y$}};
\node (v3) at (-3.3,-0.5) {\scriptsize{2}};
\draw  (v1) edge (v3);
\draw  (v3) edge (v2);
\end{tikzpicture}
\ee
which can be rewritten as
\be
\begin{tikzpicture} [scale=1.9]
\node (v1) at (-4.5,-0.5) {$\bF_0^{1+1}$};
\node (v2) at (-2.1,-0.5) {$\bF_0^{1+1}$};
\node at (-3.9,-0.4) {\scriptsize{$e$-$y,f$-$x$}};
\node at (-2.7,-0.4) {\scriptsize{$e$-$y,f$-$x$}};
\draw (v1) .. controls (-5.2,-0.1) and (-5.2,-0.9) .. (v1);
\draw (v2) .. controls (-1.4,-0.1) and (-1.4,-0.9) .. (v2);
\node at (-4.9,-0.2) {\scriptsize{$x$}};
\node at (-4.9,-0.8) {\scriptsize{$y$}};
\node at (-1.7,-0.2) {\scriptsize{$x$}};
\node at (-1.7,-0.8) {\scriptsize{$y$}};
\node (v3) at (-3.3,-0.5) {\scriptsize{2}};
\draw  (v1) edge (v3);
\draw  (v3) edge (v2);
\end{tikzpicture}
\ee
The only inequivalent $-1$ curves are $f-x$ and $x$ on both surfaces. The flop of $f-x$ does not change the geometry but flopping $x\sim y$ on both surfaces changes the geometry to
\be
\begin{tikzpicture} [scale=1.9]
\node (v1) at (-4.5,-0.5) {$\bF_0^{1}$};
\node (v2) at (-2.1,-0.5) {$\bF_0^{1}$};
\node at (-4,-0.4) {\scriptsize{$e$-$x,f$-$x$}};
\node at (-2.6,-0.4) {\scriptsize{$e$-$x,f$-$x$}};
\node (v3) at (-3.3,-0.5) {\scriptsize{2}};
\draw  (v1) edge (v3);
\draw  (v3) edge (v2);
\end{tikzpicture}
\ee
Recall from the discussion around (\ref{g2}) that flopping a self-gluing leads to the appearance of $-2x$ in the gluing curve inside the adjacent surface. This has split into two copies of $-x$ above since we have two gluing curves affected by the flop. This is consistent with the discussion around (\ref{g2}) since the above flop is not supposed to change the genus of any gluing curve.

Now the only remaining inequivalent $-1$ curve is $f-x$ in both surfaces. Blowing it down, we obtain
\be
\begin{tikzpicture} [scale=1.9]
\node (v1) at (-4.9,-0.5) {$\bF_1$};
\node (v2) at (-2.9,-0.5) {$\bF_1$};
\node at (-4.6,-0.4) {\scriptsize{$e$}};
\node at (-3.2,-0.4) {\scriptsize{$e$}};
\draw  (v1) edge (v2);
\end{tikzpicture}
\ee
which has already been accounted in (\ref{sp1LmRp}).

The geometry for the KK theory (\ref{su1su1L}) is given by \cite{Bhardwaj:2019fzv}
\be
\begin{tikzpicture} [scale=1.9]
\node (v1) at (-4.5,-0.5) {$\bF_0^{1+1}$};
\node (v2) at (-1.9,-0.5) {$\bF_1^{1+1}$};
\node at (-4,-0.4) {\scriptsize{$f$-$x$,$x$}};
\node at (-2.6,-0.4) {\scriptsize{$2h$-$x$-$2y,f$-$x$}};
\draw (v1) .. controls (-5.2,-0.1) and (-5.2,-0.9) .. (v1);
\draw (v2) .. controls (-1.2,-0.1) and (-1.2,-0.9) .. (v2);
\node at (-4.9,-0.2) {\scriptsize{$e$-$x$}};
\node at (-4.9,-0.8) {\scriptsize{$e$-$y$}};
\node at (-1.5,-0.2) {\scriptsize{$x$}};
\node at (-1.5,-0.8) {\scriptsize{$y$}};
\node (v3) at (-3.4,-0.5) {\scriptsize{2}};
\draw  (v1) edge (v3);
\draw  (v3) edge (v2);
\end{tikzpicture}
\ee
After isomorphisms on the left surface, it can rewritten as
\be
\begin{tikzpicture} [scale=1.9]
\node (v1) at (-4.5,-0.5) {$\bF_0^{1+1}$};
\node (v2) at (-1.9,-0.5) {$\bF_1^{1+1}$};
\node at (-3.9,-0.4) {\scriptsize{$e$-$y,f$-$x$}};
\node at (-2.6,-0.4) {\scriptsize{$2h$-$x$-$2y,f$-$x$}};
\draw (v1) .. controls (-5.2,-0.1) and (-5.2,-0.9) .. (v1);
\draw (v2) .. controls (-1.2,-0.1) and (-1.2,-0.9) .. (v2);
\node at (-4.9,-0.2) {\scriptsize{$x$}};
\node at (-4.9,-0.8) {\scriptsize{$y$}};
\node at (-1.5,-0.2) {\scriptsize{$x$}};
\node at (-1.5,-0.8) {\scriptsize{$y$}};
\node (v3) at (-3.3,-0.5) {\scriptsize{2}};
\draw  (v1) edge (v3);
\draw  (v3) edge (v2);
\end{tikzpicture}
\ee
Flopping $x\sim y$ living on the left surface, and then interchanging $e$ and $f$ in the left surface, we get
\be
\begin{tikzpicture} [scale=1.9]
\node (v1) at (-4.9,-0.5) {$\bF_0$};
\node (v2) at (-1.9,-0.5) {$\bF_1^{1+1+1}$};
\node at (-4.5,-0.4) {\scriptsize{$f,e$}};
\node at (-2.8,-0.4) {\scriptsize{$2h$-$x$-$2y$-$z,f$-$x$-$z$}};
\draw (v2) .. controls (-1.2,-0.1) and (-1.2,-0.9) .. (v2);
\node at (-1.5,-0.2) {\scriptsize{$x$}};
\node at (-1.5,-0.8) {\scriptsize{$y$}};
\node (v3) at (-3.7,-0.5) {\scriptsize{2}};
\draw  (v1) edge (v3);
\draw  (v3) edge (v2);
\end{tikzpicture}
\ee
Integrating out the $e$ curve of the right surface from the left side, we obtain
\be
\begin{tikzpicture} [scale=1.9]
\node (v1) at (-4.9,-0.5) {$\bF_1$};
\node (v2) at (-1.9,-0.5) {$\dP^{1+1+1}$};
\node at (-4.5,-0.4) {\scriptsize{$f,e$}};
\node at (-2.9,-0.4) {\scriptsize{$2l$-$x$-$2y$-$z,l$-$x$-$z$}};
\draw (v2) .. controls (-1.2,-0.1) and (-1.2,-0.9) .. (v2);
\node at (-1.5,-0.2) {\scriptsize{$x$}};
\node at (-1.5,-0.8) {\scriptsize{$y$}};
\node (v3) at (-3.8,-0.5) {\scriptsize{2}};
\draw  (v1) edge (v3);
\draw  (v3) edge (v2);
\end{tikzpicture}
\ee
which can be written as
\be\label{r2new}
\begin{tikzpicture} [scale=1.9]
\node (v1) at (-4.9,-0.5) {$\bF_1$};
\node (v2) at (-1.9,-0.5) {$\bF_1^{1+1}$};
\node at (-4.5,-0.4) {\scriptsize{$f,e$}};
\node at (-2.7,-0.4) {\scriptsize{$h$+$f$-$x$-$2y,f$-$x$}};
\draw (v2) .. controls (-1.2,-0.1) and (-1.2,-0.9) .. (v2);
\node at (-1.5,-0.2) {\scriptsize{$x$}};
\node at (-1.5,-0.8) {\scriptsize{$y$}};
\node (v3) at (-3.6,-0.5) {\scriptsize{2}};
\draw  (v1) edge (v3);
\draw  (v3) edge (v2);
\draw  (-5.2,0) rectangle (-1.1,-1.8);
\node at (-3,-1.1) {$M=1$};
\node at (-3,-1.5) {$6\cF=8\phi_L^3+3\phi_L\phi_R^2-9\phi_R\phi_L^2$};
\end{tikzpicture}
\ee
giving rise to a new $5d$ SCFT not discussed in the literature before. We propose that the fundamental BPS particles coming from the right surface can be identified with the curves $e$, $h+f-x-2y$, $f-x=f-y$ and $x=y$ living in the right surface.\\ One can remove the remaining self-gluing as well but that does not give rise to a $5d$ SCFT not accounted earlier.

The KK theory (\ref{su1su1T2}) has the geometry given by \cite{Bhardwaj:2019fzv}
\be
\begin{tikzpicture} [scale=1.9]
\node (v1) at (-4.5,-0.5) {$\bF_0^{1+1}$};
\node (v2) at (-2.1,-0.5) {$\bF_0^{1+1}$};
\node at (-4,-0.4) {\scriptsize{$f$-$x$,$x$}};
\node at (-2.6,-0.4) {\scriptsize{$2f$-$x$,$x$}};
\draw (v1) .. controls (-5.2,-0.1) and (-5.2,-0.9) .. (v1);
\draw (v2) .. controls (-1.4,-0.1) and (-1.4,-0.9) .. (v2);
\node at (-4.9,-0.2) {\scriptsize{$e$-$x$}};
\node at (-4.9,-0.8) {\scriptsize{$e$-$y$}};
\node at (-1.7,-0.2) {\scriptsize{$e$-$x$}};
\node at (-1.7,-0.8) {\scriptsize{$e$-$y$}};
\node (v3) at (-3.3,-0.5) {\scriptsize{2}};
\draw  (v1) edge (v3);
\draw  (v3) edge (v2);
\end{tikzpicture}
\ee
which can be rewritten as
\be
\begin{tikzpicture} [scale=1.9]
\node (v1) at (-4.5,-0.5) {$\bF_0^{1+1}$};
\node (v2) at (-1.7,-0.5) {$\bF_0^{1+1}$};
\node at (-3.9,-0.4) {\scriptsize{$e$-$y,f$-$x$}};
\node at (-2.5,-0.4) {\scriptsize{$2e$+$f$-$x$-$2y,f$-$x$}};
\draw (v1) .. controls (-5.2,-0.1) and (-5.2,-0.9) .. (v1);
\draw (v2) .. controls (-1,-0.1) and (-1,-0.9) .. (v2);
\node at (-4.9,-0.2) {\scriptsize{$x$}};
\node at (-4.9,-0.8) {\scriptsize{$y$}};
\node at (-1.3,-0.2) {\scriptsize{$x$}};
\node at (-1.3,-0.8) {\scriptsize{$y$}};
\node (v3) at (-3.3,-0.5) {\scriptsize{2}};
\draw  (v1) edge (v3);
\draw  (v3) edge (v2);
\end{tikzpicture}
\ee
Flopping $x\sim y$ living on the left surface, we obtain
\be
\begin{tikzpicture} [scale=1.9]
\node (v1) at (-4.9,-0.5) {$\bF_0$};
\node (v2) at (-1.9,-0.5) {$\bF_0^{1+1+1}$};
\node at (-4.5,-0.4) {\scriptsize{$e,f$}};
\node at (-2.9,-0.4) {\scriptsize{$2e$+$f$-$x$-$2y$-$z,f$-$x$-$z$}};
\draw (v2) .. controls (-1.2,-0.1) and (-1.2,-0.9) .. (v2);
\node at (-1.5,-0.2) {\scriptsize{$x$}};
\node at (-1.5,-0.8) {\scriptsize{$y$}};
\node (v3) at (-3.8,-0.5) {\scriptsize{2}};
\draw  (v1) edge (v3);
\draw  (v3) edge (v2);
\end{tikzpicture}
\ee
We can integrate out $e-z$ from the right surface to obtain (\ref{su2TL2}). Removing the remaining self-gluing does not lead to any new $5d$ SCFT either.

The KK theory (\ref{su1su1T3}) is described by the geometry \cite{Bhardwaj:2019fzv}
\be\label{su1su1T36d}
\begin{tikzpicture} [scale=1.9]
\node (v1) at (-4.5,-0.5) {$\bF_0^{1+1}$};
\node (v2) at (-2.1,-0.5) {$\bF_0^{1+1}$};
\node at (-4,-0.4) {\scriptsize{$f$-$x$,$x$}};
\node at (-2.6,-0.4) {\scriptsize{$3f$-$x$,$x$}};
\draw (v1) .. controls (-5.2,-0.1) and (-5.2,-0.9) .. (v1);
\draw (v2) .. controls (-1.4,-0.1) and (-1.4,-0.9) .. (v2);
\node at (-4.9,-0.2) {\scriptsize{$e$-$x$}};
\node at (-4.9,-0.8) {\scriptsize{$e$-$y$}};
\node at (-1.7,-0.2) {\scriptsize{$e$-$x$}};
\node at (-1.7,-0.8) {\scriptsize{$e$-$y$}};
\node (v3) at (-3.3,-0.5) {\scriptsize{2}};
\draw  (v1) edge (v3);
\draw  (v3) edge (v2);
\end{tikzpicture}
\ee
We claim that the above geometry should be flop equivalent to the geometry
\be\label{su1su1T3KK}
\begin{tikzpicture} [scale=1.9]
\node (v1) at (-4.9,-0.5) {$\bF_8$};
\node (v2) at (-2.9,-0.5) {$\bF_0^1$};
\node at (-4.6,-0.4) {\scriptsize{$e$}};
\node at (-3.3,-0.4) {\scriptsize{$3e$+$f$}};
\draw  (v1) edge (v2);
\end{tikzpicture}
\ee
The claim is based on the fact that the KK theory (\ref{su1su1T3}) can be described by the $5d$ $\cN=1$ gauge theory with $\fg_2$ gauge algebra and a hyper in adjoint representation \cite{Tachikawa:2011ch}. According to \cite{Bhardwaj:2019ngx}, the above geometry describes a gauge-theoretic phase of this $5d$ gauge theory. 

Now exchanging $e$ and $f$ in the above geometry (\ref{su1su1T3KK}), we can rewrite it as
\be
\begin{tikzpicture} [scale=1.9]
\node (v1) at (-4.9,-0.5) {$\bF_8$};
\node (v2) at (-2.9,-0.5) {$\bF_0^1$};
\node at (-4.6,-0.4) {\scriptsize{$e$}};
\node at (-3.3,-0.4) {\scriptsize{$e$+$3f$}};
\draw  (v1) edge (v2);
\end{tikzpicture}
\ee
We can now integrate out the blowup sitting on the right surface either from the right or from the left. Integrating it out from the right leads to a $5d$ SCFT described by
\be
\begin{tikzpicture} [scale=1.9]
\node (v1) at (-4.9,-0.5) {$\bF_8$};
\node (v2) at (-2.9,-0.5) {$\bF_0$};
\node at (-4.6,-0.4) {\scriptsize{$e$}};
\node at (-3.3,-0.4) {\scriptsize{$e$+$3f$}};
\draw  (v1) edge (v2);
\end{tikzpicture}
\ee
which has already been found as the $m=6$ case of the $5d$ SCFTs described in (\ref{su3TLm}). However, integrating the blowup from the left leads to a new $5d$ SCFT not discussed in the literature before
\be\label{su1su1T3L}
\begin{tikzpicture} [scale=1.9]
\node (v1) at (-4.9,-0.5) {$\bF_9$};
\node (v2) at (-2.9,-0.5) {$\bF_1$};
\node at (-4.6,-0.4) {\scriptsize{$e$}};
\node at (-3.3,-0.4) {\scriptsize{$h$+$3f$}};
\draw  (v1) edge (v2);
\draw  (-5.7,-0.2) rectangle (-1.9,-1.6);
\node at (-3.8,-0.9) {$M=1$};
\node at (-3.8,-1.3) {$6\cF=8\phi_L^3+8\phi_R^3-27\phi_L\phi_R^2+21\phi_R\phi_L^2$};
\end{tikzpicture}
\ee
However, this theory poses a puzzle since it can be checked that the above geometry is not shrinkable. Two possible resolutions of this puzzle are as follows. First, it could be possible that (\ref{su1su1T3L}) is an example of a $5d$ SCFT without a Coulomb branch. Second, it could be that the Coulomb branch of this $5d$ SCFT is not described by any gauge theory but rather by the phase described by the following geometry obtained by flopping the $e$ curve in the right surface of (\ref{su1su1T3L})
\be\label{triglue}
\begin{tikzpicture} [scale=1.9]
\node (v1) at (4.5,-0.5) {$\dP$};
\node (v2) at (2.2,-0.5) {$\bF_9^{1+1+1}$};
\node at (4.1,-0.4) {\scriptsize{$4l$}};
\node at (2.7,-0.4) {\scriptsize{$e$}};
\draw (v2) .. controls (1.4,0) and (1.4,-1) .. (v2);
\node at (1.6,-0.2) {\scriptsize{$x$}};
\node at (1.4,-0.5) {\scriptsize{$y$}};
\draw  (v2) edge (v1);
\node at (1.6,-0.8) {\scriptsize{$z$}};
\end{tikzpicture}
\ee
where the notation indicates that the left surface is a self-glued surface where the self-gluing $x\sim y\sim z$ is described by gluing the three blowups $x$, $y$ and $z$ all with each other. Naively, the above geometry (\ref{triglue}) is also non-shrinkable, but it is possible that there are some extra non-geometric elements introduced by such a flop so that the curve $f-x$ in the left surface of (\ref{triglue}) does not give rise to a fundamental BPS particle, but $2f-x$ does. This prescription for the BPS states ensures that the (\ref{triglue}) is a shrinkable phase and describes a sensible Coulomb branch of the $5d$ SCFT under discussion. We do not know if any of the above proposed resolutions is the correct one, and a more detailed study of this theory is required.

\subsection{$M=1$}
The KK theory (\ref{su3T}) for $k=3$ has the geometry \cite{Bhardwaj:2019fzv}
\be
\begin{tikzpicture} [scale=1.9]
\node (v1) at (-4.9,-0.5) {$\bF_{10}$};
\node (v2) at (-2.9,-0.5) {$\bF_0$};
\node at (-4.5,-0.4) {\scriptsize{$e$}};
\node at (-3.3,-0.4) {\scriptsize{$4e$+$f$}};
\draw  (v1) edge (v2);
\end{tikzpicture}
\ee
which has no $-1$ curves and hence there are no flows.

\section{Rank three}
For rank three, we have the following possibilities:
\bit
\item $T=1,G=2$
\item $T=2,G=1$
\item $T=3,G=0$
\eit
In the class $T=1,G=2$, we have the following $5d$ KK theories:
\bit
\item \be\label{sp2}
\begin{tikzpicture}
\node at (-0.5,0.4) {1};
\node at (-0.45,0.9) {$\sp(2)^{(1)}$};
\end{tikzpicture}
\ee
which describes the untwisted compactification of the $6d$ SCFT carrying $\sp(2)$ gauge algebra on a $-1$ curve. The $6d$ theory carries $12$ hypers in fundamental of $\sp(2)$. Hence there is a rank twelve flavor symmetry implying that
\be
M=13
\ee
is the number of mass parameters carried by the KK theory (\ref{sp2}).
\item \be\label{su3}
\begin{tikzpicture}
\node at (-0.5,0.4) {$k$};
\node at (-0.45,0.9) {$\su(3)^{(1)}$};
\end{tikzpicture}
\ee
for $1\le k\le 3$, which describes the untwisted compactification of the $6d$ SCFT carrying $\su(3)$ gauge algebra on a $-k$ curve. The theory carries $18-6k$ hypers in fundamental of $\su(3)$. For $k=1,2$, the $\u(1)$ subalgebra rotating all the flavors simultaneously is anomalous and hence for $k=1,2$ the theory has only $\su(18-6k)$ symmetry. Thus,
\be
M=18-6k
\ee
for $k=1,2$ and
\be
M=1
\ee
for $k=3$.
\item \be\label{g21}
\begin{tikzpicture}
\node at (-0.5,0.4) {$k$};
\node at (-0.45,0.9) {$\fg_2^{(1)}$};
\end{tikzpicture}
\ee
for $1\le k \le 3$, which describes the untwisted compactification of the $6d$ SCFT carrying $\fg_2$ on $-k$ curve. The $6d$ SCFT has $10-3k$ hypers in fundamental of $\fg_2$, implying that
\be
M=11-3k
\ee
for (\ref{g21}).
\item \be\label{su3L}
\begin{tikzpicture}
\node (v1) at (-0.5,0.4) {2};
\node at (-0.45,0.9) {$\su(3)^{(1)}$};
\draw (v1) .. controls (-1.5,-0.5) and (0.5,-0.5) .. (v1);
\end{tikzpicture}
\ee
which denotes the KK theory obtained by compactifying the $6d$ SCFT
\be
\begin{tikzpicture}
\node (v1) at (-0.5,0.45) {2};
\node at (-0.45,0.9) {$\su(3)$};
\begin{scope}[shift={(1.5,0)}]
\node (v2) at (-0.5,0.45) {2};
\node at (-0.45,0.9) {$\su(3)$};
\end{scope}
\draw  (v1) edge (v2);
\end{tikzpicture}
\ee
with an exchange of the two $-2$ curves as one goes around the circle. The matter spectrum of the $6d$ SCFT is a hyper in bifundamental plus three extra hypers in fundamental carried by each $\su(3)$. The bifundamental gives rise to a $\u(1)$ flavor symmetry and the extra fundamentals give rise to a $\su(3)\oplus\su(3)$ flavor symmetry. The discrete symmetry exchanging the two $\su(3)$ gauge algebras exchanges the two $\su(3)$ flavor symmetry algebras, while preserving the $\u(1)$ flavor symmetry. Thus, we have
\be
M=4
\ee
for (\ref{su3L}).
\item \be\label{su4T2}
\begin{tikzpicture}
\node at (-0.5,0.4) {2};
\node at (-0.45,0.9) {$\su(4)^{(2)}$};
\end{tikzpicture}
\ee
which describes the compactification of the $6d$ SCFT carrying $\su(4)$ on $-2$ curve twisted by the outer automorphism of $\su(4)$. The invariant subalgebra of $\su(4)$ under the outer automorphism is $\sp(2)$ which implies that indeed $G=2$ for (\ref{su4T2}). The $6d$ SCFT has $8$ hypers in fundamental of $\su(4)$, which are exchanged with each other in pairs under the outer automorphism \cite{Bhardwaj:2019fzv}. This means that after the reduction we obtain $4$ hypers in fundamental of $\sp(2)$ and hence
\be
M=5
\ee
for (\ref{su4T1}).
\item \be\label{su4T1}
\begin{tikzpicture}
\node at (-0.5,0.4) {2};
\node at (-0.45,0.9) {$\su(4)^{(2)}$};
\end{tikzpicture}
\ee
which describes the compactification of the $6d$ SCFT carrying $\su(4)$ on $-1$ curve twisted by the outer automorphism of $\su(4)$. The $6d$ SCFT has $12$ hypers in fundamental and one hyper in antisymmetric of $\su(4)$. The fundamentals are exchanged with each other in pairs and the antisymmetric is left invariant under the outer automorphism \cite{Bhardwaj:2019fzv}. This means that after the reduction we obtain $6$ hypers in fundamental and a hyper in antisymmetric of $\sp(2)$ and hence
\be
M=8
\ee
for (\ref{su4T1}).
\item \be\label{su5T}
\begin{tikzpicture}
\node at (-0.5,0.4) {2};
\node at (-0.45,0.9) {$\su(5)^{(2)}$};
\end{tikzpicture}
\ee
which describes the compactification of the $6d$ SCFT carrying $\su(5)$ on $-2$ curve twisted by the outer automorphism of $\su(5)$. The invariant subalgebra of $\su(5)$ under the outer automorphism is $\sp(2)$ which implies that indeed $G=2$ for (\ref{su5T}). The $6d$ SCFT has $10$ hypers in fundamental of $\su(5)$, which are exchanged with each other in pairs under the outer automorphism \cite{Bhardwaj:2019fzv}. This means that after the reduction we obtain $5$ hypers in fundamental of $\sp(2)$ and hence
\be
M=6
\ee
for (\ref{su5T}).
\item \be\label{so8}
\begin{tikzpicture}
\node at (-0.5,0.4) {$k$};
\node at (-0.45,0.9) {$\so(8)^{(3)}$};
\end{tikzpicture}
\ee
for $1\le k\le 4$, which describes the compactification of the $6d$ SCFT carrying $\so(8)$ on $-k$ curve twisted by the order three outer automorphism of $\so(8)$. The invariant subalgebra of $\so(8)$ under the outer automorphism is $\fg_2$ which implies that indeed $G=2$ for (\ref{so8}). The $6d$ SCFT has $4-k$ hypers each in fundamental, spinor and cospinor representations of $\so(8)$. These three representations are cyclically permuted under the outer automorphism and they all descend to the fundamental of $\fg_2$. This means that after the reduction we obtain $4-k$ hypers in fundamental of $\fg_2$ and hence
\be
M=5-k
\ee
for (\ref{so8}).
\eit
In the class $T=2,G=1$, we have the following KK theories:
\bit
\item \be\label{sp0su2}
\begin{tikzpicture}
\node (v1) at (-0.5,0.4) {1};
\node at (-0.45,0.9) {$\sp(0)^{(1)}$};
\begin{scope}[shift={(2,0)}]
\node (v2) at (-0.5,0.4) {2};
\node at (-0.45,0.9) {$\su(2)^{(1)}$};
\end{scope}
\draw  (v1) edge (v2);
\end{tikzpicture}
\ee
which denotes an untwisted compactification of the $6d$ SCFT arising from an empty $-1$ curve intersecting a $-2$ curve carrying $\su(2)$. The empty $-1$ carries an $\fe_8$ symmetry out of which $\su(2)$ has been gauged. The remaining flavor symmetry is $\fe_7$. Moreover, as already discussed above $\su(2)$ on $-2$ has an $\su(4)$ flavor symmetry which is left completely ungauged by $\sp(0)$. Thus,
\be
M=11
\ee
for the KK theory (\ref{sp0su2}).
\item \be\label{sp1su1}
\begin{tikzpicture}
\node (v1) at (-0.5,0.4) {1};
\node at (-0.45,0.9) {$\sp(1)^{(1)}$};
\begin{scope}[shift={(2,0)}]
\node (v2) at (-0.5,0.4) {2};
\node at (-0.45,0.9) {$\su(1)^{(1)}$};
\end{scope}
\draw  (v1) edge (v2);
\end{tikzpicture}
\ee
which denotes an untwisted compactification of the $6d$ SCFT arising from an empty $-2$ curve intersecting a $-1$ curve carrying $\sp(1)$. The empty $-2$ carries an $\su(2)$ symmetry which has been completely gauged. Moreover, as already discussed above $\sp(1)$ on $-1$ carries 10 hypers in fundamental. Out of these ten hypers, at least a half-hyper must remain localized at the intersection point of the two curves. In field theoretic terms, this half-hyper of $\sp(1)$ provides necessary degrees of freedom to complete the $\cN=(1,0)$ tensor multiplet associated to empty $-1$ curve into an $\cN=(2,0)$ tensor multiplet. This leaves only an $\so(19)$ flavor symmetry and thus,
\be
M=10
\ee
for the KK theory (\ref{sp1su1}).
\item \be\label{sp0su3T2}
\begin{tikzpicture}
\node (v1) at (-0.5,0.4) {1};
\node at (-0.45,0.9) {$\sp(0)^{(1)}$};
\begin{scope}[shift={(2,0)}]
\node (v2) at (-0.5,0.4) {2};
\node at (-0.45,0.9) {$\su(3)^{(2)}$};
\end{scope}
\draw  (v1) edge (v2);
\end{tikzpicture}
\ee
which denotes the compactification of the $6d$ SCFT arising from an empty $-1$ curve intersecting a $-2$ curve carrying $\su(3)$, twisted by the outer automorphism of $\su(3)$. Gauging an $\su(3)$ out of the $\fe_8$ symmetry of the $-1$ curve leaves an $\fe_6$ flavor symmetry. However, as explained in \cite{Bhardwaj:2019fzv}, the outer automorphism twist of $\su(3)$ is only a symmetry of the theory if this $\fe_6$ symmetry is also twisted by its outer automorphism, thus leaving only a $\ff_4$ flavor symmetry preserved after the twist. Moreover, as discussed above $\su(3)$ on $-2$ curve twisted by outer automorphism has an $\so(6)$ flavor symmetry which is left completely ungauged by $\sp(0)$. Thus,
\be
M=8
\ee
for the KK theory (\ref{sp0su3T2}).
\item \be\label{sp0su3T3}
\begin{tikzpicture}
\node (v1) at (-0.5,0.4) {1};
\node at (-0.45,0.9) {$\sp(0)^{(1)}$};
\begin{scope}[shift={(2,0)}]
\node (v2) at (-0.5,0.4) {3};
\node at (-0.45,0.9) {$\su(3)^{(2)}$};
\end{scope}
\draw  (v1) edge (v2);
\end{tikzpicture}
\ee
which denotes the compactification of the $6d$ SCFT arising from an empty $-1$ curve intersecting a $-3$ curve carrying $\su(3)$, twisted by the outer automorphism of $\su(3)$. The $\su(3)$ carries no matter content, so this twist preserves only an $\ff_4$ flavor symmetry resulting in
\be
M=5
\ee
for the above KK theory (\ref{sp0su3T3}).
\item \be\label{su2su1L}
\begin{tikzpicture}
\node (v1) at (-0.5,0.4) {$2$};
\node at (-0.45,0.9) {$\su(2)^{(1)}$};
\begin{scope}[shift={(2,0)}]
\node (v2) at (-0.5,0.4) {$2$};
\node at (-0.45,0.9) {$\su(1)^{(1)}$};
\end{scope}
\draw  (v1) -- (v2);
\draw (v1) .. controls (-1.5,-0.5) and (0.5,-0.5) .. (v1);
\end{tikzpicture}
\ee
which comes from the twisted compactification of the $6d$ SCFT
\be
\begin{tikzpicture}
\node (v1) at (-0.5,0.45) {2};
\node at (-0.45,0.9) {$\su(1)$};
\begin{scope}[shift={(1.5,0)}]
\node (v2) at (-0.5,0.45) {2};
\node at (-0.45,0.9) {$\su(2)$};
\end{scope}
\begin{scope}[shift={(3,0)}]
\node (v3) at (-0.5,0.45) {2};
\node at (-0.45,0.9) {$\su(2)$};
\end{scope}
\begin{scope}[shift={(4.5,0)}]
\node (v4) at (-0.5,0.45) {2};
\node at (-0.45,0.9) {$\su(1)$};
\end{scope}
\draw  (v1) edge (v2);
\draw  (v3) edge (v2);
\draw  (v3) edge (v4);
\end{tikzpicture}
\ee
by the twist exchanging the two $-2$ curves at the two ends with each other while simultaneously exchanging the two $-2$ curves in the middle:
\be
\begin{tikzpicture}
\node (v1) at (-0.5,0.45) {2};
\node at (-0.45,0.9) {$\su(2)$};
\begin{scope}[shift={(1.5,0)}]
\node (v2) at (-0.5,0.45) {2};
\node at (-0.45,0.9) {$\su(2)$};
\end{scope}
\begin{scope}[shift={(-1.5,0)}]
\node (v0) at (-0.5,0.45) {2};
\node at (-0.45,0.9) {$\su(1)$};
\end{scope}
\begin{scope}[shift={(3,0)}]
\node (v3) at (-0.5,0.45) {2};
\node at (-0.45,0.9) {$\su(1)$};
\end{scope}
\draw  (v1) -- (v2);
\draw  (v0) edge (v1);
\draw  (v2) edge (v3);
\draw[<->,red,thick] (v0) .. controls (-1.5,-1.5) and (2,-1.5) .. (v3);
\draw[<->,red,thick] (v1) .. controls (-0.5,-0.5) and (1,-0.5) .. (v2);
\end{tikzpicture}
\ee
The matter spectrum for the $6d$ SCFT is a bifundamental of $\su(2)\oplus\su(2)$ and two extra hypers in fundamental of each $\su(2)$. A half-hyper out of these two hypers is trapped by the neighboring $\su(1)$, thus leaving only a $\u(1)$ symmetry rotating the bifundamental. This symmetry is preserved under the twist and hence
\be
M=2
\ee
for the KK theory (\ref{su2su1L}).
\item \be\label{su3sp0T}
\begin{tikzpicture}
\node (v1) at (-0.5,0.4) {3};
\node at (-0.45,0.9) {$\su(3)^{(2)}$};
\begin{scope}[shift={(2,0)}]
\node (v2) at (-0.5,0.4) {1};
\node at (-0.45,0.9) {$\sp(0)^{1)}$};
\end{scope}
\node (v3) at (0.5,0.4) {\tiny{$2$}};
\draw  (v1) edge (v3);
\draw  [->](v3) -- (v2);
\end{tikzpicture}
\ee
This theory arises by twisting the following $6d$ SCFT by the discrete symmetry obtained by combining the following transformation
\be
\begin{tikzpicture}
\node (v1) at (-0.5,0.45) {3};
\node at (-0.45,0.9) {$\su(3)$};
\begin{scope}[shift={(1.5,0)}]
\node (v2) at (-0.5,0.45) {1};
\node at (-0.45,0.9) {$\sp(0)$};
\end{scope}
\begin{scope}[shift={(-1.5,0)}]
\node (v0) at (-0.5,0.45) {1};
\node at (-0.45,0.9) {$\sp(0)$};
\end{scope}
\draw  (v1) -- (v2);
\draw  (v0) edge (v1);
\draw[<->,red,thick] (v0) .. controls (-1.5,-1) and (0.5,-1) .. (v2);
\end{tikzpicture}
\ee
with the outer automorphism of $\su(3)$. This twist preserves an $\ff_4$ flavor symmetry and hence
\be
M=5
\ee
for it.
\item \be\label{su2su1}
\begin{tikzpicture}
\node (v1) at (-0.5,0.4) {2};
\node at (-0.45,0.9) {$\su(2)^{(1)}$};
\begin{scope}[shift={(2,0)}]
\node (v2) at (-0.5,0.4) {2};
\node at (-0.45,0.9) {$\su(1)^{(1)}$};
\end{scope}
\draw  (v1) edge (v2);
\end{tikzpicture}
\ee
which is the untwisted compactification of the $6d$ SCFT arising by an empty $-2$ curve intersecting a $-2$ curve carrying $\su(2)$. The $\su(2)$ has four hypers in fundamental out of which a half-hyper is trapped by $\su(1)$ leading to only a rank two flavor symmetry and implying that
\be
M=3
\ee
for (\ref{su2su1}).
\item \be\label{su2su1T2}
\begin{tikzpicture}
\node (v1) at (-0.5,0.4) {2};
\node at (-0.45,0.9) {$\su(2)^{(1)}$};
\begin{scope}[shift={(2,0)}]
\node (v2) at (-0.5,0.4) {2};
\node at (-0.45,0.9) {$\su(1)^{(1)}$};
\end{scope}
\node (v3) at (0.5,0.4) {\tiny{$2$}};
\draw  (v1) edge (v3);
\draw  [->](v3) -- (v2);
\end{tikzpicture}
\ee
This theory arises from the following twist of the following $6d$ SCFT
\be
\begin{tikzpicture}
\node (v1) at (-0.5,0.45) {2};
\node at (-0.45,0.9) {$\su(2)$};
\begin{scope}[shift={(1.5,0)}]
\node (v2) at (-0.5,0.45) {2};
\node at (-0.45,0.9) {$\su(1)$};
\end{scope}
\begin{scope}[shift={(-1.5,0)}]
\node (v0) at (-0.5,0.45) {2};
\node at (-0.45,0.9) {$\su(1)$};
\end{scope}
\draw  (v1) -- (v2);
\draw  (v0) edge (v1);
\draw[<->,red,thick] (v0) .. controls (-1.5,-1) and (0.5,-1) .. (v2);
\end{tikzpicture}
\ee
The $\su(2)$ has four hypers in fundamental out of which two half-hypers are trapped by the two $\su(1)$. The $6d$ SCFT thus has an $\su(3)$ flavor symmetry. Exchanging the two $\su(1)$ exchanges these two half-hypers and hence their corresponding full hypers. This is in clash with the $\su(3)$ symmetry, which can be restored only if an outer automorphism of $\su(3)$ is also performed. In any case, the maximal flavor symmetry preserved under the twist has rank one and correspondingly
\be
M=2
\ee
for the KK theory (\ref{su2su1T2}).
\item \be\label{su2su1T3}
\begin{tikzpicture}
\node (v1) at (-0.5,0.4) {2};
\node at (-0.45,0.9) {$\su(2)^{(1)}$};
\begin{scope}[shift={(2,0)}]
\node (v2) at (-0.5,0.4) {2};
\node at (-0.45,0.9) {$\su(1)^{(1)}$};
\end{scope}
\node (v3) at (0.5,0.4) {\tiny{$3$}};
\draw  (v1) edge (v3);
\draw  [->](v3) -- (v2);
\end{tikzpicture}
\ee
which arises by twisting the following $6d$ SCFT by the following action
\be\label{D42}
\begin{tikzpicture}
\node (v1) at (-0.5,0.45) {2};
\node (v4) at (-0.45,0.9) {$\su(2)$};
\begin{scope}[shift={(1.5,0)}]
\node (v2) at (-0.5,0.45) {2};
\node (v5) at (-0.45,0.9) {$\su(1)$};
\end{scope}
\begin{scope}[shift={(-1.5,0)}]
\node (v0) at (-0.5,0.45) {2};
\node (v6) at (-0.45,0.9) {$\su(1)$};
\end{scope}
\begin{scope}[shift={(0.05,1.7)}]
\node (v3) at (-0.5,0.45) {2};
\node at (-0.45,0.9) {$\su(1)$};
\end{scope}
\draw  (v1) -- (v2);
\draw  (v0) edge (v1);
\draw[->,red,thick] (v0) .. controls (-1.5,-1) and (0.5,-1) .. (v2);
\draw[->,red,thick] (v5) .. controls (1,1.7) and (0.4,2.1) .. (v3);
\draw[<-,red,thick] (v6) .. controls (-1.9,1.7) and (-1.4,2.1) .. (v3);
\draw  (v3) edge (v4);
\end{tikzpicture}
\ee
A rank one flavor symmetry can be preserved under this twist, and hence this KK theory has
\be
M=2
\ee
\eit
In the class $T=2,G=1$, we have the following KK theories:
\bit
\item \be\label{sp0su1su1}
\begin{tikzpicture}
\node (v1) at (-0.5,0.4) {1};
\node at (-0.45,0.9) {$\sp(0)^{(1)}$};
\begin{scope}[shift={(2,0)}]
\node (v2) at (-0.5,0.4) {2};
\node at (-0.45,0.9) {$\su(1)^{(1)}$};
\end{scope}
\begin{scope}[shift={(4,0)}]
\node (v3) at (-0.5,0.4) {2};
\node at (-0.45,0.9) {$\su(1)^{(1)}$};
\end{scope}
\draw  (v1) edge (v2);
\draw  (v2) edge (v3);
\end{tikzpicture}
\ee
which has an $\fe_8\oplus\su(2)$ flavor symmetry and hence
\be
M=10
\ee
for the above KK theory.
\item \be\label{su1su1su1}
\begin{tikzpicture}
\node (v1) at (-0.5,0.4) {2};
\node at (-0.45,0.9) {$\su(1)^{(1)}$};
\begin{scope}[shift={(2,0)}]
\node (v2) at (-0.5,0.4) {2};
\node at (-0.45,0.9) {$\su(1)^{(1)}$};
\end{scope}
\begin{scope}[shift={(4,0)}]
\node (v3) at (-0.5,0.4) {2};
\node at (-0.45,0.9) {$\su(1)^{(1)}$};
\end{scope}
\draw  (v1) edge (v2);
\draw  (v2) edge (v3);
\end{tikzpicture}
\ee
which has an $\su(2)$ flavor symmetry and hence
\be
M=2
\ee
for the above KK theory.
\item \be\label{su1su1su1L}
\begin{tikzpicture}
\node (v1) at (-0.5,0.4) {$2$};
\node at (-0.45,0.9) {$\su(1)^{(1)}$};
\begin{scope}[shift={(2,0)}]
\node (v2) at (-0.5,0.4) {$2$};
\node at (-0.45,0.9) {$\su(1)^{(1)}$};
\end{scope}
\begin{scope}[shift={(4,0)}]
\node (v3) at (-0.5,0.4) {$2$};
\node at (-0.45,0.9) {$\su(1)^{(1)}$};
\end{scope}
\draw  (v1) -- (v2);
\draw (v1) .. controls (-1.5,-0.5) and (0.5,-0.5) .. (v1);
\draw  (v2) edge (v3);
\end{tikzpicture}
\ee
which arises from the $A_6$ $(2,0)$ theory by the following twist
\be
\begin{tikzpicture}
\node (v1) at (-0.5,0.45) {2};
\node at (-0.45,0.9) {$\su(1)$};
\begin{scope}[shift={(1.5,0)}]
\node (v2) at (-0.5,0.45) {2};
\node at (-0.45,0.9) {$\su(1)$};
\end{scope}
\begin{scope}[shift={(-1.5,0)}]
\node (v0) at (-0.5,0.45) {2};
\node at (-0.45,0.9) {$\su(1)$};
\end{scope}
\begin{scope}[shift={(-3,0)}]
\node (v-1) at (-0.5,0.45) {2};
\node at (-0.45,0.9) {$\su(1)$};
\end{scope}
\begin{scope}[shift={(3,0)}]
\node (v3) at (-0.5,0.45) {2};
\node at (-0.45,0.9) {$\su(1)$};
\end{scope}
\begin{scope}[shift={(4.5,0)}]
\node (v4) at (-0.5,0.45) {2};
\node at (-0.45,0.9) {$\su(1)$};
\end{scope}
\draw  (v1) -- (v2);
\draw  (v0) edge (v1);
\draw  (v2) edge (v3);
\draw[<->,red,thick] (v0) .. controls (-1.5,-1.5) and (2,-1.5) .. (v3);
\draw[<->,red,thick] (v1) .. controls (-0.5,-0.5) and (1,-0.5) .. (v2);
\draw  (v3) edge (v4);
\draw  (v-1) edge (v0);
\draw[<->,red,thick] (v-1) .. controls (-3.5,-2.4) and (4,-2.4) .. (v4);
\end{tikzpicture}
\ee
and carries
\be
M=2
\ee 
number of mass parameters.
\item \be\label{su1su1su1TL}
\begin{tikzpicture}
\node (v1) at (-0.5,0.4) {2};
\node at (-0.45,0.9) {$\su(1)^{(1)}$};
\begin{scope}[shift={(2,0)}]
\node (v2) at (-0.5,0.4) {2};
\node at (-0.45,0.9) {$\su(1)^{(1)}$};
\end{scope}
\begin{scope}[shift={(4,0)}]
\node (v4) at (-0.5,0.4) {2};
\node at (-0.45,0.9) {$\su(1)^{(1)}$};
\end{scope}
\node (v3) at (0.5,0.4) {\tiny{$2$}};
\draw  (v1) edge (v3);
\draw  [->](v3) -- (v2);
\draw  (v2) edge (v4);
\end{tikzpicture}
\ee
which arises from the following twist of $A_5$ $(2,0)$ theory
\be
\begin{tikzpicture}
\node (v1) at (-0.5,0.45) {2};
\node at (-0.45,0.9) {$\su(1)$};
\begin{scope}[shift={(1.5,0)}]
\node (v2) at (-0.5,0.45) {2};
\node at (-0.45,0.9) {$\su(1)$};
\end{scope}
\begin{scope}[shift={(3,0)}]
\node (v3) at (-0.5,0.45) {2};
\node at (-0.45,0.9) {$\su(1)$};
\end{scope}
\begin{scope}[shift={(-1.5,0)}]
\node (v0) at (-0.5,0.45) {2};
\node at (-0.45,0.9) {$\su(1)$};
\end{scope}
\begin{scope}[shift={(-3,0)}]
\node (v-1) at (-0.5,0.45) {2};
\node at (-0.45,0.9) {$\su(1)$};
\end{scope}
\draw  (v1) -- (v2);
\draw  (v0) edge (v1);
\draw[<->,red,thick] (v0) .. controls (-1.5,-1) and (0.5,-1) .. (v2);
\draw[<->,red,thick] (v-1) .. controls (-2.2,-1.8) and (1.2,-1.8) .. (v3);
\draw  (v-1) edge (v0);
\draw  (v2) edge (v3);
\end{tikzpicture}
\ee
and carries
\be
M=2
\ee 
number of mass parameters.
\item \be\label{su1su1su1TR}
\begin{tikzpicture}
\node (v1) at (-0.5,0.4) {2};
\node at (-0.45,0.9) {$\su(1)^{(1)}$};
\begin{scope}[shift={(2,0)}]
\node (v2) at (-0.5,0.4) {2};
\node at (-0.45,0.9) {$\su(1)^{(1)}$};
\end{scope}
\begin{scope}[shift={(-2,0)}]
\node (v0) at (-0.5,0.4) {2};
\node at (-0.45,0.9) {$\su(1)^{(1)}$};
\end{scope}
\node (v3) at (0.5,0.4) {\tiny{$2$}};
\draw  (v1) edge (v3);
\draw  [->](v3) -- (v2);
\draw  (v0) edge (v1);
\end{tikzpicture}
\ee
which arises from the following twist of $D_4$ $(2,0)$ theory
\be
\begin{tikzpicture}
\node (v1) at (-0.5,0.45) {2};
\node (v4) at (-0.45,0.9) {$\su(1)$};
\begin{scope}[shift={(1.5,0)}]
\node (v2) at (-0.5,0.45) {2};
\node (v5) at (-0.45,0.9) {$\su(1)$};
\end{scope}
\begin{scope}[shift={(-1.5,0)}]
\node (v0) at (-0.5,0.45) {2};
\node (v6) at (-0.45,0.9) {$\su(1)$};
\end{scope}
\begin{scope}[shift={(0.05,1.7)}]
\node (v3) at (-0.5,0.45) {2};
\node at (-0.45,0.9) {$\su(1)$};
\end{scope}
\draw  (v1) -- (v2);
\draw  (v0) edge (v1);
\draw[<->,red,thick] (v5) .. controls (1,1.7) and (0.4,2.1) .. (v3);
\draw  (v3) edge (v4);
\end{tikzpicture}
\ee
and carries
\be
M=2
\ee 
number of mass parameters.
\eit
Now we will move onto a study of RG flows of these KK theories.

\subsection{$M=13$}\label{sp2S}
The geometry associated to the KK theory (\ref{sp2}) is \cite{Bhardwaj:2018yhy,Bhardwaj:2018vuu,Bhardwaj:2019fzv}
\be
\begin{tikzpicture} [scale=1.9]
\node (v1) at (-4.6,-0.5) {$\bF_1^{12}$};
\node (v2) at (-2.8,-0.5) {$\bF_6$};
\node (v3) at (-1.3,-0.5) {$\bF_1$};
\draw  (v1) edge (v2);
\draw  (v2) edge (v3);
\node at (-4.1,-0.4) {\scriptsize{$2h$-$\sum x_i$}};
\node at (-3.1,-0.4) {\scriptsize{$h$}};
\node at (-2.5,-0.4) {\scriptsize{$e$}};
\node at (-1.6,-0.4) {\scriptsize{$2h$}};
\end{tikzpicture}
\ee
which we can rewrite in our desired isomorphism frame
\be\label{sp2KK}
\begin{tikzpicture} [scale=1.9]
\node (v1) at (-4.6,-0.5) {$\bF_8^{12}$};
\node (v2) at (-2.8,-0.5) {$\bF_6$};
\node (v3) at (-1.3,-0.5) {$\bF_1$};
\draw  (v1) edge (v2);
\draw  (v2) edge (v3);
\node at (-4.2,-0.4) {\scriptsize{$e$}};
\node at (-3.1,-0.4) {\scriptsize{$h$}};
\node at (-2.5,-0.4) {\scriptsize{$e$}};
\node at (-1.6,-0.4) {\scriptsize{$2h$}};
\end{tikzpicture}
\ee
Notice that it is not possible to integrate out any blowup from the middle surface. Only left and right surfaces allow a blowup in (\ref{sp2KK}) to be integrated out. For instance, blowing down an $f-x_i$ in the left surface yields
\be
\begin{tikzpicture} [scale=1.9]
\node (v1) at (-4.6,-0.5) {$\bF_7^{11}$};
\node (v2) at (-2.8,-0.5) {$\bF^1_6$};
\node (v3) at (-1.3,-0.5) {$\bF_1$};
\draw  (v1) edge (v2);
\draw  (v2) edge (v3);
\node at (-4.2,-0.4) {\scriptsize{$e$}};
\node at (-3.2,-0.4) {\scriptsize{$h$-$x$}};
\node at (-2.5,-0.4) {\scriptsize{$e$}};
\node at (-1.6,-0.4) {\scriptsize{$2h$}};
\end{tikzpicture}
\ee
To integrate out the blowup $x$ on the middle surface, we would like to absorb it into the surface such that it does not appear in any gluing curves. To remove it from the gluing curve $h-x$ we have to perform the isomorphism $\cI_5^{-1}$ on the middle surface, and then we obtain
\be
\begin{tikzpicture} [scale=1.9]
\node (v1) at (-4.6,-0.5) {$\bF_7^{11}$};
\node (v2) at (-2.8,-0.5) {$\bF^1_5$};
\node (v3) at (-1.3,-0.5) {$\bF_1$};
\draw  (v1) edge (v2);
\draw  (v2) edge (v3);
\node at (-4.2,-0.4) {\scriptsize{$e$}};
\node at (-3.1,-0.4) {\scriptsize{$h$}};
\node at (-2.4,-0.4) {\scriptsize{$e$-$x$}};
\node at (-1.6,-0.4) {\scriptsize{$2h$}};
\end{tikzpicture}
\ee
So, instead of absorbing the blowup, we have only managed to move it to the other gluing curve.

Removing $m$ blowups from the left surface in (\ref{sp2KK}), we obtain a series of $5d$ SCFTs
\be\label{sp2Lm}
\begin{tikzpicture} [scale=1.9]
\node (v1) at (-4.6,-0.5) {$\bF_8^{12-m}$};
\node (v2) at (-2.8,-0.5) {$\bF_6$};
\node (v3) at (-1.3,-0.5) {$\bF_1$};
\draw  (v1) edge (v2);
\draw  (v2) edge (v3);
\node at (-4.1,-0.4) {\scriptsize{$e$}};
\node at (-3.1,-0.4) {\scriptsize{$h$}};
\node at (-2.5,-0.4) {\scriptsize{$e$}};
\node at (-1.6,-0.4) {\scriptsize{$2h$}};
\node at (0.1,-0.5) {$1\le m\le 12$};
\draw  (-5.9,-0.2) rectangle (1.5,-1.6);
\node at (-2.2,-0.9) {$M=13-m$};
\node at (-2.2,-1.3) {$6\cF=(m-4)\phi_L^3+8\phi_M^3+8\phi_R^3
-24\phi_L\phi_M^2+18\phi_M\phi_L^2-18\phi_M\phi_R^2
+12\phi_R\phi_M^2$};
\end{tikzpicture}
\ee
where we denote the Coulomb branch parameters corresponding to left, middle and right surfaces respectively as $\phi_L$, $\phi_M$ and $\phi_R$. Using what we have already discussed in Sections \ref{R1} and \ref{R2} we can compute all the coefficients in $6\cF$ for a rank three theory except for the coefficient of $\phi_L\phi_M\phi_R$ which arises only when the graph associated to the geometry is cyclic and has a loop-like structure. We will discuss how to compute this term when such a geometry arises.

Integrating out a blowup from the right surface and integrating out $m$ blowups from the left surface in (\ref{sp2KK}), we obtain another series of $5d$ SCFTs
\be\label{sp2LmR}
\begin{tikzpicture} [scale=1.9]
\node (v1) at (-4.6,-0.5) {$\bF_7^{11-m}$};
\node (v2) at (-2.8,-0.5) {$\bF_5$};
\node (v3) at (-1.3,-0.5) {$\bF_1$};
\draw  (v1) edge (v2);
\draw  (v2) edge (v3);
\node at (-4.1,-0.4) {\scriptsize{$e$}};
\node at (-3.1,-0.4) {\scriptsize{$h$}};
\node at (-2.5,-0.4) {\scriptsize{$e$}};
\node at (-1.7,-0.4) {\scriptsize{$h$+$f$}};
\node at (0.1,-0.5) {$1\le m\le 11$};
\draw  (-5.8,-0.2) rectangle (1.4,-1.6);
\node at (-2.2,-0.9) {$M=12-m$};
\node at (-2.2,-1.3) {$6\cF=(m-3)\phi_L^3+8\phi_M^3+8\phi_R^3
-21\phi_L\phi_M^2+15\phi_M\phi_L^2-15\phi_M\phi_R^2
+9\phi_R\phi_M^2$};
\end{tikzpicture}
\ee
Integrating out the $e$ curve from the right surface in (\ref{sp2Lm}), we obtain the $5d$ SCFT described by the geometry
\be\label{sp2L12R}
\begin{tikzpicture} [scale=1.9]
\node (v1) at (-4.6,-0.5) {$\bF_8$};
\node (v2) at (-2.8,-0.5) {$\bF_6$};
\node (v3) at (-1.3,-0.5) {$\dP$};
\draw  (v1) edge (v2);
\draw  (v2) edge (v3);
\node at (-4.3,-0.4) {\scriptsize{$e$}};
\node at (-3.1,-0.4) {\scriptsize{$h$}};
\node at (-2.5,-0.4) {\scriptsize{$e$}};
\node at (-1.6,-0.4) {\scriptsize{$2l$}};
\draw  (-6.2,-0.2) rectangle (0.6,-1.6);
\node at (-2.8,-0.9) {$M=0$};
\node at (-2.8,-1.3) {$6\cF=8\phi_L^3+8\phi_M^3+9\phi_R^3
-24\phi_L\phi_M^2+18\phi_M\phi_L^2-18\phi_M\phi_R^2
+12\phi_R\phi_M^2$};
\end{tikzpicture}
\ee
The geometry (\ref{sp2L12R}) is also the result we obtain by integrating out the $e$ curve of the right surface in (\ref{sp2LmR}).

Now, integrating out two blowups from the right surface and $m$ blowups from the left surface in (\ref{sp2KK}) we obtain yet another series of $5d$ SCFTs described by
\be
\begin{tikzpicture} [scale=1.9]
\node (v1) at (-4.6,-0.5) {$\bF_6^{10-m}$};
\node (v2) at (-2.8,-0.5) {$\bF_4$};
\node (v3) at (-1.3,-0.5) {$\bF_0$};
\draw  (v1) edge (v2);
\draw  (v2) edge (v3);
\node at (-4.1,-0.4) {\scriptsize{$e$}};
\node at (-3.1,-0.4) {\scriptsize{$h$}};
\node at (-2.5,-0.4) {\scriptsize{$e$}};
\node at (-1.7,-0.4) {\scriptsize{$e$+$f$}};
\node at (0.1,-0.5) {$2\le m\le 10$};
\draw  (-5.7,-0.2) rectangle (1.5,-1.6);
\node at (-2.1,-0.9) {$M=11-m$};
\node at (-2.1,-1.3) {$6\cF=(m-2)\phi_L^3+8\phi_M^3+8\phi_R^3
-18\phi_L\phi_M^2+12\phi_M\phi_L^2-12\phi_M\phi_R^2
+6\phi_R\phi_M^2$};
\end{tikzpicture}
\ee
The lower bound on $m$ has been put in the above equation so that we do not overcount $5d$ SCFTs.

Another series of $5d$ SCFTs is obtained by integrating out $p\ge3$ blowups from the right surface and $m$ blowups from the left surface in (\ref{sp2KK})
\be\label{sp2LmRp}
\begin{tikzpicture} [scale=1.9]
\node (v1) at (-4.6,-0.5) {$\bF_{8-p}^{12-m-p}$};
\node (v2) at (-2.8,-0.5) {$\bF_{6-p}$};
\node (v3) at (-1.3,-0.5) {$\bF_{p-4}$};
\draw  (v1) edge (v2);
\draw  (v2) edge (v3);
\node at (-4,-0.4) {\scriptsize{$e$}};
\node at (-3.2,-0.4) {\scriptsize{$h$}};
\node at (-2.4,-0.4) {\scriptsize{$e$}};
\node at (-1.7,-0.4) {\scriptsize{$e$}};
\node at (0.1,-0.3) {$p\le m\le 12-p$};
\draw  (-5.4,0) rectangle (1.4,-2.1);
\node at (0.1,-0.7) {$3\le p\le 6$};
\node at (-2,-1.1) {$M=13-m-p$};
\node at (-2,-1.5) {$6\cF=(m+p-4)\phi_L^3+8\phi_M^3+8\phi_R^3
+3(p-8)\phi_L\phi_M^2+3(6-p)\phi_M\phi_L^2$};
\node at (-2,-1.8) {$+3(p-6)\phi_M\phi_R^2
+3(4-p)\phi_R\phi_M^2$};
\end{tikzpicture}
\ee

Consider the $m=11$ case of (\ref{sp2Lm})
\be
\begin{tikzpicture} [scale=1.9]
\node (v1) at (-4.6,-0.5) {$\bF_8^{1}$};
\node (v2) at (-2.8,-0.5) {$\bF_6$};
\node (v3) at (-1.3,-0.5) {$\bF_1$};
\draw  (v1) edge (v2);
\draw  (v2) edge (v3);
\node at (-4.3,-0.4) {\scriptsize{$e$}};
\node at (-3.1,-0.4) {\scriptsize{$h$}};
\node at (-2.5,-0.4) {\scriptsize{$e$}};
\node at (-1.6,-0.4) {\scriptsize{$2h$}};
\end{tikzpicture}
\ee
As we know from earlier discussions, we can send the remaining blowup to the right surface and then bring it back to the left surface to obtain the following flop-equivalent frame of the above geometry
\be
\begin{tikzpicture} [scale=1.9]
\node (v1) at (-4.6,-0.5) {$\bF_8^{1}$};
\node (v2) at (-2.8,-0.5) {$\bF_6$};
\node (v3) at (-1.2,-0.5) {$\bF_0$};
\draw  (v1) edge (v2);
\draw  (v2) edge (v3);
\node at (-4.3,-0.4) {\scriptsize{$e$}};
\node at (-3.1,-0.4) {\scriptsize{$h$}};
\node at (-2.5,-0.4) {\scriptsize{$e$}};
\node at (-1.6,-0.4) {\scriptsize{$2e$+$f$}};
\end{tikzpicture}
\ee
Blowing down the blowup on the left surface in the above geometry yields the $5d$ SCFT
\be
\begin{tikzpicture} [scale=1.9]
\node (v1) at (-4.6,-0.5) {$\bF_8$};
\node (v2) at (-2.8,-0.5) {$\bF_6$};
\node (v3) at (-1.2,-0.5) {$\bF_0$};
\draw  (v1) edge (v2);
\draw  (v2) edge (v3);
\node at (-4.3,-0.4) {\scriptsize{$e$}};
\node at (-3.1,-0.4) {\scriptsize{$h$}};
\node at (-2.5,-0.4) {\scriptsize{$e$}};
\node at (-1.6,-0.4) {\scriptsize{$2e$+$f$}};
\draw  (-6.2,-0.2) rectangle (0.6,-1.6);
\node at (-2.8,-0.9) {$M=1$};
\node at (-2.8,-1.3) {$6\cF=8\phi_L^3+8\phi_M^3+8\phi_R^3
-24\phi_L\phi_M^2+18\phi_M\phi_L^2-18\phi_M\phi_R^2
+12\phi_R\phi_M^2$};
\end{tikzpicture}
\ee

Consider the $p=3$ case of (\ref{sp2LmRp})
\be
\begin{tikzpicture} [scale=1.9]
\node (v1) at (-4.6,-0.5) {$\bF_5^{9-m}$};
\node (v2) at (-2.8,-0.5) {$\bF_3$};
\node (v3) at (-1.3,-0.5) {$\bF_1$};
\draw  (v1) edge (v2);
\draw  (v2) edge (v3);
\node at (-4.1,-0.4) {\scriptsize{$e$}};
\node at (-3.1,-0.4) {\scriptsize{$h$}};
\node at (-2.5,-0.4) {\scriptsize{$e$}};
\node at (-1.6,-0.4) {\scriptsize{$h$}};
\end{tikzpicture}
\ee
Blowing down the $e$ curve in the right surface, we obtain the following series of $5d$ SCFTs
\be
\begin{tikzpicture} [scale=1.9]
\node (v1) at (-4.6,-0.5) {$\bF_5^{9-m}$};
\node (v2) at (-2.8,-0.5) {$\bF_3$};
\node (v3) at (-1.3,-0.5) {$\dP$};
\draw  (v1) edge (v2);
\draw  (v2) edge (v3);
\node at (-4.1,-0.4) {\scriptsize{$e$}};
\node at (-3.1,-0.4) {\scriptsize{$h$}};
\node at (-2.5,-0.4) {\scriptsize{$e$}};
\node at (-1.6,-0.4) {\scriptsize{$l$}};
\node at (0.1,-0.5) {$0\le m\le 9$};
\draw  (-5.5,-0.2) rectangle (1.5,-1.6);
\node at (-2,-0.9) {$M=9-m$};
\node at (-2,-1.3) {$6\cF=(m-1)\phi_L^3+8\phi_M^3+9\phi_R^3
-15\phi_L\phi_M^2+9\phi_M\phi_L^2-9\phi_M\phi_R^2
+3\phi_R\phi_M^2$};
\end{tikzpicture}
\ee
Consider $m=3$ case of the above series of geometries
\be
\begin{tikzpicture} [scale=1.9]
\node (v1) at (-4.4,-0.5) {$\bF_5^{6}$};
\node (v2) at (-2.8,-0.5) {$\bF_3$};
\node (v3) at (-1.3,-0.5) {$\dP$};
\draw  (v1) edge (v2);
\draw  (v2) edge (v3);
\node at (-4.1,-0.4) {\scriptsize{$e$}};
\node at (-3.1,-0.4) {\scriptsize{$h$}};
\node at (-2.5,-0.4) {\scriptsize{$e$}};
\node at (-1.6,-0.4) {\scriptsize{$l$}};
\end{tikzpicture}
\ee
We can blow down $h-\sum x_i$ in the left surface to obtain the $5d$ SCFT
\be\label{sp2L3LtR3Rt}
\begin{tikzpicture} [scale=1.9]
\node (v1) at (-4.4,-0.5) {$\dP^{6}$};
\node (v2) at (-2.8,-0.5) {$\bF_3$};
\node (v3) at (-1.3,-0.5) {$\dP$};
\draw  (v1) edge (v2);
\draw  (v2) edge (v3);
\node at (-3.9,-0.4) {\scriptsize{$l$-$\sum x_i$}};
\node at (-3.1,-0.4) {\scriptsize{$h$}};
\node at (-2.5,-0.4) {\scriptsize{$e$}};
\node at (-1.6,-0.4) {\scriptsize{$l$}};
\draw  (-6.1,-0.2) rectangle (0.4,-1.6);
\node at (-2.8,-0.9) {$M=5$};
\node at (-2.8,-1.3) {$6\cF=3\phi_L^3+8\phi_M^3+9\phi_R^3
-15\phi_L\phi_M^2+9\phi_M\phi_L^2-9\phi_M\phi_R^2
+3\phi_R\phi_M^2$};
\end{tikzpicture}
\ee

\subsection{$M=12$}\label{su3S}
The geometry for the KK theory (\ref{su3}) for $k=1$ can be written as \cite{Bhardwaj:2018yhy,Bhardwaj:2018vuu,Bhardwaj:2019fzv}
\be\label{su36d}
\begin{tikzpicture} [scale=1.9]
\node (v1) at (-4.6,-0.5) {$\bF_1^{12}$};
\node (v2) at (-2.8,-0.5) {$\bF_7$};
\node (v3) at (-1.3,-0.5) {$\bF_3$};
\draw  (v1) edge (v2);
\draw  (v2) edge (v3);
\node at (-4.1,-0.4) {\scriptsize{$h$-$\sum x_i$}};
\node at (-3.2,-0.4) {\scriptsize{$h$+$f$}};
\node at (-2.5,-0.4) {\scriptsize{$e$}};
\node at (-1.7,-0.4) {\scriptsize{$h$+$f$}};
\draw (v1) .. controls (-4.6,-1.3) and (-1.3,-1.3) .. (v3);
\node at (-4.7,-0.8) {\scriptsize{$h$}};
\node at (-1.2,-0.8) {\scriptsize{$e$}};
\end{tikzpicture}
\ee
The way this geometry is written, it is not possible to integrate out any blowup through any surface due to the same reasons as explained in the last subsection. The reason this happens is that every single gluing curve contains the $e$ curve of that surface. However, we can perform some flops and isomorphisms to remove $e$ from one of the gluing curves as follows. First move one blowup from the left surface to the middle surface, then from the middle surface to the right surface, and then from the right surface back to the left surface, such that the blowup goes around the whole geometry in a loop. At the end of this process, one ends up with the following geometry
\be
\begin{tikzpicture} [scale=1.9]
\node (v1) at (-4.6,-0.5) {$\bF_1^{11+1}$};
\node (v2) at (-2.8,-0.5) {$\bF_6$};
\node (v3) at (-1.3,-0.5) {$\bF_2$};
\draw  (v1) edge (v2);
\draw  (v2) edge (v3);
\node at (-4,-0.4) {\scriptsize{$h$-$\sum x_i$}};
\node at (-3.2,-0.4) {\scriptsize{$h$+$f$}};
\node at (-2.5,-0.4) {\scriptsize{$e$}};
\node at (-1.7,-0.4) {\scriptsize{$h$+$f$}};
\draw (v1) .. controls (-4.6,-1.3) and (-1.3,-1.3) .. (v3);
\node at (-4.8,-0.8) {\scriptsize{$h$-$y$}};
\node at (-1.2,-0.8) {\scriptsize{$e$}};
\end{tikzpicture}
\ee
Now applying $\cI_0^{-1}$ on the left surface using the blowup $y$, we can rewrite the above as
\be
\begin{tikzpicture} [scale=1.9]
\node (v1) at (-4.6,-0.5) {$\bF_0^{12}$};
\node (v2) at (-2.8,-0.5) {$\bF_6$};
\node (v3) at (-1.3,-0.5) {$\bF_2$};
\draw  (v1) edge (v2);
\draw  (v2) edge (v3);
\node at (-4,-0.4) {\scriptsize{$e$+$f$-$\sum x_i$}};
\node at (-3.2,-0.4) {\scriptsize{$h$+$f$}};
\node at (-2.5,-0.4) {\scriptsize{$e$}};
\node at (-1.7,-0.4) {\scriptsize{$h$+$f$}};
\draw (v1) .. controls (-4.6,-1.3) and (-1.3,-1.3) .. (v3);
\node at (-4.7,-0.8) {\scriptsize{$e$}};
\node at (-1.2,-0.8) {\scriptsize{$e$}};
\end{tikzpicture}
\ee
Now exchanging $e$ and $f$ achieves our goal as the geometry at hand becomes
\be
\begin{tikzpicture} [scale=1.9]
\node (v1) at (-4.6,-0.5) {$\bF_0^{12}$};
\node (v2) at (-2.8,-0.5) {$\bF_6$};
\node (v3) at (-1.3,-0.5) {$\bF_2$};
\draw  (v1) edge (v2);
\draw  (v2) edge (v3);
\node at (-4,-0.4) {\scriptsize{$e$+$f$-$\sum x_i$}};
\node at (-3.2,-0.4) {\scriptsize{$h$+$f$}};
\node at (-2.5,-0.4) {\scriptsize{$e$}};
\node at (-1.7,-0.4) {\scriptsize{$h$+$f$}};
\draw (v1) .. controls (-4.6,-1.3) and (-1.3,-1.3) .. (v3);
\node at (-4.7,-0.8) {\scriptsize{$f$}};
\node at (-1.2,-0.8) {\scriptsize{$e$}};
\end{tikzpicture}
\ee
Now we can absorb all the twelve blowups into the left surface to obtain the frame
\be\label{su3KK}
\begin{tikzpicture} [scale=1.9]
\node (v1) at (-4.6,-0.5) {$\bF_{10}^{12}$};
\node (v2) at (-2.8,-0.5) {$\bF_6$};
\node (v3) at (-1.3,-0.5) {$\bF_2$};
\draw  (v1) edge (v2);
\draw  (v2) edge (v3);
\node at (-4.2,-0.4) {\scriptsize{$e$}};
\node at (-3.2,-0.4) {\scriptsize{$h$+$f$}};
\node at (-2.5,-0.4) {\scriptsize{$e$}};
\node at (-1.7,-0.4) {\scriptsize{$h$+$f$}};
\draw (v1) .. controls (-4.6,-1.3) and (-1.3,-1.3) .. (v3);
\node at (-4.7,-0.8) {\scriptsize{$f$}};
\node at (-1.2,-0.8) {\scriptsize{$e$}};
\end{tikzpicture}
\ee
In the above frame, we can only remove blowups from the left surface. Removing $m$ blowups gives us a series of $5d$ SCFTs described by geometries
\be\label{su3Lm}
\begin{tikzpicture} [scale=1.9]
\node (v1) at (-4.6,-0.5) {$\bF_{10}^{12-m}$};
\node (v2) at (-2.8,-0.5) {$\bF_6$};
\node (v3) at (-1.3,-0.5) {$\bF_2$};
\draw  (v1) edge (v2);
\draw  (v2) edge (v3);
\node at (-4.1,-0.4) {\scriptsize{$e$}};
\node at (-3.2,-0.4) {\scriptsize{$h$+$f$}};
\node at (-2.5,-0.4) {\scriptsize{$e$}};
\node at (-1.7,-0.4) {\scriptsize{$h$+$f$}};
\draw (v1) .. controls (-4.6,-1.3) and (-1.3,-1.3) .. (v3);
\node at (-4.7,-0.8) {\scriptsize{$f$}};
\node at (-1.2,-0.8) {\scriptsize{$e$}};
\node at (0.1,-0.5) {$1\le m\le 12$};
\draw  (-5.8,-0.2) rectangle (1.5,-2.4);
\node at (-2.1,-1.4) {$M=12-m$};
\node at (-2.1,-1.8) {$6\cF=(m-4)\phi_L^3+8\phi_M^3+8\phi_R^3
-30\phi_L\phi_M^2+24\phi_M\phi_L^2-18\phi_M\phi_R^2
+12\phi_R\phi_M^2$};
\node at (-2.1,-2.1) {$-6\phi_R\phi_L^2+6\phi_L\phi_M\phi_R$};
\end{tikzpicture}
\ee
where the coefficient of $\phi_L\phi_M\phi_R$ in $6\cF$ is computed by $6C_{L;M}\cdot C_{L;R}$. It can also be computed using $6C_{M;L}\cdot C_{M;R}$ or $6C_{R;M}\cdot C_{R;L}$. This can also be viewed as a consistency condition on the geometries since all these three expressions must match.

We can also consider first sending a blowup around the whole geometry to obtain
\be
\begin{tikzpicture} [scale=1.9]
\node (v1) at (-4.6,-0.5) {$\bF_{9}^{11+1}$};
\node (v2) at (-2.8,-0.5) {$\bF_5$};
\node (v3) at (-1.3,-0.5) {$\bF_1$};
\draw  (v1) edge (v2);
\draw  (v2) edge (v3);
\node at (-4.2,-0.4) {\scriptsize{$e$}};
\node at (-3.2,-0.4) {\scriptsize{$h$+$f$}};
\node at (-2.5,-0.4) {\scriptsize{$e$}};
\node at (-1.7,-0.4) {\scriptsize{$h$+$f$}};
\draw (v1) .. controls (-4.6,-1.3) and (-1.3,-1.3) .. (v3);
\node at (-4.8,-0.8) {\scriptsize{$f$-$y$}};
\node at (-1.2,-0.8) {\scriptsize{$e$}};
\end{tikzpicture}
\ee
and then flopping the gluing curve $f-y$ in the left surface (which also flops $e$ in the right surface) to obtain the following geometry in which a blowup is available to be integrated out from the middle surface
\be
\begin{tikzpicture} [scale=1.9]
\node (v1) at (-4.6,-0.5) {$\bF_{8}^{11}$};
\node (v2) at (-2.8,-0.5) {$\bF_6^1$};
\node (v3) at (-1.3,-0.5) {$\dP$};
\draw  (v1) edge (v2);
\draw  (v2) edge (v3);
\node at (-4.2,-0.4) {\scriptsize{$e$}};
\node at (-3.1,-0.4) {\scriptsize{$h$}};
\node at (-2.5,-0.4) {\scriptsize{$e$}};
\node at (-1.6,-0.4) {\scriptsize{$2l$}};
\end{tikzpicture}
\ee
Removing this blowup gives rise to a flop equivalent frame of the geometry having $m=1$ in the series of geometries (\ref{sp2LmR}) presented above. Similar remarks will hold true in whatever follows in this subsection: Every time we integrate out a $-1$ curve from the middle surface such that the resulting geometry has no edge between the left and right surfaces, then that geometry has already been discussed in Section \ref{sp2S}.

To obtain more RG flows, we can play the same game as above by removing the presence of $e$ curve from one more gluing curve. Notice that this gluing curve cannot be the second gluing curve on the left surface. For, if it was this curve then the intersection of the two gluing curves in the left surface cannot be positive, while the intersection of the two gluing curves in the left surface in all of the geometries above is $+1$. So, let us choose one of the gluing curves in the right surface\footnote{Note that this can be done without loss of generality. Choosing the right surface or the middle surface gives the same results, essentially because the original geometry (\ref{su36d}) has cyclic symmetry upto flops. We encourage any interested reader to check this fact in more detail.} to perform this transformation. We have two choices: we can either choose the gluing curve for the left surface or the gluing curve for the middle surface from which we will remove the $e$ curve. Let us first choose the gluing curve for the left surface. To carry out this procedure, we have to convert the right surface to $\bF_0$, which can be done by sending two blowups to the right surface in (\ref{su3KK}) and performing some isomorphisms to rewrite the geometry as
\be\label{su3KK2}
\begin{tikzpicture} [scale=1.9]
\node (v1) at (-4.6,-0.5) {$\bF_{8}^{10}$};
\node (v2) at (-2.8,-0.5) {$\bF_4$};
\node (v3) at (-1.3,-0.5) {$\bF^2_0$};
\draw  (v1) edge (v2);
\draw  (v2) edge (v3);
\node at (-4.2,-0.4) {\scriptsize{$e$}};
\node at (-3.2,-0.4) {\scriptsize{$h$+$f$}};
\node at (-2.5,-0.4) {\scriptsize{$e$}};
\node at (-1.7,-0.4) {\scriptsize{$e$+$f$}};
\draw (v1) .. controls (-4.6,-1.3) and (-1.3,-1.3) .. (v3);
\node at (-4.7,-0.8) {\scriptsize{$f$}};
\node at (-1,-0.8) {\scriptsize{$e$-$\sum x_i$}};
\end{tikzpicture}
\ee
Now interchanging $e$ and $f$ in the right surface we obtain the geometry
\be
\begin{tikzpicture} [scale=1.9]
\node (v1) at (-4.6,-0.5) {$\bF_{8}^{10}$};
\node (v2) at (-2.8,-0.5) {$\bF_4$};
\node (v3) at (-1.3,-0.5) {$\bF^2_0$};
\draw  (v1) edge (v2);
\draw  (v2) edge (v3);
\node at (-4.2,-0.4) {\scriptsize{$e$}};
\node at (-3.2,-0.4) {\scriptsize{$h$+$f$}};
\node at (-2.5,-0.4) {\scriptsize{$e$}};
\node at (-1.7,-0.4) {\scriptsize{$e$+$f$}};
\draw (v1) .. controls (-4.6,-1.3) and (-1.3,-1.3) .. (v3);
\node at (-4.7,-0.8) {\scriptsize{$f$}};
\node at (-1,-0.8) {\scriptsize{$f$-$\sum x_i$}};
\end{tikzpicture}
\ee
which allows us to integrate out blowups either from the left surface or from the right surface. To obtain any new $5d$ SCFT not accounted above, we must integrate out at least one blowup from the right surface, otherwise we can simply reverse the isomorphisms performed on the right surface to go back to geometries of the form already considered before. Thus, the new $5d$ SCFTs we obtain are
\be\label{su3LmRp}
\begin{tikzpicture} [scale=1.9]
\node (v1) at (-4.6,-0.5) {$\bF_{8-p}^{10-m-p}$};
\node (v2) at (-2.8,-0.5) {$\bF_{4-p}$};
\node (v3) at (-1.3,-0.5) {$\bF^2_{p-2}$};
\draw  (v1) edge (v2);
\draw  (v2) edge (v3);
\node at (-4,-0.4) {\scriptsize{$e$}};
\node at (-3.3,-0.4) {\scriptsize{$h$+$f$}};
\node at (-2.4,-0.4) {\scriptsize{$e$}};
\node at (-1.7,-0.4) {\scriptsize{$e$}};
\draw (v1) .. controls (-4.6,-1.3) and (-1.3,-1.3) .. (v3);
\node at (-4.7,-0.8) {\scriptsize{$f$}};
\node at (-1,-0.8) {\scriptsize{$f$-$\sum x_i$}};
\node at (0.8,-0.5) {$p\le m\le10-p$};
\node at (0.8,-0.9) {$1\le p\le 5$};
\draw  (-5.2,-0.2) rectangle (1.8,-2.4);
\node at (-1.6,-1.4) {$M=12-m-p$};
\node at (-1.6,-1.8) {$6\cF=(m+p-2)\phi_L^3+8\phi_M^3+6\phi_R^3
+3(p-8)\phi_L\phi_M^2+3(6-p)\phi_M\phi_L^2$};
\node at (-1.4,-2.1) {$+3(p-4)\phi_M\phi_R^2+3(2-p)\phi_R\phi_M^2
-6\phi_R\phi_L^2+6\phi_L\phi_M\phi_R$};
\end{tikzpicture}
\ee
where we have integrated out $p$ blowups from the right surface and $m$ blowups from the left surface.

Now let us instead choose to remove the $e$ curve from the gluing curve for the middle surface living in the right surface. To do this, we must send two more blowups onto the right surface in (\ref{su3KK2}) to obtain
\be
\begin{tikzpicture} [scale=1.9]
\node (v1) at (-4.6,-0.5) {$\bF_{6}^{8}$};
\node (v2) at (-2.8,-0.5) {$\bF_2$};
\node (v3) at (-1.3,-0.5) {$\bF^4_0$};
\draw  (v1) edge (v2);
\draw  (v2) edge (v3);
\node at (-4.2,-0.4) {\scriptsize{$e$}};
\node at (-3.2,-0.4) {\scriptsize{$h$+$f$}};
\node at (-2.5,-0.4) {\scriptsize{$e$}};
\node at (-1.6,-0.4) {\scriptsize{$e$}};
\draw (v1) .. controls (-4.6,-1.3) and (-1.3,-1.3) .. (v3);
\node at (-4.7,-0.8) {\scriptsize{$f$}};
\node at (-0.9,-0.8) {\scriptsize{$e$+$f$-$\sum x_i$}};
\end{tikzpicture}
\ee
Interchanging $e$ and $f$ on the right surface now achieves our goal
\be
\begin{tikzpicture} [scale=1.9]
\node (v1) at (-4.6,-0.5) {$\bF_{6}^{8}$};
\node (v2) at (-2.8,-0.5) {$\bF_2$};
\node (v3) at (-1.3,-0.5) {$\bF^4_0$};
\draw  (v1) edge (v2);
\draw  (v2) edge (v3);
\node at (-4.2,-0.4) {\scriptsize{$e$}};
\node at (-3.2,-0.4) {\scriptsize{$h$+$f$}};
\node at (-2.5,-0.4) {\scriptsize{$e$}};
\node at (-1.6,-0.4) {\scriptsize{$f$}};
\draw (v1) .. controls (-4.6,-1.3) and (-1.3,-1.3) .. (v3);
\node at (-4.7,-0.8) {\scriptsize{$f$}};
\node at (-0.9,-0.8) {\scriptsize{$e$+$f$-$\sum x_i$}};
\end{tikzpicture}
\ee
We can now rewrite the above as
\be\label{su3KK3}
\begin{tikzpicture} [scale=1.9]
\node (v1) at (-4.6,-0.5) {$\bF_{6}^{8}$};
\node (v2) at (-2.8,-0.5) {$\bF_2$};
\node (v3) at (-1.3,-0.5) {$\bF^4_2$};
\draw  (v1) edge (v2);
\draw  (v2) edge (v3);
\node at (-4.2,-0.4) {\scriptsize{$e$}};
\node at (-3.2,-0.4) {\scriptsize{$h$+$f$}};
\node at (-2.5,-0.4) {\scriptsize{$e$}};
\node at (-1.6,-0.4) {\scriptsize{$f$}};
\draw (v1) .. controls (-4.6,-1.3) and (-1.3,-1.3) .. (v3);
\node at (-4.7,-0.8) {\scriptsize{$f$}};
\node at (-1.2,-0.8) {\scriptsize{$e$}};
\end{tikzpicture}
\ee
Integrating out $m$ blowups from the left surface and $p$ blowups from the right surface, we obtain
\be\label{su3LmRtp}
\begin{tikzpicture} [scale=1.9]
\node (v1) at (-4.6,-0.5) {$\bF_{6}^{8-m}$};
\node (v2) at (-2.8,-0.5) {$\bF_2$};
\node (v3) at (-1.3,-0.5) {$\bF^{4-p}_2$};
\draw  (v1) edge (v2);
\draw  (v2) edge (v3);
\node at (-4.2,-0.4) {\scriptsize{$e$}};
\node at (-3.2,-0.4) {\scriptsize{$h$+$f$}};
\node at (-2.5,-0.4) {\scriptsize{$e$}};
\node at (-1.7,-0.4) {\scriptsize{$f$}};
\draw (v1) .. controls (-4.6,-1.3) and (-1.3,-1.3) .. (v3);
\node at (-4.7,-0.8) {\scriptsize{$f$}};
\node at (-1.2,-0.8) {\scriptsize{$e$}};
\end{tikzpicture}
\ee
However, notice that if $p\le 2$, then we do not obtain any new geometries since we can relate the above series of geometries to (\ref{su3LmRp}) by isomorphisms as described below. We use two blowups on the right surface to convert the description to
\be
\begin{tikzpicture} [scale=1.9]
\node (v1) at (-4.6,-0.5) {$\bF_{6}^{8-m}$};
\node (v2) at (-2.8,-0.5) {$\bF_2$};
\node (v3) at (-1.3,-0.5) {$\bF^{2+(2-p)}_2$};
\draw  (v1) edge (v2);
\draw  (v2) edge (v3);
\node at (-4.2,-0.4) {\scriptsize{$e$}};
\node at (-3.2,-0.4) {\scriptsize{$h$+$f$}};
\node at (-2.5,-0.4) {\scriptsize{$e$}};
\node at (-1.8,-0.4) {\scriptsize{$f$}};
\draw (v1) .. controls (-4.6,-1.3) and (-1.3,-1.3) .. (v3);
\node at (-4.7,-0.8) {\scriptsize{$f$}};
\node at (-1,-0.8) {\scriptsize{$e$-$\sum x_i$}};
\end{tikzpicture}
\ee
and then interchange $e$ and $f$ on the right surface to obtain
\be
\begin{tikzpicture} [scale=1.9]
\node (v1) at (-4.6,-0.5) {$\bF_{6}^{8-m}$};
\node (v2) at (-2.8,-0.5) {$\bF_2$};
\node (v3) at (-1.3,-0.5) {$\bF^{2+(2-p)}_2$};
\draw  (v1) edge (v2);
\draw  (v2) edge (v3);
\node at (-4.2,-0.4) {\scriptsize{$e$}};
\node at (-3.2,-0.4) {\scriptsize{$h$+$f$}};
\node at (-2.5,-0.4) {\scriptsize{$e$}};
\node at (-1.8,-0.4) {\scriptsize{$e$}};
\draw (v1) .. controls (-4.6,-1.3) and (-1.3,-1.3) .. (v3);
\node at (-4.7,-0.8) {\scriptsize{$f$}};
\node at (-1,-0.8) {\scriptsize{$f$-$\sum x_i$}};
\end{tikzpicture}
\ee
which is flop equivalent to geometries (\ref{su3LmRp}). So first substituting $p=3$, we obtain a series of $5d$ SCFTs not described above
\be
\begin{tikzpicture} [scale=1.9]
\node (v1) at (-4.6,-0.5) {$\bF_{6}^{8-m}$};
\node (v2) at (-2.8,-0.5) {$\bF_2$};
\node (v3) at (-1.3,-0.5) {$\bF^{1}_2$};
\draw  (v1) edge (v2);
\draw  (v2) edge (v3);
\node at (-4.2,-0.4) {\scriptsize{$e$}};
\node at (-3.2,-0.4) {\scriptsize{$h$+$f$}};
\node at (-2.5,-0.4) {\scriptsize{$e$}};
\node at (-1.6,-0.4) {\scriptsize{$f$}};
\draw (v1) .. controls (-4.6,-1.3) and (-1.3,-1.3) .. (v3);
\node at (-4.7,-0.8) {\scriptsize{$f$}};
\node at (-1.2,-0.8) {\scriptsize{$e$}};
\node at (0.4,-0.7) {$1\le m\le 8$};
\draw  (-5.8,-0.2) rectangle (1.8,-2.1);
\node at (-2,-1.4) {$M=9-m$};
\node at (-2,-1.8) {$6\cF=m\phi_L^3+8\phi_M^3+7\phi_R^3
-18\phi_L\phi_M^2+12\phi_M\phi_L^2-6\phi_M\phi_R^2
-6\phi_R\phi_L^2+6\phi_L\phi_M\phi_R$};
\end{tikzpicture}
\ee
Substituting $p=4$, we obtain a series of $5d$ SCFTs
\be
\begin{tikzpicture} [scale=1.9]
\node (v1) at (-4.6,-0.5) {$\bF_{6}^{8-m}$};
\node (v2) at (-2.8,-0.5) {$\bF_2$};
\node (v3) at (-1.3,-0.5) {$\bF_2$};
\draw  (v1) edge (v2);
\draw  (v2) edge (v3);
\node at (-4.2,-0.4) {\scriptsize{$e$}};
\node at (-3.2,-0.4) {\scriptsize{$h$+$f$}};
\node at (-2.5,-0.4) {\scriptsize{$e$}};
\node at (-1.6,-0.4) {\scriptsize{$f$}};
\draw (v1) .. controls (-4.6,-1.3) and (-1.3,-1.3) .. (v3);
\node at (-4.7,-0.8) {\scriptsize{$f$}};
\node at (-1.2,-0.8) {\scriptsize{$e$}};
\node at (0.4,-0.7) {$1\le m\le 8$};
\draw  (-5.8,-0.2) rectangle (1.8,-2.1);
\node at (-1.9,-1.4) {$M=8-m$};
\node at (-2,-1.8) {$6\cF=m\phi_L^3+8\phi_M^3+8\phi_R^3
-18\phi_L\phi_M^2+12\phi_M\phi_L^2-6\phi_M\phi_R^2
-6\phi_R\phi_L^2+6\phi_L\phi_M\phi_R$};
\end{tikzpicture}
\ee

Just like we saw that we can exchange $e$ and $f$ on the right surface in (\ref{su3KK3}), we can also exchange $e$ and $f$ on the left surface by using six of the blowups and rewrite (\ref{su3KK3}) as
\be
\begin{tikzpicture} [scale=1.9]
\node (v1) at (-4.6,-0.5) {$\bF_{0}^{6+2}$};
\node (v2) at (-2.8,-0.5) {$\bF_2$};
\node (v3) at (-1.3,-0.5) {$\bF^4_2$};
\draw  (v1) edge (v2);
\draw  (v2) edge (v3);
\node at (-4,-0.4) {\scriptsize{$f$-$\sum x_i$}};
\node at (-3.2,-0.4) {\scriptsize{$h$+$f$}};
\node at (-2.5,-0.4) {\scriptsize{$e$}};
\node at (-1.6,-0.4) {\scriptsize{$f$}};
\draw (v1) .. controls (-4.6,-1.3) and (-1.3,-1.3) .. (v3);
\node at (-4.7,-0.8) {\scriptsize{$e$}};
\node at (-1.2,-0.8) {\scriptsize{$e$}};
\end{tikzpicture}
\ee
Transferring the four blowups on the right surface to the left surface, we obtain
\be\label{su3KK4}
\begin{tikzpicture} [scale=1.9]
\node (v1) at (-4.6,-0.5) {$\bF_{4}^{6+6}$};
\node (v2) at (-2.8,-0.5) {$\bF_2$};
\node (v3) at (-1.3,-0.5) {$\bF_0$};
\draw  (v1) edge (v2);
\draw  (v2) edge (v3);
\node at (-4,-0.4) {\scriptsize{$f$-$\sum x_i$}};
\node at (-3.2,-0.4) {\scriptsize{$h$+$f$}};
\node at (-2.5,-0.4) {\scriptsize{$e$}};
\node at (-1.6,-0.4) {\scriptsize{$f$}};
\draw (v1) .. controls (-4.6,-1.3) and (-1.3,-1.3) .. (v3);
\node at (-4.7,-0.8) {\scriptsize{$e$}};
\node at (-1.1,-0.8) {\scriptsize{$e$+$f$}};
\end{tikzpicture}
\ee
Integrating out blowups from the left surface in the above geometry does not yield new $5d$ SCFTs since we can exchange $e$ and $f$ on the right surface to map the resulting geometries to geometries of the form (\ref{su3Lm}). So we must integrate out at least one blowup from the right surface. Suppose furthermore we integrate out $m$ blowups from the left surface
\be
\begin{tikzpicture} [scale=1.9]
\node (v1) at (-4.8,-0.5) {$\bF_{3}^{6+(5-m)}$};
\node (v2) at (-2.8,-0.5) {$\bF_2$};
\node (v3) at (-1.3,-0.5) {$\bF_1$};
\draw  (v1) edge (v2);
\draw  (v2) edge (v3);
\node at (-4,-0.4) {\scriptsize{$f$-$\sum x_i$}};
\node at (-3.2,-0.4) {\scriptsize{$h$+$f$}};
\node at (-2.5,-0.4) {\scriptsize{$e$}};
\node at (-1.6,-0.4) {\scriptsize{$f$}};
\draw (v1) .. controls (-4.8,-1.3) and (-1.3,-1.3) .. (v3);
\node at (-4.9,-0.8) {\scriptsize{$e$}};
\node at (-1.2,-0.8) {\scriptsize{$h$}};
\end{tikzpicture}
\ee
If $m\le4$, then we can write the above geometry as
\be
\begin{tikzpicture} [scale=1.9]
\node (v1) at (-4.8,-0.5) {$\bF_{2}^{6+(4-m)}$};
\node (v2) at (-2.8,-0.5) {$\bF_2$};
\node (v3) at (-1.3,-0.5) {$\bF^1_0$};
\draw  (v1) edge (v2);
\draw  (v2) edge (v3);
\node at (-4,-0.4) {\scriptsize{$f$-$\sum x_i$}};
\node at (-3.2,-0.4) {\scriptsize{$h$+$f$}};
\node at (-2.5,-0.4) {\scriptsize{$e$}};
\node at (-1.6,-0.4) {\scriptsize{$f$}};
\draw (v1) .. controls (-4.8,-1.3) and (-1.3,-1.3) .. (v3);
\node at (-4.9,-0.8) {\scriptsize{$e$}};
\node at (-1.2,-0.8) {\scriptsize{$e$}};
\end{tikzpicture}
\ee
which can be mapped to geometries of the form (\ref{su3LmRtp}) upto flops and exchange of surfaces. Thus the only way to obtain a new SCFT is to susbstitute $m=5$
\be
\begin{tikzpicture} [scale=1.9]
\node (v1) at (-4.6,-0.5) {$\bF_{3}^{6}$};
\node (v2) at (-2.8,-0.5) {$\bF_2$};
\node (v3) at (-1.3,-0.5) {$\bF_1$};
\draw  (v1) edge (v2);
\draw  (v2) edge (v3);
\node at (-4.1,-0.4) {\scriptsize{$f$-$\sum x_i$}};
\node at (-3.2,-0.4) {\scriptsize{$h$+$f$}};
\node at (-2.5,-0.4) {\scriptsize{$e$}};
\node at (-1.6,-0.4) {\scriptsize{$f$}};
\draw (v1) .. controls (-4.6,-1.3) and (-1.3,-1.3) .. (v3);
\node at (-4.7,-0.8) {\scriptsize{$e$}};
\node at (-1.2,-0.8) {\scriptsize{$h$}};
\draw  (-7.1,-0.2) rectangle (1.3,-2.1);
\node at (-2.9,-1.4) {$M=6$};
\node at (-2.9,-1.8) {$6\cF=2\phi_L^3+8\phi_M^3+8\phi_R^3
-18\phi_L\phi_M^2+12\phi_M\phi_L^2-6\phi_M\phi_R^2
+3\phi_R\phi_L^2-9\phi_L\phi_R^2+6\phi_L\phi_M\phi_R$};
\end{tikzpicture}
\ee
Integrating out two blowups from the right surface and $m$ blowups from the left surface in (\ref{su3KK4}) leads to
\be
\begin{tikzpicture} [scale=1.9]
\node (v1) at (-4.7,-0.5) {$\bF_{2}^{6+(4-m)}$};
\node (v2) at (-2.8,-0.5) {$\bF_2$};
\node (v3) at (-1.3,-0.5) {$\bF_0$};
\draw  (v1) edge (v2);
\draw  (v2) edge (v3);
\node at (-4,-0.4) {\scriptsize{$f$-$\sum x_i$}};
\node at (-3.2,-0.4) {\scriptsize{$h$+$f$}};
\node at (-2.5,-0.4) {\scriptsize{$e$}};
\node at (-1.6,-0.4) {\scriptsize{$f$}};
\draw (v1) .. controls (-4.7,-1.3) and (-1.3,-1.3) .. (v3);
\node at (-4.8,-0.8) {\scriptsize{$e$}};
\node at (-1.2,-0.8) {\scriptsize{$e$}};
\end{tikzpicture}
\ee
which can always be made equivalent to geometries of the form (\ref{su3LmRtp}) irrespective of the value of $m$. Integrating out three blowups from the right and $m$ blowups from the left in (\ref{su3KK4}) yields
\be\label{su3LtmRt3}
\begin{tikzpicture} [scale=1.9]
\node (v1) at (-4.7,-0.5) {$\bF_{1}^{6+(3-m)}$};
\node (v2) at (-2.8,-0.5) {$\bF_2$};
\node (v3) at (-1.3,-0.5) {$\bF_1$};
\draw  (v1) edge (v2);
\draw  (v2) edge (v3);
\node at (-4,-0.4) {\scriptsize{$f$-$\sum x_i$}};
\node at (-3.2,-0.4) {\scriptsize{$h$+$f$}};
\node at (-2.5,-0.4) {\scriptsize{$e$}};
\node at (-1.6,-0.4) {\scriptsize{$f$}};
\draw (v1) .. controls (-4.7,-1.3) and (-1.3,-1.3) .. (v3);
\node at (-4.8,-0.8) {\scriptsize{$e$}};
\node at (-1.2,-0.8) {\scriptsize{$e$}};
\end{tikzpicture}
\ee
which can be converted into geometries of the form (\ref{su3LmRtp}) upto flops as long as $m\le2$, since the above geometry is isomorphic to the geometry
\be
\begin{tikzpicture} [scale=1.9]
\node (v1) at (-4.7,-0.5) {$\bF_{0}^{6+1+(2-m)}$};
\node (v2) at (-2.8,-0.5) {$\bF_2$};
\node (v3) at (-1.3,-0.5) {$\bF_1$};
\draw  (v1) edge (v2);
\draw  (v2) edge (v3);
\node at (-3.9,-0.4) {\scriptsize{$f$-$\sum x_i$}};
\node at (-3.2,-0.4) {\scriptsize{$h$+$f$}};
\node at (-2.5,-0.4) {\scriptsize{$e$}};
\node at (-1.6,-0.4) {\scriptsize{$f$}};
\draw (v1) .. controls (-4.7,-1.3) and (-1.3,-1.3) .. (v3);
\node at (-4.9,-0.8) {\scriptsize{$e$-$y$}};
\node at (-1.2,-0.8) {\scriptsize{$e$}};
\end{tikzpicture}
\ee
in which we can exchange $e$ and $f$ on the first surface. Thus the only new $5d$ SCFT we obtain from geometries of the form (\ref{su3LtmRt3}) is
\be
\begin{tikzpicture} [scale=1.9]
\node (v1) at (-4.6,-0.5) {$\bF_{1}^{6}$};
\node (v2) at (-2.8,-0.5) {$\bF_2$};
\node (v3) at (-1.3,-0.5) {$\bF_1$};
\draw  (v1) edge (v2);
\draw  (v2) edge (v3);
\node at (-4.1,-0.4) {\scriptsize{$f$-$\sum x_i$}};
\node at (-3.2,-0.4) {\scriptsize{$h$+$f$}};
\node at (-2.5,-0.4) {\scriptsize{$e$}};
\node at (-1.6,-0.4) {\scriptsize{$f$}};
\draw (v1) .. controls (-4.6,-1.3) and (-1.3,-1.3) .. (v3);
\node at (-4.7,-0.8) {\scriptsize{$e$}};
\node at (-1.2,-0.8) {\scriptsize{$e$}};
\draw  (-7.1,-0.2) rectangle (1.3,-2.1);
\node at (-2.9,-1.4) {$M=6$};
\node at (-2.9,-1.8) {$6\cF=2\phi_L^3+8\phi_M^3+8\phi_R^3
-18\phi_L\phi_M^2+12\phi_M\phi_L^2-6\phi_M\phi_R^2
-3\phi_R\phi_L^2-3\phi_L\phi_R^2+6\phi_L\phi_M\phi_R$};
\end{tikzpicture}
\ee
Cases involving integrating out of more than three blowups from the right surface and $m$ blowups from the left surface in (\ref{su3KK4}) have already been discussed earlier since the geometry (\ref{su3KK4}) is left-right symmetric upto flops.

To obtain even more RG flows, we can continue playing the same game as above by removing the presence of $e$ curve from yet another gluing curve. Due to reasons already explained before, this time we must choose a gluing curve living in the middle surface to perform this operation. Without loss of generality, since the geometry for KK theory is left-right symmetric, we can choose it to be the gluing curve for the left surface inside the middle surface. We obtain the following new $5d$ SCFTs this way:
\be
\begin{tikzpicture} [scale=1.9]
\node (v1) at (-4.6,-0.5) {$\bF_{2}^{4-m}$};
\node (v2) at (-2.8,-0.5) {$\bF^{6-n-p}_{4-p}$};
\node (v3) at (-1.3,-0.5) {$\bF^2_{p-2}$};
\draw  (v1) edge (v2);
\draw  (v2) edge (v3);
\node at (-4.2,-0.4) {\scriptsize{$e$}};
\node at (-3.3,-0.4) {\scriptsize{$f$}};
\node at (-2.3,-0.4) {\scriptsize{$e$}};
\node at (-1.7,-0.4) {\scriptsize{$e$}};
\draw (v1) .. controls (-4.6,-1.3) and (-1.3,-1.3) .. (v3);
\node at (-4.7,-0.8) {\scriptsize{$f$}};
\node at (-1,-0.8) {\scriptsize{$f$-$\sum x_i$}};
\node at (0.8,-0.5) {$3\le p\le m\le4$};
\node at (0.8,-0.9) {$1\le n\le3$};
\draw  (-5,-0.2) rectangle (1.8,-2.4);
\node at (-1.6,-1.4) {$M=12-m-n-p$};
\node at (-1.6,-1.8) {$6\cF=(m+4)\phi_L^3+(n+p+2)\phi_M^3+6\phi_R^3
-6\phi_L\phi_M^2+3(p-4)\phi_M\phi_R^2$};
\node at (-1.6,-2.1) {$+3(2-p)\phi_R\phi_M^2-6\phi_R\phi_L^2+6\phi_L\phi_M\phi_R$};
\end{tikzpicture}
\ee
\be\label{su3LmM5R}
\begin{tikzpicture} [scale=1.9]
\node (v1) at (-4.6,-0.5) {$\bF_{2}^{4-m}$};
\node (v2) at (-2.8,-0.5) {$\bF_{3}$};
\node (v3) at (-1.3,-0.5) {$\bF^2_{1}$};
\draw  (v1) edge (v2);
\draw  (v2) edge (v3);
\node at (-4.2,-0.4) {\scriptsize{$e$}};
\node at (-3.1,-0.4) {\scriptsize{$f$}};
\node at (-2.5,-0.4) {\scriptsize{$e$}};
\node at (-1.6,-0.4) {\scriptsize{$h$}};
\draw (v1) .. controls (-4.6,-1.3) and (-1.3,-1.3) .. (v3);
\node at (-4.7,-0.8) {\scriptsize{$f$}};
\node at (-1,-0.8) {\scriptsize{$f$-$\sum x_i$}};
\node at (0.8,-0.7) {$1\le m\le4$};
\draw  (-5.8,-0.2) rectangle (2.2,-2.1);
\node at (-1.8,-1.4) {$M=6-m$};
\node at (-1.8,-1.8) {$6\cF=(m+4)\phi_L^3+8\phi_M^3+6\phi_R^3
-6\phi_L\phi_M^2-9\phi_M\phi_R^2
+3\phi_R\phi_M^2-6\phi_R\phi_L^2+6\phi_L\phi_M\phi_R$};
\end{tikzpicture}
\ee
\be
\begin{tikzpicture} [scale=1.9]
\node (v1) at (-4.6,-0.5) {$\bF_{2}^{4-m}$};
\node (v2) at (-2.8,-0.5) {$\bF_{1}$};
\node (v3) at (-1.3,-0.5) {$\bF^2_{3}$};
\draw  (v1) edge (v2);
\draw  (v2) edge (v3);
\node at (-4.2,-0.4) {\scriptsize{$e$}};
\node at (-3.1,-0.4) {\scriptsize{$f$}};
\node at (-2.5,-0.4) {\scriptsize{$h$}};
\node at (-1.6,-0.4) {\scriptsize{$e$}};
\draw (v1) .. controls (-4.6,-1.3) and (-1.3,-1.3) .. (v3);
\node at (-4.7,-0.8) {\scriptsize{$f$}};
\node at (-1,-0.8) {\scriptsize{$f$-$\sum x_i$}};
\node at (0.8,-0.7) {$3\le m\le4$};
\draw  (-5.8,-0.2) rectangle (2.2,-2.1);
\node at (-1.7,-1.4) {$M=6-m$};
\node at (-1.8,-1.8) {$6\cF=(m+4)\phi_L^3+8\phi_M^3+6\phi_R^3
-6\phi_L\phi_M^2+3\phi_M\phi_R^2
-9\phi_R\phi_M^2-6\phi_R\phi_L^2+6\phi_L\phi_M\phi_R$};
\end{tikzpicture}
\ee
\be\label{su3L5MR5}
\begin{tikzpicture} [scale=1.9]
\node (v1) at (-4.6,-0.5) {$\bF_{3}$};
\node (v2) at (-2.8,-0.5) {$\dP$};
\node (v3) at (-1.3,-0.5) {$\bF^2_{3}$};
\draw  (v1) edge (v2);
\draw  (v2) edge (v3);
\node at (-4.3,-0.4) {\scriptsize{$e$}};
\node at (-3.1,-0.4) {\scriptsize{$l$}};
\node at (-2.5,-0.4) {\scriptsize{$l$}};
\node at (-1.6,-0.4) {\scriptsize{$e$}};
\draw (v1) .. controls (-4.6,-1.3) and (-1.3,-1.3) .. (v3);
\node at (-4.7,-0.8) {\scriptsize{$f$}};
\node at (-1,-0.8) {\scriptsize{$f$-$\sum x_i$}};
\draw  (-6.8,-0.2) rectangle (1.4,-2.1);
\node at (-2.7,-1.4) {$M=1$};
\node at (-2.7,-1.8) {$6\cF=8\phi_L^3+9\phi_M^3+6\phi_R^3
-9\phi_L\phi_M^2+3\phi_M\phi_L^2+3\phi_M\phi_R^2
-9\phi_R\phi_M^2-6\phi_R\phi_L^2+6\phi_L\phi_M\phi_R$};
\end{tikzpicture}
\ee
\be
\begin{tikzpicture} [scale=1.9]
\node (v1) at (-4.6,-0.5) {$\bF_{2}^{4-m}$};
\node (v2) at (-2.8,-0.5) {$\bF^{4-n}_2$};
\node (v3) at (-1.3,-0.5) {$\bF^{4-p}_2$};
\draw  (v1) edge (v2);
\draw  (v2) edge (v3);
\node at (-4.2,-0.4) {\scriptsize{$e$}};
\node at (-3.2,-0.4) {\scriptsize{$f$}};
\node at (-2.4,-0.4) {\scriptsize{$e$}};
\node at (-1.7,-0.4) {\scriptsize{$f$}};
\draw (v1) .. controls (-4.6,-1.3) and (-1.3,-1.3) .. (v3);
\node at (-4.7,-0.8) {\scriptsize{$f$}};
\node at (-1.2,-0.8) {\scriptsize{$e$}};
\node at (0.4,-0.7) {$1\le p\le n\le m\le4$};
\draw  (-5.7,-0.2) rectangle (2.5,-2.1);
\node at (-1.6,-1.4) {$M=12-m-n-p$};
\node at (-1.6,-1.8) {$6\cF=(m+4)\phi_L^3+(n+4)\phi_M^3+(p+4)\phi_R^3
-6\phi_L\phi_M^2-6\phi_M\phi_R^2
-6\phi_R\phi_L^2+6\phi_L\phi_M\phi_R$};
\end{tikzpicture}
\ee

\subsection{$M=11$}
The KK theory (\ref{sp0su2}) has geometry \cite{Bhardwaj:2018vuu,Bhardwaj:2019fzv}
\be
\begin{tikzpicture} [scale=1.9]
\node (v1) at (-4.6,-0.5) {$\bF_0^{4}$};
\node (v2) at (-2.6,-0.5) {$\dP^{7+2}$};
\node (v3) at (-1,-0.5) {$\bF_2$};
\draw  (v1) edge (v2);
\draw  (v2) edge (v3);
\node at (-4.3,-0.4) {\scriptsize{$f$}};
\node at (-3.4,-0.4) {\scriptsize{$3l$-$\sum x_i$-$2y_1$}};
\node at (-2,-0.4) {\scriptsize{$y_1$-$y_2$}};
\node at (-1.3,-0.4) {\scriptsize{$f$}};
\node at (-4.9,-0.8) {\scriptsize{$e,e$-$\sum x_i$}};
\node at (-0.8,-0.8) {\scriptsize{$e,h$}};
\node (v4) at (-2.8,-1.1) {\scriptsize{2}};
\draw (v1) .. controls (-4.6,-1) and (-3.5,-1.1) .. (v4);
\draw (v3) .. controls (-1,-1) and (-2.1,-1.1) .. (v4);
\end{tikzpicture}
\ee
where the label $2$ in the middle of the edge between the left and the right surface signifies that there are two gluing curves between these two surfaces. It can be easily seen that the above is equivalent to
\be
\begin{tikzpicture} [scale=1.9]
\node (v1) at (-4.6,-0.5) {$\bF_0^{2}$};
\node (v2) at (-2.6,-0.5) {$\bF_1^{7+1}$};
\node (v3) at (-1,-0.5) {$\bF^2_0$};
\draw  (v1) edge (v2);
\draw  (v2) edge (v3);
\node at (-4.3,-0.4) {\scriptsize{$f$}};
\node at (-3.3,-0.4) {\scriptsize{$h$+$2f$-$\sum x_i$}};
\node at (-2.1,-0.4) {\scriptsize{$e$-$y$}};
\node at (-1.3,-0.4) {\scriptsize{$f$}};
\node at (-4.9,-0.8) {\scriptsize{$e,e$-$\sum x_i$}};
\node at (-0.7,-0.8) {\scriptsize{$e$-$\sum x_i,e$}};
\node (v4) at (-2.8,-1.1) {\scriptsize{2}};
\draw (v1) .. controls (-4.6,-1) and (-3.5,-1.1) .. (v4);
\draw (v3) .. controls (-1,-1) and (-2.1,-1.1) .. (v4);
\end{tikzpicture}
\ee
which can be rewritten as
\be
\begin{tikzpicture} [scale=1.9]
\node (v1) at (-4.6,-0.5) {$\bF_0^{2}$};
\node (v2) at (-2.5,-0.5) {$\bF_2^{8}$};
\node (v3) at (-1,-0.5) {$\bF^2_0$};
\draw  (v1) edge (v2);
\draw  (v2) edge (v3);
\node at (-4.3,-0.4) {\scriptsize{$e$}};
\node at (-3.1,-0.4) {\scriptsize{$h$+$2f$-$\sum x_i$}};
\node at (-2.2,-0.4) {\scriptsize{$e$}};
\node at (-1.3,-0.4) {\scriptsize{$e$}};
\node at (-4.9,-0.8) {\scriptsize{$f,f$-$\sum x_i$}};
\node at (-0.7,-0.8) {\scriptsize{$f$-$\sum x_i,f$}};
\node (v4) at (-2.8,-1.1) {\scriptsize{2}};
\draw (v1) .. controls (-4.6,-1) and (-3.5,-1.1) .. (v4);
\draw (v3) .. controls (-1,-1) and (-2.1,-1.1) .. (v4);
\end{tikzpicture}
\ee
which is equivalent to
\be\label{sp0su2KK}
\begin{tikzpicture} [scale=1.9]
\node (v1) at (-4.6,-0.5) {$\bF^{8}_{8}$};
\node (v2) at (-2.7,-0.5) {$\bF_2$};
\node (v3) at (-1,-0.5) {$\bF^{2+2}_0$};
\draw  (v1) edge (v2);
\draw  (v2) edge (v3);
\node at (-4.3,-0.4) {\scriptsize{$e$}};
\node at (-3.1,-0.4) {\scriptsize{$h$+$2f$}};
\node at (-2.4,-0.4) {\scriptsize{$e$}};
\node at (-1.4,-0.4) {\scriptsize{$e$}};
\node at (-4.8,-0.8) {\scriptsize{$f,f$}};
\node at (-0.4,-0.8) {\scriptsize{$f$-$\sum x_i,f$-$\sum y_i$}};
\node (v4) at (-2.8,-1.1) {\scriptsize{2}};
\draw (v1) .. controls (-4.6,-1) and (-3.5,-1.1) .. (v4);
\draw (v3) .. controls (-1,-1) and (-2.1,-1.1) .. (v4);
\end{tikzpicture}
\ee
Notice that the above geometry is left-right symmetric upto flop equivalences.

Integrating out $m$ blowups from the left surface we obtain the following $5d$ SCFTs
\be\label{sp0su2Lm}
\begin{tikzpicture} [scale=1.9]
\node (v1) at (-4.6,-0.5) {$\bF^{8-m}_{8}$};
\node (v2) at (-2.7,-0.5) {$\bF_2$};
\node (v3) at (-1,-0.5) {$\bF^{2+2}_0$};
\draw  (v1) edge (v2);
\draw  (v2) edge (v3);
\node at (-4.2,-0.4) {\scriptsize{$e$}};
\node at (-3.1,-0.4) {\scriptsize{$h$+$2f$}};
\node at (-2.4,-0.4) {\scriptsize{$e$}};
\node at (-1.4,-0.4) {\scriptsize{$e$}};
\node at (-4.8,-0.8) {\scriptsize{$f,f$}};
\node at (-0.4,-0.8) {\scriptsize{$f$-$\sum x_i,f$-$\sum y_i$}};
\node (v4) at (-2.8,-1.1) {\scriptsize{2}};
\draw (v1) .. controls (-4.6,-1) and (-3.5,-1.1) .. (v4);
\draw (v3) .. controls (-1,-1) and (-2.1,-1.1) .. (v4);
\node at (1.1,-0.6) {$1\le m\le 8$};
\draw  (-5.4,-0.2) rectangle (2.4,-2.1);
\node at (-1.5,-1.4) {$M=11-m$};
\node at (-1.5,-1.8) {$6\cF=m\phi_L^3+8\phi_M^3+4\phi_R^3
-24\phi_L\phi_M^2+18\phi_M\phi_L^2-6\phi_M\phi_R^2
-12\phi_R\phi_L^2+12\phi_L\phi_M\phi_R$};
\end{tikzpicture}
\ee
Integrating out $1\le p\le3$ blowups from the right surface and $m$ blowups from the left surface in (\ref{sp0su2KK}), we obtain
\be
\begin{tikzpicture} [scale=1.9]
\node (v1) at (-4.6,-0.5) {$\bF^{8-m-p}_{8-p}$};
\node (v2) at (-2.7,-0.5) {$\bF_{2-p}$};
\node (v3) at (-1,-0.5) {$\bF^{2+2}_p$};
\draw  (v1) edge (v2);
\draw  (v2) edge (v3);
\node at (-4.1,-0.4) {\scriptsize{$e$}};
\node at (-3.2,-0.4) {\scriptsize{$h$+$2f$}};
\node at (-2.3,-0.4) {\scriptsize{$e$}};
\node at (-1.4,-0.4) {\scriptsize{$e$}};
\node at (-4.8,-0.8) {\scriptsize{$f,f$}};
\node at (-0.4,-0.8) {\scriptsize{$f$-$\sum x_i,f$-$\sum y_i$}};
\node (v4) at (-2.8,-1.1) {\scriptsize{2}};
\draw (v1) .. controls (-4.6,-1) and (-3.5,-1.1) .. (v4);
\draw (v3) .. controls (-1,-1) and (-2.1,-1.1) .. (v4);
\node at (1,-0.6) {$p\le m\le 8-p$};
\draw  (-5.1,-0.2) rectangle (1.8,-2.4);
\node at (1,-1) {$1\le p\le 3$};
\node at (-1.5,-1.4) {$M=11-m-p$};
\node at (-1.5,-1.8) {$6\cF=(m+p)\phi_L^3+8\phi_M^3+4\phi_R^3
+3(p-8)\phi_L\phi_M^2+3(6-p)\phi_M\phi_L^2$};
\node at (-1.3,-2.1) {$+3(p-2)\phi_M\phi_R^2-3p\phi_R\phi_M^2-12\phi_R\phi_L^2+12\phi_L\phi_M\phi_R$};
\end{tikzpicture}
\ee
We can also integrate out four blowups each from the left and the right surfaces to obtain
\be
\begin{tikzpicture} [scale=1.9]
\node (v1) at (-4.6,-0.5) {$\bF_{4}$};
\node (v2) at (-2.8,-0.5) {$\bF_{0}$};
\node (v3) at (-1,-0.5) {$\bF^{2+2}_4$};
\draw  (v1) edge (v2);
\draw  (v2) edge (v3);
\node at (-4.3,-0.4) {\scriptsize{$e$}};
\node at (-3.2,-0.4) {\scriptsize{$e$+$f$}};
\node at (-2.4,-0.4) {\scriptsize{$e$+$f$}};
\node at (-1.4,-0.4) {\scriptsize{$e$}};
\node at (-4.8,-0.8) {\scriptsize{$f,f$}};
\node at (-0.4,-0.8) {\scriptsize{$f$-$\sum x_i,f$-$\sum y_i$}};
\node (v4) at (-2.8,-1.1) {\scriptsize{2}};
\draw (v1) .. controls (-4.6,-1) and (-3.5,-1.1) .. (v4);
\draw (v3) .. controls (-1,-1) and (-2.1,-1.1) .. (v4);
\draw  (-5.6,-0.2) rectangle (1,-2.4);
\node at (-2.3,-1.4) {$M=3$};
\node at (-2.3,-1.8) {$6\cF=8\phi_L^3+8\phi_M^3+4\phi_R^3
-12\phi_L\phi_M^2+6\phi_M\phi_L^2+6\phi_M\phi_R^2
-12\phi_R\phi_M^2$};
\node at (-2.3,-2.1) {$-12\phi_R\phi_L^2+12\phi_L\phi_M\phi_R$};
\end{tikzpicture}
\ee

We can also consider first sending a blowup from the right surface to the left surface in (\ref{sp0su2KK}) to obtain
\be
\begin{tikzpicture} [scale=1.9]
\node (v1) at (-4.6,-0.5) {$\bF^{1+8}_{8}$};
\node (v2) at (-2.7,-0.5) {$\bF_2$};
\node (v3) at (-1,-0.5) {$\bF^{1+2}_0$};
\draw  (v1) edge (v2);
\draw  (v2) edge (v3);
\node at (-4.2,-0.4) {\scriptsize{$e$}};
\node at (-3.1,-0.4) {\scriptsize{$h$+$2f$}};
\node at (-2.4,-0.4) {\scriptsize{$e$}};
\node at (-1.4,-0.4) {\scriptsize{$e$}};
\node at (-4.9,-0.8) {\scriptsize{$f$-$x,f$}};
\node at (-0.5,-0.8) {\scriptsize{$f$-$x,f$-$\sum y_i$}};
\node (v4) at (-2.8,-1.1) {\scriptsize{2}};
\draw (v1) .. controls (-4.6,-1) and (-3.5,-1.1) .. (v4);
\draw (v3) .. controls (-1,-1) and (-2.1,-1.1) .. (v4);
\end{tikzpicture}
\ee
and then flopping the gluing curve $f-x$ in the left surface (which also flops $f-x$ in the right surface) to obtain the following geometry in which a blowup is available to be integrated out from the middle surface
\be
\begin{tikzpicture} [scale=1.9]
\node (v1) at (-4.6,-0.5) {$\bF_{7}^{8}$};
\node (v2) at (-2.9,-0.5) {$\bF^1_3$};
\node (v3) at (-1.3,-0.5) {$\bF_1^2$};
\draw  (v1) edge (v2);
\draw  (v2) edge (v3);
\node at (-4.3,-0.4) {\scriptsize{$e$}};
\node at (-3.3,-0.4) {\scriptsize{$h$+$f$}};
\node at (-2.6,-0.4) {\scriptsize{$e$}};
\node at (-1.6,-0.4) {\scriptsize{$h$}};
\draw (v1) .. controls (-4.6,-1.3) and (-1.3,-1.3) .. (v3);
\node at (-4.7,-0.8) {\scriptsize{$f$}};
\node at (-1,-0.8) {\scriptsize{$f$-$\sum x_i$}};
\end{tikzpicture}
\ee
Removing this blowup gives rise to the geometry having $m=p=1$ in the series of geometries (\ref{su3LmRp}) presented above. Similar remarks will hold true in whatever follows in this subsection: Every time we integrate out a $-1$ curve from the middle surface such that the resulting geometry has only a single gluing curve between the left and right surfaces, then that geometry has already been discussed in Section \ref{su3S}. Thus, we can focus our attention exclusively on only geometries that contain two gluing curves between the left and the right surfaces.

We have learned in the previous subsection that new RG flows appear when we change the degree of one of the Hirzebruch surfaces to zero and then exchange $e$ with $f$ in that surface. Exchange $e$ and $f$ on the right surface in (\ref{sp0su2KK}) would not be of any help since we will obtain two gluing curves in the right surface containing the $e$ curve. Next, we could try to perform an automorphism on the right surface and then exchange $x$ and $y$. The only possible non-trivial automorphism involves applying $\cI_0$ using $y_1$ and then $\cI_0^{-1}$ using $x_1$ which gives
\be

\ee
But interchanging $e$ and $f$ on the right surface now leaves the geometry invariant, so this is also of no help. There is a another way that involves first sending a blowup from the left surface to the right surface to obtain
\be
%
\ee
and then applying $\cI_0^{-1}$ on the right surface using the blowup $y_1$ to rewrite the above as
\be
%
\ee
After interchanging $e$ and $f$, we can now rewrite it as
\be\label{sp0su2KK2}
%
\ee
Notice the key point that this operation has changed which gluing curve on the right surface carries the $e$ curve. Sending $y_2$ onto the left surface and further integrating out $m$ blowups from the left surface and $p$ blowups from the right surface, we obtain the following series of $5d$ SCFTs
\be\label{sp0su2LmRtp}
%
\ee

Consider the geometry (\ref{sp0su2LmRtp}) for $m=p=0$. In this geometry, we can change which gluing curve in the left surface carries the $e$ curve, obtaining
\be
%
\ee
Sending the three available blowups on the right surface onto the left surface, we obtain
\be
%
\ee
Now, integrating out $m$ blowups from the left surface and $p$ blowups from the right surface, we obtain
\be
%
\ee
But only the cases $m=p=2$ and $m=4,p=0$ lead to new $5d$ SCFTs not accounted earlier:
\be
%
\ee

Sending two blowups from left surface to right surface in $m=5$ version of (\ref{sp0su2Lm}) changes the middle surface to $\bF_0$ in which we can exchange $e$ and $f$ to rewrite the geometry in the following frame
\be
%
\ee
We can take the remaining blowup onto the middle surface and back to the left surface to rewrite the above geometry as
\be
%
\ee
Removing the blowup on the left surface and integrating out $p$ blowups from the right surface, we obtain the following series of $5d$ SCFTs
\be
%
\ee
From the $p=2$ geometry above, we can integrate out the $e$ curve of the middle surface towards right to obtain the $5d$ SCFT
\be
%
\ee

Similarly, we can obtain the following $5d$ SCFTs
\be
%
\ee

\subsection{$M=10$}
The KK theory (\ref{sp1su1}) has the geometry \cite{Bhardwaj:2018vuu,Bhardwaj:2019fzv}
\be\label{sp1su16d}
%
\ee
which can be rewritten as
\be
%
\ee
Flopping $x\sim y$ in the right surface, we get
\be
%
\ee
Sending all the blowups except $y$ from the left surface to the right surface, we obtain
\be
%
\ee
Let us rewrite the above by performing a left-right exchange
\be\label{sp1su1KK}
%
\ee
which is our desired frame to express the geometry (\ref{sp1su16d}). It can be checked that every time we integrate out a $-1$ curve from the middle surface such that the resulting geometry has only a single gluing curve between the left and right surfaces, then that geometry has already been discussed in Section \ref{su3S}. Thus, we can focus our attention exclusively on the geometries that contain two gluing curves between the left and the right surfaces.

We can integrate out $m$ blowups from the left surface now to obtain the following series of $5d$ SCFTs
\be
%
\ee

It can be checked that removing $e$ curve from one of the gluing curves in the right surface of (\ref{sp1su1KK}) and then integrating out the blowups does not lead to any new $5d$ SCFTs. However, removing $e$ curve from one of the gluing curves on the right surface and one of the gluing curves on the middle surface leads to the following $5d$ SCFT
\be
%
\ee

The KK theory (\ref{sp0su1su1}) has the following geometry \cite{Bhardwaj:2018vuu,Bhardwaj:2019fzv}
\be\label{sp0su1su16d}
%
\ee
which can be rewritten as
\be
%
\ee
Let us now flop $x\sim y$ in the middle surface. Since these $-1$ curves intersect the gluing curves for both the left and the right surfaces, the flop creates blowups in both the left and the right surfaces, but these blowups must be identified with each other as shown below
\be
%
\ee
Flopping $x\sim y$ in the right surface, we obtain
\be
%
\ee
Applying $\cI_0^{-1}$ using $y$ on the left surface, and then exchanging $e$ with $f$ on the left surface, we get
\be
%
\ee
Cyclically rotating the surfaces by one unit, we prefer to write the above as
\be
%
\ee
Moving the blowups on the middle surface to left surface through the right surface, we obtain
\be\label{sp0su1su1KK}
%
\ee
which is our preferred frame for the geometry (\ref{sp0su1su16d}).

Now integrating out $m$ blowups from the left surface in (\ref{sp0su1su1KK}), we obtain the following series of $5d$ SCFTs
\be\label{sp0su1su1Ltm}
%
\ee

The only other $5d$ SCFT arising from (\ref{sp0su1su1KK}) and not accounted earlier in this paper is obtained by removing $e$ curve from the gluing curve for the left surface in the middle surface and removing some blowups
\be
%
\ee

\subsection{$M=8$}
The KK theory (\ref{su4T1}) has the geometry \cite{Bhardwaj:2019fzv}
\be\label{su4T16d}
%
\ee
where the labels at the ends of the edge between the left and the right surface are displayed as $f$ and $f-x_i-y_i$ respectively while leaving the value of $i$ unspecified. This notation means that  a copy of $f$ in the left surface is glued to $f-x_i-y_i$ in the right surface for each value of $i$. Hence, the edge between the left and right surfaces carries the label $6$ in its middle denoting the fact that there are six gluings between the two surfaces. We will continue using this notation in what follows.

Flopping $z\sim w$ in the right surface leads to
\be
%
\ee
which can be rewritten (upto flops) as
\be
%
\ee
where the notation dictates that, for each $i$, the $-1$ curve $f-x_i$ in the left surface is glued to the $-1$ curve $f-x_i$ in the right surface. Flopping three of these $f-x_i$ and the curve $x$ in the middle surface, we obtain
\be
%
\ee
which can be rewritten as
\be
%
\ee
Moving blowups $x_i$ living on the middle surface and the blowup $y$ living on the right surface onto the left surface, we obtain
\be\label{su4T1KK}
%
\ee
which is the our desired frame for (\ref{su4T16d}). It is possible to flow to geometries that carry two gluing curves between the left and right surfaces, but we claim that all such geometries have been accounted earlier in previous subsection. The reader can easily check this claim. Thus, we focus our attention only on those geometries that carry three gluing curves between the left and the right surfaces.

Integrating out $m$ blowups from the left surface in (\ref{su4T1KK}) gives rise to
\be
%
\ee
Integrating out a blowup from the right surface and $m$ blowups from the left surface gives rise to
\be
%
\ee
The final case involves integrating out two blowups each from the left and the right surfaces
\be
%
\ee

The KK theory (\ref{sp0su3T2}) has the following geometry \cite{Bhardwaj:2019fzv}
\be
%
\ee
which can be shown to be flop equivalent to
\be\label{sp0su3T2KK}
%
\ee
Integrating out $m$ blowups from the left gives us the following series of $5d$ SCFTs
\be
%
\ee
Integrating out a blowup from right and $m$ blowups from left gives the following $5d$ SCFTs
\be\label{sp0su3T2LmR}
%
\ee
It can be shown that other ways of integrating out $-1$ curves discussed earlier do not give rise to any new $5d$ SCFTs. However, a new way of integrating out $-1$ curves opens up in this theory. First, sending one blowup from the left surface to the right surface in (\ref{sp0su3T2KK}) results in
\be
%
\ee
which is isomorphic to
\be
%
\ee
Now exchanging $e$ and $f$ on the right surface yields
\be
%
\ee
which after flopping $x_3$ in the left surface and applying isomorphisms can be written as
\be
%
\ee
Now we can integrate out $m$ blowups from the left surface and $p$ blowups from the right surface to obtain the following $5d$ SCFTs
\be
%
\ee

The KK theory (\ref{g21}) for $k=1$ has the geometry \cite{Bhardwaj:2018yhy,Bhardwaj:2018vuu,Bhardwaj:2019fzv}
\be
%
\ee
which is equivalent to
\be
%
\ee
and only produces the following new $5d$ SCFTs
\be
%
\ee

\subsection{$M=6$}
The KK theory (\ref{su3}) for $k=2$ has the geometry \cite{Bhardwaj:2018yhy,Bhardwaj:2018vuu,Bhardwaj:2019fzv}
\be
%
\ee
Sending two blowups onto the right surface, we can write the above geometry as
\be
%
\ee
We can send one of the blowups on the right surface onto the left surface to get
\be
%
\ee
Blowing down the gluing curve $f-y$ in the left surface (which also blows down the gluing curve $f-x$ in the right surface) leads to the following $5d$ SCFT
\be
%
\ee
which is same as (\ref{sp2L3LtR3Rt}). There are no other RG flows possible for this KK theory.

The KK theory (\ref{su5T}) has the geometry \cite{Bhardwaj:2019fzv}
\be
%
\ee
which is equivalent to
\be
%
\ee
from which we can obtain the following rank three $5d$ SCFTs
\be\label{su5TLm}
%
\ee
At $m=3$, we can also integrate out $h-\sum x_i$ in the left surface to obtain
\be
%
\ee

\subsection{$M=5$}
The KK theory (\ref{g21}) for $k=2$ has the geometry \cite{Bhardwaj:2018yhy,Bhardwaj:2018vuu,Bhardwaj:2019fzv}
\be
%
\ee
which is equivalent to
\be
%
\ee
This frame allows us to easily integrate out $-1$ curves and it can be seen that, in doing so, no $5d$ SCFTs can be obtained which have not been discussed above.

The KK theory (\ref{su4T2}) has geometry \cite{Bhardwaj:2019fzv}
\be
%
\ee
and can be seen to not give rise to any new $5d$ SCFTs.

The KK theory (\ref{sp0su3T3}) has the following geometry \cite{Bhardwaj:2019fzv}
\be
%
\ee
and can be seen not to lead to any $5d$ SCFTs not discussed earlier in this paper.

The KK theory (\ref{su3sp0T}) has the geometry \cite{Bhardwaj:2019fzv}
\be
%
\ee
We can integrate out blowups sitting on the left surface to obtain the following $5d$ SCFTs
\be
%
\ee

\subsection{$M=4$}
The KK theory (\ref{so8}) for $k=1$ has the geometry \cite{Bhardwaj:2019fzv}
\be
%
\ee
and gives rise to the SCFTs
\be
%
\ee

The KK theory (\ref{su3L}) has the geometry \cite{Bhardwaj:2019fzv}
\be
%
\ee
Moving one blowup from the left surface to the middle surface, then to the right surface and then back to the left surface,we can write the above as
\be\label{su3LKK}
%
\ee
Removing $m$ blowups from the left surface, we obtain the following $5d$ SCFTs
\be\label{su3LLm}
%
\ee
We propose that the fundamental BPS particles arising from the middle surface are associated to the curves $e+f$, $e+2f-x-2y$, $e+2f-z$, $f-x-z=f-y-z$, $x=y$ and $z$ living in the middle surface.

Now, moving the blowup $z$ on the middle surface and one blowup on the left surface onto the right surface in (\ref{su3LKK}), we obtain
\be
%
\ee
which can be rewritten, by performing isomorphisms on the right surface, as
\be
%
\ee
Exchanging $e$ and $f$ on the right surface, we obtain
\be
%
\ee
Now we can integrate out blowups (living on the left surface) from both the left and the right surface. The only possibility giving rise to a new $5d$ SCFT not discussed above is obtained by integrating out one blowup from the left and one blowup from the right:
\be
%
\ee
We propose that the fundamental BPS particles arising from the middle surface are associated to the curves $e+f$, $e+4f-x-2y$, $f-x=f-y$ and $x=y$ living in the middle surface. Removing the self-gluing leads to $5d$ SCFTs already discussed before.

\subsection{$M=3$}
The KK theory (\ref{so8}) for $k=2$ has the geometry \cite{Bhardwaj:2019fzv}
\be
%
\ee
which is flop equivalent to
\be
%
\ee
from which we can obtain the $5d$ SCFT described by the geometry
\be
%
\ee

The KK theory (\ref{su2su1}) is described by the geometry \cite{Bhardwaj:2018yhy,Bhardwaj:2018vuu,Bhardwaj:2019fzv}
\be
%
\ee
Our desired frame for this geometry is
\be
%
\ee
from which it is easily shown that no new $5d$ SCFTs arise. For example, integrating out $f-y$ in middle surface and $x$ in left surface leads to the $5d$ SCFT
\be
%
\ee
which has already been discussed in Section \ref{su3S}. To see this, notice that flopping $x$ on the middle surface leads to
\be
%
\ee
which is clearly flop equivalent to (\ref{su3LmM5R}) for $m=4$.

\subsection{$M=2$}
The KK theory (\ref{g21}) for $k=3$ has the geometry \cite{Bhardwaj:2018yhy,Bhardwaj:2018vuu,Bhardwaj:2019fzv}
\be
%
\ee
and produces no new $5d$ SCFTs.

Similarly, the KK theory (\ref{so8}) for $k=3$ is associated the geometry \cite{Bhardwaj:2019fzv}
\be
%
\ee
which can be written in our desired frame as
\be
%
\ee
The only possible blowdown is the blowdown of a blowup in the left surface, which induces a flop transition leading to
\be
%
\ee
Removing the blowup from the middle surface leads to (\ref{sp0su1su1Ltm}) for $m=9$ and hence we don't obtain any new $5d$ SCFT.

The KK theory (\ref{su1su1su1}) is described by the geometry \cite{Bhardwaj:2018vuu,Bhardwaj:2019fzv}
\be
%
\ee
from which we can blowdown $x$ to obtain a $5d$ SCFT discussed in Section \ref{sp2S}. Thus, there are no new $5d$ SCFTs arising from this KK theory.

Now, consider the KK theory (\ref{su1su1su1TL}), which is described by the following geometry easily seen to be equivalent to the geometry assigned to the KK theory by \cite{Bhardwaj:2019fzv}
\be
%
\ee
Flopping $x\sim y$ in the left surface and $x\sim y$ in the right surface, we obtain
\be
%
\ee
Adding some decoupled states onto the right surface and flopping $x\sim y$ in the middle surface, we obtain
\be
%
\ee
where the blowup on the left surface must glue to both $x$ and $y$ in the right surface, and to account for that the blowup on the left surface appears twice in the gluing rules with the right surface. Flopping $f-x$ and $f-y$ in the right surface (which flops $f-y$ in the middle surface as well) leads to
\be
%
\ee
Now flop $f-x$ in the left surface which is glued to $f-x$ in the middle surface
\be
%
\ee
Flopping $f-x$ in the left surface (and other curves in the threefold glued to it) leads to
\be
%
\ee
Now flopping $x\sim y$ in the right surface leads to
\be
%
\ee
Finally, flopping $x\sim y$ in the right surface leads us to our desired frame for this KK theory
\be
%
\ee
which indeed, as expected \cite{Tachikawa:2011ch}, describes the $5d$ gauge theory having gauge algebra $\so(7)$ and a hyper in adjoint. See \cite{Bhardwaj:2019ngx} for more details on the correspondence between geometries and $5d$ gauge theory descriptions. Removing the blowup from the middle surface integrates out the adjoint matter leading to the geometry for pure $\cN=1$ $\so(7)$ theory in $5d$, but this was already accounted as $m=5$ case of (\ref{su5TLm}). Thus there are no new $5d$ SCFTs arising from this KK theory.

For the KK theory (\ref{su2su1T3}), the associated geometry is \cite{Bhardwaj:2019fzv}
\be

\ee
By doing similar flop gymnastics as above, it can be shown that it is equivalent to the following geometry
\be
%
\ee
from which it can be seen that no new flows can be produced.

For the KK theory (\ref{su2su1T2}), the associated geometry is \cite{Bhardwaj:2019fzv}
\be
%
\ee
and turns out to be equivalent to
\be
%
\ee
from which the only flow arises by blowing down the blowup $y$ in the left surface leading to the $5d$ SCFT described by the geometry
\be
%
\ee
which was already found earlier (\ref{su3L5MR5}).

The KK theory (\ref{su2su1L}) is described by \cite{Bhardwaj:2019fzv}
\be
%
\ee
which is equivalent to
\be\label{su2su1LKK}
%
\ee
Removing the blowup on the left surface leads to a $5d$ SCFT described by
\be
%
\ee
We propose that the fundamental BPS particles arising from the middle surface are identified with the curves $e$, $h-x-2y$, $f-x=f-y$ and $x=y$ living in the middle surface. We can also move the blowup on the left surface of (\ref{su2su1LKK}) to the middle surface, then to the right surface and then back to the left surface to obtain
\be
%
\ee
Flopping $f-x$ in the left surface and $h-x-y$ in the right surface, we obtain
\be
%
\ee
which is the same as $m=3$ case of (\ref{su3LLm}) upto decoupled states.

The KK theory (\ref{su1su1su1L}) is described by the geometry \cite{Bhardwaj:2019fzv}
\be
%
\ee
which can be rewritten as
\be
%
\ee
Flopping $x\sim y$ in the middle and right surfaces leads to
\be\label{ana}
%
\ee
which is isomorphic to
\be
%
\ee
Rotating the above configuration clockwise by one unit, we get
\be
%
\ee
We claim that the blowup $z$ on the middle surface can actually be integrated out from the left surface. To see this, let us exchange $e,f$ on the right and left surfaces to rewrite the above geometry as
\be
%
\ee
Now it is clear that the blowup $z$ on the middle surface can be taken to right surface and then to the left surface. Performing this action leads to
\be
%
\ee
which can be rewritten to produce our desired frame for this KK theory
\be
%
\ee
Removing the extra blowup on the left surface not participating in any of the gluing curves leads to the following $5d$ SCFT
\be
%
\ee
We propose that the fundamental BPS particles arising from the middle surface are identified with the curves $e$, $h+f-x-2y$, $f-x=f-y$ and $x=y$ living in the middle surface.

The KK theory (\ref{su1su1su1L}) is described by the geometry \cite{Bhardwaj:2019fzv}
\be
%
\ee
which is flop equivalent to
\be
%
\ee
Notice that the above configuration is analogous to the configuration (\ref{ana}), but it does not lead to any flows, though the configuration (\ref{ana}) does. There are thus no new $5d$ SCFTs obtained from this KK theory.

\subsection{$M=1$}
The KK theory (\ref{so8}) for $k=4$ has the geometry \cite{Bhardwaj:2019fzv}
\be
%
\ee
which has no $-1$ curves and hence no RG flows.

The KK theory (\ref{su3}) for $k=3$ has the geometry \cite{DelZotto:2017pti,Bhardwaj:2018yhy,Bhardwaj:2018vuu,Bhardwaj:2019fzv}
\be
%
\ee
The only $-1$ curve is the $e$ curve in each surface which are all glued to each other. Integrating it out leads to three copies of the rank one SCFT given by (\ref{sp09}), but not to a rank three SCFT.

\section*{Acknowledgements}
The author thanks Hee-Cheol Kim and Gabi Zafrir for useful discussions. This work is supported by NSF grant PHY-1719924.

\bibliographystyle{ytphys}
\let\bbb\bibitem\def\bibitem{\itemsep4pt\bbb}
\bibliography{ref}

\end{document}